\documentclass[12pt,cite,epsfig]{article}
\usepackage{times,epsfig}
\newcommand{\nc}{\newcommand}
\nc{\bb}{\bibitem}
\nc{\be}{\begin{equation}}
\nc{\ee}{\end{equation}}
\nc{\pa}{\partial}
\nc{\parsym} {\stackrel{\leftrightarrow}{\pa}}
\nc{\ra}{\rightarrow}
\nc{\la}{\leftarrow}
\nc{\etp}{{\eta^\prime}}
\nc{\omg}{\omega}
\nc{\ggam}{\gamma \gamma}
\nc{\gam}{\gamma }
\nc{\I}{{\rm i} }
\nc{\beas}{\begin{eqnarray*}}
\nc{\eeas}{\end{eqnarray*}}
\nc{\ba}{\begin{eqnarray}}
\nc{\ea}{\end{eqnarray}}
\nc{\non}{\nonumber}
\nc{\mv}{{\rm MeV }}
\nc{\second}{{\prime\prime}}
\def\hhht{\rule[ 0.mm]{0.mm}{6.mm}}

\def\hhhc{\rule[-3.mm]{0.mm}{3.mm}}
\def\hhhd{\rule[-3.mm]{0.mm}{2.mm}}
\def\hhhe{\rule[-3.mm]{0.mm}{4.mm}}

\def\hhhv{\rule[-3.mm]{0.mm}{9.mm}}
\def\hhhw{\rule[-1.mm]{0.mm}{5.mm}}

\newcommand{\amuh}{a_\mu^{\rm had}}

\newcommand{\gv}{\rm GeV }
\newcommand{\D}{\rm d}

\renewcommand{\D}{{\rm d} }
\nc{\E}{{\rm e} }
\nc{\cF}{{\cal F} }
\nc{\cM}{{\cal M} }

\begin{document}
\begin{titlepage}
\vbox{~~~ \\
                                  \null \hfill \small{LPNHE/2015--01}\\
                                   \null \hfill \small{HU-EP--15/32}\\
                                   \null \hfill \small{DESY 15--116}\\
				       
\title{ Muon $g-2$ Estimates~:
 \\ Can One Trust Effective Lagrangians and Global Fits?
   }
\author{
M.~Benayoun$^a$, P.~David$^{a,b}$, L.~DelBuono$^a$, F.~Jegerlehner$^{c,d}$ \\
\small{$^a$ LPNHE des Universit\'es Paris VI et Paris VII, IN2P3/CNRS, F--75252 Paris, France }\\
\small{$^b$ LIED, Universit\'e Paris-Diderot/CNRS UMR 8236, F--75013 PARIS, France  } \\
\small{$^c$ Humboldt--Universit\"at zu Berlin, Institut f\"ur Physik, Newtonstrasse 15, D--12489 Berlin,
Germany }\\
\small{$^d$ Deutsches  Elektronen--Synchrotron (DESY), Platanenallee 6, D--15738 Zeuthen, Germany}
}
\date{\today}
\maketitle
\begin{abstract}
Previous studies have shown that the  Hidden Local
Symmetry (HLS) Model, supplied with appropriate
symmetry breaking mechanisms, provides an Effective Lagrangian
(BHLS) which encompasses a large number of processes within a unified
framework. Based on it, a global fit procedure 
allows for a simultaneous description of the $e^+ e^-$
annihilation  into 6 final states -- $\pi^+\pi^-$, $\pi^0\gamma$, 
$\eta \gamma$, $\pi^+\pi^-\pi^0$, $K^+K^-$, $K_L K_S$ --
and includes the dipion spectrum in the $\tau$ decay and some more
light meson decay partial widths. The  contribution 
to the muon anomalous magnetic moment  $a_\mu^{\rm th}$ of these
annihilation channels over the range of validity  of the HLS model 
(up to 1.05 GeV) 
is found much improved in comparison to the standard approach of
integrating the measured spectra directly.
However, because most spectra for the annihilation process 
$e^+e^- \to \pi^+\pi^-$ undergo overall scale uncertainties
which dominate the other sources, one may suspect some bias in
the dipion contribution to $a_\mu^{\rm th}$, which could question the reliability
of the global fit method.
However, an iterated global fit
algorithm, shown  to lead to unbiased results by a Monte Carlo study,
is defined and applied succesfully to the $e^+e^- \to \pi^+\pi^-$ 
data samples from CMD2, SND, KLOE, BaBar and BESSIII.
The iterated fit solution is shown to further improve the prediction for $a_\mu$, 
which we find to deviate from
its experimental value above the $4\sigma$ level.
The contribution to $a_\mu$
of the $\pi^+\pi^-$ intermediate state up to 1.05 GeV  has an uncertainty
about 3 times smaller than the corresponding usual estimate.
 Therefore,
global fit techniques are shown to work and lead to improved unbiased results.
\end{abstract}
}
\end{titlepage}

\section{Introduction}
\label{introduction}
\indent \indent As is well known, the Standard Model is the gauge theory which covers 
 the realm of weak, electromagnetic and strong interactions among
quarks, leptons and the various gauge bosons (gluons, photons, $W^\pm$, $Z^0$).
In energy regions where perturbative methods apply, the  Standard Model (SM) allows to yield
precise estimates for several physical effects, sometimes with accuracies
of the order of a few 10$^{-12}$. In contrast, in energy regions where the non--perturbative 
regime of QCD is involved, getting similar precision  may become challenging.
This is the case for the low energy part of the photon hadronic vacuum polarization (HVP);
this HVP plays a crucial role in determining the theoretical value for the muon 
anomalous moment $a_\mu$, one of the best measured  particle properties. 
 
Fortunately, getting precise estimates in the low energy hadron SM sector is not completely
out of reach as exemplified by the Chiral Perturbation Theory (ChPT)\cite{GL1,GL2} 
which is rigorously the low energy limit of QCD,  valid
 up to  400 $\div$ 500 MeV but lets the resonance region  
 outside  its scope. Lattice QCD (LQCD) is also a
promising method  under rapid development which already allows to 
perform precise computations at low (and very low) energies
\cite{Colangelo_final}.
Interesting LQCD estimates for the  HVP's of the three leptons have already been 
produced  \cite{Lattgm2-1,Lattgm2-2}  which clearly show 
that LQCD reaches results in accord with  expectations; this is especially 
striking for $a_\mu$ with, however,  still unsatisfactory uncertainties \cite{Lattgm2-1}.

So, much progress remains to be done before LQCD evaluations
can compete with the accuracy of the experimental measurements already
 available \cite{BNL,BNL2} or, {\it a fortiori}, with those expected in a near 
 future at Fermilab \cite{LeeRoberts,Fermilab_gm2} 
 or, slightly later,  at J--PARC \cite{Iinuma}. Since lattice QCD is intrinsically 
 an Euclidean approach, it is intrinsically unable 
 to account for the existing rich amount of low energy hadronic data in the 
 non-perturbative time-like region, {\it i.e.}
from thresholds to 2 $\div$ 3 GeV. Therefore, other methods, able to encompass large
fractions of the physics from this important energy region, are valuable.

A natural approach to this issue is provided by  Effective Lagrangians  
which cover the resonance region. Such Effective Lagrangians should be
constructed so as to  preserve the symmetry properties of QCD as already done by
standard ChPT, however only valid  up to the $\eta$ mass region. As it
includes meson resonances, the Resonance Chiral Perturbation Theory  (R$\chi$PT) \cite{Ecker2}
is an appropriate framework to study  $e^+e^- $ annihilations from their respective
thresholds up to the intermediate energy region.

It has been proven  \cite{Ecker1} that the coupling constants occuring 
at order $p^4$ in ChPT are saturated by low lying meson resonances of 
various kinds as soon as they can contribute. This emphasizes  the role of the fundamental 
vector meson nonet (V) and confirms the relevance  of the Vector Meson  
Dominance (VMD) concept  in low energy physics. 

On the other hand, it has been proven \cite{Ecker2} that the Hidden Local Symmetry (HLS) 
Model \cite{HLSRef} and R$\chi$PT  are equivalent provided consistency with the QCD asymptotic 
behavior is incorporated. It thus follows that the HLS model is also a 
motivated and constraining QCD rooted framework. As the original HLS Model only deals with
the lowest mass resonances, it provides a framework for the $e^+e^- $ annihilations 
naturally bounded by the $\phi$ mass region -- {\it i.e.}  up to $\simeq 1.05$ GeV.

The non--anomalous \cite{HLSOrigin} and anomalous \cite{FKTUY} 
sectors of the HLS Model open a  wide scope and  can deal 
with a large corpus of  physics processes in a unified way. However, as such, 
HLS cannot precisely reach   the numerical precision requested 
by the wide ensemble of high statistics  data samples collected by 
several sophisticated 
experiments on several annihilation channels.
In order to achieve such a program,  the HLS Lagrangian  must be supplied
with appropriate symmetry breaking mechanisms not present in its original
formulation \cite{HLSRef}.

This was soon recognized by the  HLS Model authors who first proposed the mechanism 
to break SU(3) symmetry \cite{BKY} named BKY according to its author
names. Its success was illustrated by several phenomenological studies
based on the BKY breaking scheme \cite{BGP,BGPbis,Heath}.  It was also soon extended 
to SU(2)/Isospin symmetry breaking \cite{Hashimoto}. 
However, in order to account simultaneously for all the radiative decays of 
the light flavor mesons, the additional step of breaking the nonet symmetry for light 
pseudoscalar mesons was required; based on the heuristic formulation of
the $VP\gamma$ couplings by O'Donnell \cite{ODonnell} which includes 
nonet symmetry breaking in the pseudoscalar ($P$) sector in a specific way, a global
and successfull account of all $VP\gamma$ and $P\gamma\gamma$ couplings has been reached 
\cite{rad}. The BKY SU(3) breaking and this nonet symmetry breaking included
within the HLS Model was shown  \cite{WZWChPT} to meet the requirements of Extended
Chiral Perturbation Theory \cite{leutw,leutwb}. Finally, introducing the physical
vector meson fields as the eigenstates of the loop modified vector meson mass matrix
provided a mixing scheme of the $\rho^0-\omg-\phi$ system which together with
the $V-\gamma$ loop transitions implied by the HLS model at one 
loop\footnote{See also  \cite{Fred11} where the 
role of the $\rho^0-\gamma$ mixing is especially emphasized.} leads to a  
satisfactory solution \cite{taupaper} of the long--standing $\tau- e^+e^-$ puzzle  
\cite{DavierPrevious1,DavierHoecker,Eidelman,Fred09}.

Therefore, the approach just sketched is a $global$  framework aiming at
accounting for the
largest possible ensemble of data spectra collected in the largest possible number 
of low energy physics channels. As this $global$ model is an Effective Lagrangian
constructed from the ($P$ and $V$) fields relevant in the low energy regime  of QCD
and because it is 
consistent with the symmetries of QCD, one naturally expects their 
low energy results to be consistent with the Standard Model.

It was then shown, that the Effective Lagrangian constructed from the
original HLS Model supplemented with the breaking schemes 
listed above was able to provide a satisfactory 
simultaneous description of the $e^+e^-$ annihilations into the
 $\pi^+\pi^-$, $\pi^0\gamma$, $\eta \gamma$, $\pi^+\pi^-\pi^0$
  final states and of the dipion spectrum in the decay
of the  $\tau$ lepton \cite{ExtMod1,ExtMod2}. This tended to indicate that
the  $\tau- e^+e^-$ puzzle just referred  was related to an incomplete incorporation 
of isospin symmetry breaking effects within models. 

Slightly extending these breaking schemes,  one is led to the Broken HLS (BHLS) Model 
\cite{ExtMod3} which  provides a fully consistent picture of
 all examined $e^+e^-$ annihilation cross 
 sections\footnote{Specifically the 6  $e^+e^-$ annihilation channels to 
 $\pi^+\pi^-$, $\pi^0\gamma$, $\eta \gamma$, $\pi^+\pi^-\pi^0$, $K^+K^-$,
$K^0\overline{K^0}$, each from its threshold up to 1.05 GeV, {\it i.e.}
including the $\phi$ signal region.}, the $\tau$ dipion spectrum 
 and, additionally,  some light meson decay information
with a limited number of free parameters to be extracted from data. 
An interesting outcome of the BHLS based fit framework was a novel
evaluation of the dominating low energy piece of the HVP, leading to
an improved estimate of the muon anomalous magnetic moment at more
than $4\sigma$ from its measured value\footnote{One should note that
the BHLS evaluation for the muon HVP is the closest to the central value
preferred by the Lattice QCD study \cite{Lattgm2-1}. }
 \cite{BNL,BNL2}.
  
\vspace{0.5cm}

Introducing the  dipion spectra collected in the ISR mode 
 confirmed that the muon $g-2$ departs from expectation by more than $4\sigma$
 \cite{ExtMod4} . One should note that the 
 high statistics ISR dipion spectra recently published by the 
 KLOE \cite{KLOE08,KLOE10,KLOE12}, BaBar \cite{BaBar,BaBar2} and BESSIII 
 \cite{BESS-III} Collaborations 
 are strongly dominated by overall scale ({\it i.e.}
 normalization) uncertainties; additionally  the KLOE and BaBar normalization
 uncertainties are energy dependent.  However, sizeable overall scale uncertainties
raise an important issue related with their possibly biasing the physics
quantity values extracted from their spectra. This issue has been identified in the
reference work of G. D'Agostini \cite{D'Agostini} where a very simple case 
is proposed  which illustrates that biasing effects can be 
dramatic\footnote{The issue raised by G. D'Agostini in this paper has also been met
formerly
in the context of Nuclear Physics where it is referred to as the "Peelle's Pertinent Puzzle"
(PPP) \cite{Peelle} which is examined thoroughly in \cite{Chiba}.}. 
Of course,
for a key quantity like the muon $g-2$, the problem should be explored and
possible biases identified and fixed. The way 
out is already mentioned in \cite{D'Agostini} and further emphasized
in other studies
\cite{Blobel_2003,Blobel_2006, Chiba}; the exact solution exhibits
a delicate issue  as the removal of the bias on some quantity supposes
to know its exact value. Nevertheless, as already suggested in \cite{D'Agostini}
and emphasized in  \cite{Chiba},
iterative methods can be defined and are expected to be bias free; this has been applied
successfully to the derivation of parton density functions in \cite{Ball}. 

\vspace{0.5cm}

The present work mostly aims at reexamining the results provided in \cite{ExtMod3,ExtMod4}
concerning the muon HVP using an appropriately  defined iterative fit method 
adapted to the dealing  with form factors or
cross sections in such a way that fit results and derived quantities -- like
the HVP, but not only -- could be ascertained to be bias free. In this way, one can 
positively answer the question raised in the title of this study at the methodological 
level.

 The real issue of the physics model dependence can only be answered
by having at disposal results derived from several independent model frameworks, 
all successfully (undoubtfully) accounting for the largest possible corpus of data.
Indeed, the physics
correlations relating the {\it different}  physics processes encompassed within a
given framework cannot easily accomodate a model--independent approach. 
Moreover, several issues within the global fit approach are related with the
formulation of isospin symmetry breaking (IB) which can hardly be made model independent,
especially in a global framework.

The paper is organized as follows. In Section \ref{EF_method}, one briefly reminds 
the concern of using Effective Lagrangian global frameworks in order to strengthen
the constraints on the parameters to be derived from global fits. As our 
HLS Lagrangian framework has a range limited upward to 1.05 GeV, the brief Section
\ref{gM2} reminds how the full HVP is derived from fit results and from additional information.

Section \ref{GF_Methods} is, actually,  the center piece of the present paper as its purpose
is to define the fit method when one should deal with samples affected by strong overall scale
 uncertainties. This firstly turns out to 
precisely define the $\chi^2$ functions to be minimized, depending on the specific properties
of the spectra considered and, secondly, to set up and justify the iterative procedure we 
propose\footnote{After completion of this
work, we found that \cite{Ball_2} applies a method similar to ours to derive
unbiased parton density functions from various kinds of measured spectra.}.  
Subsection \ref{scale_err} puts
special emphasis on the specific $\chi^2$ function associated with samples affected
by overall scale uncertainties besides a more usual experimental error matrix.
The iterative fit procedure  to deal with biases is formulated therein. 

Most of the ISR data samples exhibit $s$--dependent overall scale uncertainties, which
are certainly a novel feature in our field; Subsection  \ref{scale_err_2} defines an 
appropriate $\chi^2$ function suitable for such a case. 
Finally, Subsection \ref{scale_err_3} reports on the main features of 
the iterative global fit method when fitting sets of data samples containing samples with
overall scale uncertainties of various magnitudes compared to statistical errors.
The  conclusions reported here rely  on a Monte Carlo study outlined and illustrated
in Appendix \ref{MC_test}.

Section \ref{Status} reminds the data samples used within the BHLS procedure and
reports for a (minor) correction affecting the amplitudes for the
annihilation channels $\pi^0\gam$ and $\eta\gam$. Section  \ref{nsk_iter} 
reports on the updated results of the fits performed using only the scan data
and discarding all ISR data samples; the effects of the iterative method is
illustrated here and it is shown that the needed number of  iterations
in the global fit procedure does not exceed 1. The more general running
is the subject of Section  \ref{global_iter} where updated results are given to
correct for coding bugs affecting some of the numbers given in our \cite{ExtMod3,ExtMod4}.
The properties of the recently published KLOE12   \cite{KLOE12} and BESSIII \cite{BESS-III} 
data samples are examined.
The evaluation of the muon $g-2$ based on the iterated fits of various combinations
of data samples is the subject  of Section  \ref{global_iter_hvp}, where the HVP slope
at $s=0$ is also computed within BHLS and compared to its value directly derived from
experimental data.
 Finally, Section \ref{conclusion} is devoted to conclusions and remarks. 

\section{Effective Lagrangian Frameworks And Global Fits}
\label{EF_method}
\indent \indent
As reminded in the Introduction, it is a common approach to rely on the
Effective Lagrangian (EL) method to cover the low energy region  where
QCD exhibits its non--perturbative regime and where the quark and gluon degrees of freedom 
are replaced by hadron fields. Each EL of practical use generally depends on  parameters originating
from the starting Lagrangians (like the pion decay constant $f_\pi$ or 
the universal vector coupling $g$) and  on parameters generated by
the unavoidable symmetry breaking effects (like quark mass differences);
all such parameters are determined from data with various precisions.
 
Needless to say that any  (broken) Effective Lagrangian provides amplitudes expected 
to account simultaneously for several different processes. 
 This has a trivial consequence which, nevertheless, deserves to be stressed~: 
{\it All the Effective Lagrangians predict physics correlations among the different 
physical  processes they can encompass~:  ${\cal H} \equiv \{H_i, i=1, \cdots~p\}$}. 

Therefore, having plugged from start the  physics correlations inside
the (broken) Lagrangian, the amplitudes derived herefrom should allow
for  a global, simultaneous and constrained fit of all available data samples covering
all the channels in ${\cal H}$. Provided the global fit is clearly successful, 
the parameter central values and uncertainties returned can be considered as the
 optimal values accounting for all the processes in ${\cal H}$ $simultaneously$. 
Therefore, one can consider that the fit information -- parameter central values and 
error covariance matrix
--  exhausts the experimental information contained in the data samples covering 
all the processes in ${\cal H}$. 

\vspace{0.5cm}
From now on, one specializes to the Broken HLS (BHLS) model as defined
and used in \cite{ExtMod3}. All data samples used in the global fit procedure 
defined in this paper have already been listed  and analyzed in this 
Reference\footnote{ \label{samples_used}
 Concerning the non--$\pi^+ \pi^-$ channels, all 
existing data samples collected in scan mode at Novosibirsk are considered. 
 The $\tau$ data included in the global fit procedure
are the samples collected by  ALEPH \cite{Aleph}, CLEO
\cite{Cleo} and Belle \cite{Belle}.}; this will not be repeated here.
As for the $\pi^+ \pi^-$ annihilation
final state, which is a central piece of HVP studies, this Reference dealt with only 
the available scan data which are dominated 
by the samples from CMD2 \cite{CMD2-1998-1,CMD2-1998-2} and SND \cite{SND-1998}. 
The  samples collected in the ISR mode  by Babar \cite{DavierHoecker2}
as well as the former  KLOE data samples (KLOE08 \cite{KLOE08} and KLOE10 \cite{KLOE10})
have been considered in  \cite{ExtMod4}. Preliminary  results  including also 
the most recent KLOE sample (KLOE12)
\cite{KLOE12} have been given in \cite{BM_paris_2013,BM_roma2013}. The BESSIII spectrum
\cite{BESS-III},
published by mid of 2015,  is also included within our analysis.

\section{Estimating the Muon Non--Perturbative HVP}
\label{gM2}
\indent \indent 
The issue raised in this paper is whether Effective Lagrangian methods  really improve 
the evaluation of the dominating non--perturbative  part of the HVP \cite{ExtMod3,ExtMod4}
compared to a  direct integration of experimental data (see \cite{Fred11,DavierHoecker3,Teubner2}
for instance). As we are working within the original HLS framework \cite{HLSRef},
 what is discussed is  the HVP fraction  associated with the 
$\pi^+\pi^-$, $\pi^0\gamma$, $\eta \gamma$, $\pi^+\pi^-\pi^0$, $K^+K^-$,
$K^0\overline{K^0}$ intermediate states -- covered by BHLS -- up to $\simeq 1.05$ GeV;
this  represents more than 80\% of the total LO--HVP . 

Basically, the leading order (LO) non--perturbative QCD contribution to the muon HVP 
is estimated separately for each intermediate hadronic state $H_i$ via~:
\be
\displaystyle a_\mu(H_i)=\frac{1}{4 \pi^3}
\int^{s_{cut}}_{s_{H_i}} K(s) \sigma_{H_i}(s)
\label{Eq1}
\ee 
and the total non--pertubative HVP component is the sum of all the possible $a_\mu(H_i) $.
The function  $K(s)$ in Eq. (\ref{Eq1})
 is a known kernel \cite{Fred09} enhancing the threshold regions ($s_{H_i}$)
for any channel $H_i$ and  $\sigma_{H_i}(s)$ is the undressed 
cross section \footnote{Final state radiation (FSR) effects also contribute and
are estimated  as in \cite{Fred09}.}  for the  $e^+e^- \ra H_i$ annhilation;
 $s_{cut}$ is an energy limit above which  perturbative  expansions 
are supposed to become valid. BHLS permits to evaluate the 6
integrals   $\{a_\mu(H_i), i=1, \cdots~6\}$ up to $s_\phi \simeq 1.05 $ GeV.
As the  energy interval  $[s_\phi,s_{cut}]$ contribution to
$a_\mu(H_i)$ is  beyond the BHLS energy range of validity, 
 it is estimated using customary methods (like those defined in  
\cite{DavierHoecker3,Teubner2,Teubner}, for instance), as also the  full contributions 
of the channels outside the present BHLS scope, like the 4 pion final states. As already stated, these pieces 
represent altogether about 20\% of the muon LO--HVP contribution to $a_\mu$.

\vspace{0.5cm}

As can be checked by looking at the cross section formulae given in 
\cite{ExtMod3}, most parameters to be fitted 
 appear simultaneously in the 6 different cross sections 
 $\{\sigma_{H_i}(s), i=1, \cdots~6\}$
 and each annihilation channel $H_i$ comes in
with several experimental data samples\footnote{An 
experimental data sample is defined as  the measured
spectrum $m$  and all the uncertainties which affect it.}.  Therefore,
for instance, the data samples covering any of
the $\pi^0\gamma$, $\eta \gamma$, $\pi^+\pi^-\pi^0$,
$K^+K^-$, $K^0\overline{K^0}$ annihilation channels play as additional constraints 
on the $\pi^+\pi^-$ cross section and are treated
on the same footing than the $\pi^+\pi^-$ annihilation data themselves. On the other hand, the
constraints carried by the dipion $\tau$ decay spectrum data \cite{Aleph,Cleo,Belle} 
influence the fit and allow
to reduce the BHLS parameter uncertainties in a consistent way\footnote{So also do
the decay partial widths of the form $P \ra \gamma \gamma $ and
$V \ra P \gamma$ (or $\eta^\prime \ra \omg \gamma$) extracted from the Review of
Particle Properties (RPP) \cite{RPP2012} and implemented within BHLS.}. 
This explains why the global fit method is  expected to improve each $a_\mu(H_i)$ contribution
compared to more traditional methods -- those from \cite{Fred11,DavierHoecker3,Teubner2}
for instance --  as these ignore the inter--channel correlations revealed by
the BHLS Effective Lagrangian and validated by satisfactory global fits.
Of course, inter--channel correlations are a general feature of Effective Lagrangians, 
and not particular for the BHLS implementation.

As any method, the BHLS based global fit method carries specific
systematics which have been examined in great details in \cite{ExtMod4}.
It is worth remarking, to avoid ambiguities, that the isospin breaking effects 
specific of the $\tau$ dipion spectra are introduced in the dipion spectrum
 \cite{ExtMod4} as commonly done in the literature 
 \cite{Marciano,Cirigliano1,Cirigliano2,Cirigliano3,Mexico2,Mexico1,Mexico4,Ghozzi}
(see also \cite{Fred11});  
 they are totally independent of the isospin breaking
schemes involved in the BHLS Lagrangian and, actually, come supplementing 
these \cite{ExtMod4}. 

\section{Can One Trust Global Fit Results~?}
\label{GF_Methods}
\indent \indent The global fit method previously used in  
\cite{ExtMod3,ExtMod4} defines a so--called VMD strategy which can be
phrased in the following way~:

\begin{itemize}
 
 \item {\bf 1/} If the physics correlations  predicted by a given Effective  Lagrangian  Model
  are supported by the experimental data they encompass,
 they can be considered as exact {\it at the  accuracy level reported for the data}.
 
  \item {\bf 2/} Whenever the description -- global fit -- provided by
  a given Effective  Lagrangian  is satisfactory, 
  the model cross sections,  the fit parameter values and the
  parameter error covariance matrix  exhaust reliably the physics information contained
  in the fitted data samples.  
 \end{itemize}
 
 In the present case where the BHLS model is concerned, and focusing on
 the muon LO--HVP, Statement \# {\bf 2} means that the improvements for
 the 6 accessibles $a_\mu(H_i)$ derived from  Eq. (\ref{Eq1}) by integrating 
 from $s_{H_i}$ to 1.05 GeV/c are legitimately valid and conceptually supported.
 
 On the other hand, Statement \# {\bf 1} does not mean that  
the importance of the word "Effective" is forgotten, as clear from
 the italic sentence it carries~: Its validity might have to be 
 revised if the experimental context evolves towards a degraded 
 account of the  data\footnote{However, if an ensemble of
 data is $internally$  conflicting within a given Effective Lagrangian
   framework, as the fit results can be affected
 in an unpredictable way, some action has to be taken. The simplest solution
 is certainly to discard the faulty data samples; however,
 as suggested by \cite{Blobel_2006}, 
 a down--weighting of the outlier contributions to the
 minimized $\chi^2$ might also be considered. This could be a way to reconcile
 the preservation of the fit information quality with the use of all available samples.}. 
 
 \vspace{0.5cm}

Obviously, a VMD strategy heavily relies  on the statistical methods used to
 analyze and fit the data; thus, one should ascertain that all aspects of the data
 handling are taken into account as they should.  In particular, all  features of 
 the experimental uncertainties should be implemented canonically within the
 minimized global $\chi^2$  and in the fitting procedure. 
 Indeed, as  remarked in \cite{Blobel_2003,Fruehwirth}, incorrect fit results are 
 more frequently due to an incorrect dealing with the experimental errors 
 (and correlations) rather than to
  the minimization  procedure itself. 
 Therefore, special care is requested in dealing with experimental uncertainties
 and in choosing the appropriate $\chi^2$ expression adapted to each data sample.

 It is the purpose of this Section to address this issue and
 check whether the procedure defined in \cite{ExtMod3,ExtMod4} fulfills
 this statement;   this will lead us to complement the fitting procedure
 by an iterative method. 
 
 \subsection{The Basic $\chi^2$/Least Square Method}
\label{least_sq}
\indent \indent 
Usually, performing a  fit -- global or not -- requires to minimize
a  $\chi^2$ function\footnote{Which is a true $\chi^2$ if the errors are
gaussian.} 
relating  the differences between the
 measurements ($m=\{m_i,i=1,\cdots~n\}$) and the corresponding
 model (theoretical) expections ($M(\vec{a})=\{M_i(\vec{a}),i=1,\cdots~n\}$) 
 weighted by the  error covariance matrix $V$ provided together with
 the data spectrum. Leaving aside for now possible $global$ (additive or mutiplicative) 
 systematic  uncertainties,  the error matrix  $V$ provided by experimental groups gathers 
 the statistical  and systematic errors and, thus, is not necessarily diagonal.
 The vector  $\vec{a}$ denoting  the unknown internal model parameter list,  
  minimizing~:  
   \be
  \chi^2=[m-M(\vec{a})]^TV^{-1}[m-M(\vec{a})]
  \label{Eq2}
  \ee
  with respect to $\vec{a}$ 
allows to derive its optimum value $\vec{a}_0$. When several independent data samples
are to be treated simultaneously, the minimized $\chi^2$ is a sum of terms like Eq. (\ref{Eq2}),
one for each data sample.

As reminded in \cite{Blobel_2003}, if the model $M(\vec{a})$ is linear in the 
parameters\footnote{Actually, fitting is generally performed in the neighborhood 
of some given solution; this makes the linearity condition less constraining in practice. }
and if the error covariance matrix is  correct, the estimated parameter vector  
$\vec{a}_0$ has unbiased components and this estimator $\vec{a}_0$ has the
smallest variance. As illustration, in the case of a straight line fit ($M=q+px$),
 V. Blobel  \cite{Blobel_2003} produced the residual plots for the model parameters
 using several kinds of error distributions for the generated
 data points   (each with the same standard deviation)
 and showed that these plots  are  always gaussian distributions, as expected from
 the Central Limit Theorem. Of course, the probability distribution is  flat only if
 the error distributions are gaussian, {\it i.e.} if the effective  $\chi^2$
 function is actually  a real  $\chi^2$. 
  
  When analyzing (a collection of) actual spectra 
 obtained by various  groups, nothing better can be done and the
 derived fit solution faithfully reflects the whole data information on which it relies~: 
 It corresponds, at worst, to the least square solution
 and, at best, to the minimum  $\chi^2$ solution, depending on the functional nature of
 the true experimental error distributions.
 
  \subsection{Iterative Treatment of Global Scale Uncertainties}
\label{scale_err}
\indent \indent In the Subsection just above we have briefly summarized the traditional method which
applies when the handled spectra are not significantly affected by (correlated)  global
uncertainties. These can be of either kinds~: additive (offset error)
or multiplicative (scale/normalization error).  As  no offset error issue is reported for
the spectra we analyze within BHLS \cite{ExtMod3,ExtMod4}, we skip this case 
and let  the interested readers  refer to
suitable references \cite{D'Agostini,Blobel_2003,Blobel_2006}.  In contrast,
multiplicative (global scale) uncertainties are reported for most experimental spectra;
when they are  non--negligible compared with the other (more standard) kinds of errors, they
should be specifically accounted for within the global fit procedure. This is
of special concern  for the important $e^+e^- \ra \pi^+ \pi^-$  data samples collected
in scan mode \cite{CMD2-1998-1,CMD2-1998-2,SND-1998}, and even more for those
collected using the Initial State Radiation (ISR) mode by KLOE \cite{KLOE08,KLOE10,KLOE12}, 
BaBar \cite{BaBar,BaBar2} or BESSIII \cite{BESS-III}; furthermore, the normalization uncertainties
 reported for each of the  ISR data samples have all a peculiar structure  
 which deserves each a specific treatment -- this is the subject of the next Subsection.
  
 \vspace{0.5cm}

A constant global scale uncertainty, as those affecting the data samples from CMD2, SND
or BESSIII, 
 can be written $\beta=1 + \lambda$, where $\lambda$ is a random variable with range on $]-1,+\infty[$.
 As $E(\lambda)=0$ and $E(\lambda^2)=\sigma^2$ with $\sigma <<1$,  the gaussian  approximation
 for $\lambda$ is safe \cite{Blobel_2003,Blobel_2006}.
A  data sample subject to such a global scale uncertainty  provides an individual contribution
 to an effective global $\chi^2_{glob.}$ which should  {\it a priori} be written~:
\be
\displaystyle
\chi^2=[m-M(\vec{a})-\lambda  A]^TV^{-1}[m-M(\vec{a})-\lambda  A] + \frac{\lambda^2}{\sigma^2}
\label{Eq3}
\ee
where $m$, $M$, $V$ and $\vec{a}$ carry the  same definitions as in Subsection \ref{least_sq}
while $ \lambda$ and $\sigma$ have just been defined. As for $A$, even if intuitively
one may prefer  $A=m$, the choice $A=M(\vec{a})$ has been shown to drop out any biasing 
issue\footnote{This does not mean that the choice $A=m$
necessarily leads to a significantly biased solution as  shown below. }
 \cite{D'Agostini,Blobel_2003,Fruehwirth}.

Assuming that the unknown scale factor $\lambda$ is solely of experimental origin -- and, then,
independent of the model parameters  $\vec{a}$ -- the solution to  
$\partial \chi^2 /\partial \lambda=0$   provides its most probable value $\lambda_0$ \cite{ExtMod3}. 
After substitution, Eq. (\ref{Eq3})  becomes~: 
\be
\begin{array}{lll}
\displaystyle
\chi^2=[m-M(\vec{a})]^TW^{-1}[m-M(\vec{a})] & ~~~{\rm with}~~~&
W=V+\sigma^2 AA^T
\end{array}
\label{Eq4}
\ee
which exhibits a modified error covariance matrix $W$ 
and only depends on the (physics) model parameters. 
More precisely, the single recollection  
of the scale uncertainty  $\lambda$  is the occurence of its variance $\sigma^2$ in the
modified covariance matrix $W$.

However, Eq. (\ref{Eq4}) clearly points toward a difficulty if the model is not numerically
known  beforehand as the modified  covariance matrix becomes $\vec{a}$--dependent when setting
the unbiasing choice $A=M$. In this case, the 
parameter error covariance matrix provided by the $\chi^2$ minimization might be uneasy to interpret.

\vspace{0.5cm}

The way out is to define  iterative procedures; this is allusively stated 
in \cite{D'Agostini}, but explicitly considered in \cite{Chiba}
as solution to the so--called "Peelle's Pertinent 
Puzzle"\footnote{Peelle's reference is no longer of common access, but its
main content -- which closely resembles the  D'Agostini issue raised
in  \cite{D'Agostini} -- is reproduced in  \cite{Chiba}.} \cite{Peelle},
provided a good starting approximate solution is known beforehand; however, 
defining such a tool might be a delicate task if the underlying
model is non--linear, as quite usual in particle physics.
 Such a procedure has already been followed
and successfully worked out  in \cite{Ball} in order to derive through a minimization
procedure the parton density functions from several measured spectra. 
When dealing with samples of form factor and/or cross section data, 
other appropriate iterative methods should be defined.

The starting step of the iteration implies choosing some initial value for $A$,
 say $A=A_0$.
Without further information, the best approximation one can choose is obviously
$A_0\equiv m$, the experimental spectrum itself. Quite interestingly, this turns out to
start iterating with $\lambda=0$ ($\sigma=0$ in Eq. (\ref{Eq4})),  {\it i.e.} 
$\beta=1$, a unit scale factor; this makes the connexion with the iterative 
method followed in \cite{Ball}.

Then, the minimization of the $\chi^2$ in Eq. 
(\ref{Eq4}) with  $A=A_0\equiv m$  is 
performed using the {\sc minuit} procedure \cite{minuit} which yields the (step \# 0) 
solution\footnote{The analysis method in \cite{ExtMod3,ExtMod4} actually stops
there; the present analysis aims at going beyond.}
$M_0$ via the fitted parameter vector value $\vec{a}_0$. The next  step (\# 1)
consists in minimizing Eq. (\ref{Eq4}) using $A=M_0\equiv M(\vec{a}_0)$ which is easily
implemented in the procedure and, at convergence,  {\sc minuit}  provides the 
step \# 1 solution 
$M(\vec{a}_1)$. This stepwise procedure\footnote{Each such step is defined
as a  full ({\sc minuit}) minimization procedure where
 the covariance matrix is  unchanged until 
convergence is reached.}.  
is followed until  some convergence criterium  is met. 
As in each minimization procedure the covariance matrix is constant, the interpretation
of the parameter error covariance matrix is canonical.

The  convergence speed of the iterative procedure cannot be guessed {\it ab initio}
 but may be expected fast, referring to the fit of the parton density functions
 where the convergence is essentially reached at the first iteration \cite{Ball}. 
 This is confirmed by the Monte Carlo studies reported in Appendix \ref{MC_test}.
 
 Nevertheless, one may infer that the number of iteration steps is smaller
 for a starting guess for $A$ close to the actual model than for an arbitrary
 choice; clearly, as the choice  $A= m$ (the experimental spectrum) should
 be the closest to the actual model, one may think that it should minimize
 the number  of interations needed to reach convergence. Additionally, this
 choice does not imply any {\it a priori} assumption on the parameter vector to be fitted.
 
Among the data samples one deals within the BHLS based global fit method, most 
have been collected in scan mode, essentially at Novosibirsk, and carry a constant
scale uncertainty merging several effects. This is especially the case for the 
$e^+e^- \ra \pi^+ \pi^-$  data samples collected by the CMD2 
\cite{CMD2-1998-1,CMD2-1998-2}
and SND \cite{SND-1998} detectors; this also covers the case of the BESSIII data sample
\cite{BESS-III}.

In order to simplify and unify the notations in the following discussion, it is suitable 
to perform the change of 
random variable $\lambda=\sigma \mu$.  Then, the statistical properties for $\lambda$ propagate
to $E(\mu)=0$ and $E(\mu^2)=1$ and, defining in addition $B=\sigma A$, Eq. (\ref{Eq3})
above becomes~:
 
\be
\displaystyle
\chi^2=[m-M(\vec{a})-\mu  B]^TV^{-1}[m-M(\vec{a})-\mu  B] +\mu^2
\label{Eq5}
\ee
The condition $\partial \chi^2/\partial \mu =0$ provides the most probable value for $\mu$~:
\be
\displaystyle
\mu=\frac{B^T V^{-1} [m-M(\vec{a})]}{B^T V^{-1} B +1}
\label{Eq6}
\ee
and, substituting this into Eq. (\ref{Eq5}), one gets~:
\be
\begin{array}{lll}
\displaystyle
\chi^2=[m-M(\vec{a})]^TW^{-1}[m-M(\vec{a})] & ~~~{\rm with}~~~&
W=V+BB^T
\end{array}
\label{Eq7}
\ee 
Stated otherwise, from the point of view of the physics model,
the minimization procedure keeps track of the scale dependence by
a modified covariance matrix which, in turn, influences the fit. A faithful graphical 
comparison of data and model -- like the usual fit residual plots -- should
take into account   the fitted scale,  as illustrated
in \cite{ExtMod4} for instance.

\subsection{Global Scale Uncertainties Effects in ISR Experiments}
\label{scale_err_2}
\indent \indent 
With the advent of the $\Phi$ factory in Frascati, of the $J/\psi$ factory in Beijing
 and of the $B$ factories at SLAC and KEK,
the possibility opened to get large data samples for the various $e^+e^-$ annihilation channels
in the region of interest of the BHLS model, namely,  from the thresholds to the $\phi$ meson mass
energy region  ($\sqrt{s} \leq 1.05$ GeV). The production mechanism
involved  is the emission 
of a hard photon in the initial state \cite{Benayoun_isr}, 
the so--called Initial State Radiation (ISR) phenomenon. This ISR production mode has been used 
 to collect high statistics data samples for 
 the $e^+e^- \ra \pi^+\pi^- $ channel covering the low energies by the 
 KLOE \cite{KLOE08,KLOE10,KLOE12}, BaBar \cite{BaBar,BaBar2} 
 and BESSIII \cite{BESS-III} Collaborations. 

However, it is a common feature of the KLOE  and BaBar (ISR) data samples to carry
non--trivial error structures. Beside a non--diagonal statistical error
covariance matrix ($V$), they exhibit a  large
number of (statistically independent) bin--to--bin correlated uncertainties, 
most of these being additionally $s$--dependent. As far as we know, this seems to be 
a premi\`ere in particle physics and how this is dealt  with  inside minimization procedures
deserves to be clarified and explicitely stated (see also  \cite{ExtMod4}).

Let us consider a given experimental data sample  $E$, a spectrum $m$ function of $s$,
for which the (given) statistical error covariance
matrix is $V$; the information provided for the bin--to--bin correlated uncertainties
defines  several independent scale uncertainties $\lambda_\alpha$ ($\alpha=1,\cdots n_{scale}$)
and should be understood as follows~: each of the scale uncertainty $\lambda_\alpha$
 is a random variable of zero mean and 
carrying a $s$--dependent
standard deviation $\sigma_\alpha(s)$ as tabulated by each experiment.
It  is clearer to make the change of (random) variables  
$\lambda_\alpha=\sigma_\alpha (s) \mu_\alpha$  ($\alpha=1,\cdots n_{scale}$)  and assume
that all the random variables  $\mu_\alpha$ fulfill  $E(\mu_\alpha)=0$ and
$E(\mu_\alpha \mu_\beta)=\delta_{\alpha \beta}$.

Then, the  other notations being identical to those previously defined, the $\chi^2$ in
Eq. (\ref{Eq5})  generalizes to~:
\be
\displaystyle
\chi^2=[m-M(\vec{a})-\mu_\alpha  B_\alpha]^TV^{-1}[m-M(\vec{a})-\mu_\beta  B_\beta]
 +\mu_\alpha \mu_\beta \delta_{\alpha \beta}
\label{Eq8}
\ee
where implicit sum over  repeated greek indices  is understood. One has defined
$B_\alpha(s)=\sigma_\alpha(s) A(s)$, $A$ being the $s$--dependent vector already defined.  $A$  is  
 iteratively redefined as emphasized in the previous Subsection.
Using the minimum $\chi^2$ conditions  $\partial \chi^2/\partial \mu_\alpha =0$ and
the  independence conditions of the various
 sources of scale uncertainty $\partial \mu_\alpha/\partial \mu_\beta=\delta_{\alpha \beta}$,
 the most probable values for the $\mu_\alpha$'s
can be derived \cite{ExtMod4}. A recursion can be defined and allows to derive\footnote{
For clarity, defining $Z=A$ or $B$,  $Z_k$ denotes the quantity $Z(s_k)$ for short. }
from Eq. (\ref{Eq8})~:
\be
\left\{
\begin{array}{l}
\displaystyle
\chi^2=[m-M(\vec{a})]^TW^{-1}[m-M(\vec{a})] \\[0.5cm]
W_{ij}=V_{ij}+B_i B_j
=V_{ij}+ \left[\sum_{\alpha=1}^{n_{scale}} \sigma_\alpha(s_i)\sigma_\alpha(s_j)\right ] A_iA_j
~~~,~~~(\forall [i,~j])
\end{array}
\right .
\label{Eq9}
\ee 
in close correspondence with Eq. (\ref{Eq7}).

A specific feature of Eq. (\ref{Eq9}) deserves to be noted. As each experimental
group reports {\it separately} on each identified {\it independent} source of (scale) uncertainty, 
these should indeed be fitted {\it separately} as stated just above to go from
Eq. (\ref{Eq8}) to Eq. (\ref{Eq9}). More precisely, for the experiment $E$, 
we are not using the quadratic sum $(\sigma_E(s))^2=\sum_{\alpha} [\sigma_\alpha(s)]^2$
for its partial $\chi^2$, which would have given $\sigma_E(s_i) \sigma_E(s_j)A_iA_j$
inside the full error covariance matrix instead of what is shown in Eq. (\ref{Eq9}).
Stated otherwise, the various sources of normalization uncertainties are not summed in quadrature
but really treated as statistically independent.

\subsection{Numerical Tests of the  Global Fit Iterative Method}
\label{scale_err_3}
\indent \indent As stated in the header of the present Section, if the 
physics correlations predicted by the Effective Lagrangian (here BHLS)
are fulfilled by the data,   the estimate
of the model parameters and the parameter error covariance matrix are
legitimate tools to serve  evaluating related physics quantities.

As in the previous studies relying on the HLS model, at the early
stages \cite{ExtMod1,ExtMod2}  or more recently
\cite{ExtMod3,ExtMod4,BM_paris_2013,BM_roma2013}, the method is
to minimize a global $\chi^2$ expression taking into account
the largest possible number of data samples and using appropriately
all information provided by the experimentalists concerning
all kinds of uncertainties which affect their data samples. The aim
of Subsections \ref{least_sq} -- \ref{scale_err_2}  was to
detail how the $\chi^2$ piece associated with each data sample
should be constructed, depending on its reported error structure.  

In contrast with previous references (including ours), the fit procedure will be 
adapted in the present study in order to examine and cure possible biases produced
by having stopped the fit procedure at the $A=m$ step instead of iterating 
further on as suggested in \cite{D'Agostini}, explicitly proposed in \cite{Chiba} and 
performed in \cite{Ball}.

In order to check whether estimates based on  global fit results can be trusted 
as, for instance, the muon HVP central value and its uncertainty derived from the
fit information returned by
{\sc minuit}, some additional checks on the fitting method and its iterative aspect
deserve to be performed, at least to control that, indeed~:
\begin{itemize}
\item The fit parameter residuals $\Delta_i= a_i^{fit}-a_i^{true}$ are 
unbiased gaussians,
\item  The parameter pulls are centered gaussians of unit standard deviations.
\end{itemize}
One should also check that the fit probabilities distributions are uniformly 
distributed on $[0,1]$
when the measurements are indeed true unbiased gaussian distributions.

This condition list  can be supplemented with some examination of 
the  effects due to non--linear dependences upon the parameters to be fitted.

\vspace{0.5cm}

However,  checking this list of properties obviously implies that the
true parameter values are known, that the measurements are indeed sampled 
on truely centered gaussian distributions and that their errors are
indeed the true standard deviations of the measured spectrum.
 Stated otherwise, this exercise goes beyond
using actual  measured experimental data samples as, then, truth is unknown~:
The global fit method --as any other method -- should be evaluated using data samples
generated  by Monte Carlo techniques; in this case, the true parameter values and
their uncertainties are known at the sample generation level and can reliably be compared
to the fit results. The detailed study is transferred to Appendix \ref{MC_test};
the most involved results are summarized right now~:
\begin{itemize}
\item The effects of non--linear parameter dependence within  models used to
fit data spectra (see Subsection \ref{Shape})
 are likely to be marginal for the kind of experimental
distributions we are dealing with. This
should be related with the local minimum finding structure of the algorithms gathered
within the {\sc minuit} package.
\item When scale uncertainties dominate the sets of spectra globally submitted to fit, 
using\footnote{ $m_E$ being the experimental spectrum in the expression for the $\chi^2$ 
(see Eqs. (\ref{Eq3}) or (\ref{Eq4})).} $A=m_E$  gives a solution which
 can exhibit strong biases,  but this solution  is the start of
 an iterative procedure 
which leads rapidly to the unbiased solution to the minimization problem.
The biases occuring at start of the procedure can be very large, but they are
observed to practically vanish already at the first iteration step (the previously called
$M_1$ solution).
\item  When performing a global fit of some data samples dominated by global scale uncertainties 
together with others where the statistical errors 
 (e.g. affecting {\it randomly} each bin) dominate, 
the iterative method obviously
works as well as just stated. In this case, however, the presence of
some samples free from scale errors exhibits an unexpected pattern~:
Even if the data samples  free from scale uncertainties are affected by enlarged 
statistical errors, they strongly reduce the biases  generated by the $A=m_E$ choice. 
Stated otherwise, the effects of data samples
where the normalization errors are dominated by the (random) statistical errors
is to favor the smearing out of the biases in the parameter value estimations. 

\end{itemize}
 
 The properties just listed concerning the unbiasing of the fit parameters
 extend to the estimates of physics quantitites derived from using the
 fit result information (parameter values and error covariance matrix).
 Additionally, as the parameter pulls are observed as centered gaussians of unit
 standard deviation, the calculated uncertainties relying on Monte Carlo sampling
 of the fit parameter distributions should also be reliable. This is of special
 relevance for the evaluation of the various contributions to the muon LO--HVP
 discussed in Section \ref{gM2}.
 
 The last item in the list just above has important consequences
 while working with real (and so, not really perfect) experimental data. However,
 even if the fraction of  data samples
 free from -- or marginally affected by -- scale uncertainties  may look large enough, it is 
 nevertheless cautious to ascertain that the fit solution
 is indeed unbiased by performing one or two additional iterations. Indeed, the studies reported
 in Appendix \ref{MC_test} tell that, anyway, the iterated fit solutions are always unbiased.
 
 Therefore, one may conclude from this Section and from the simulation studies
 reported in Appendix \ref{MC_test} that global fit methods can indeed be trusted. The single
 proviso is that iterating the fit procedure as explained above is mandatory
 or, at least, cautious.  
 
 The issue is now to examine how the results given in \cite{ExtMod3} and
 \cite{ExtMod4} are modified when iterating beyond the approximation $A_E=m_E$
 for all data samples significantly affected by scale uncertainties, constant
 (as, mostly, the spectra reported in  \cite{CMD2-1998-1,CMD2-1998-2,SND-1998} ) 
 or $s$--dependent (as all the ISR spectra reported in \cite{KLOE08,KLOE10,KLOE12,BaBar}). 
 Observing  the stabilizing effect of the data samples dominated by statistical errors
 (like the $\gamma \pi^0$ and $\gamma \eta$ final states) is also  methodologically relevant.
 
\section{BHLS Global Fit Method~: Present Status and Corrigendum} 
\label{Status}
\indent \indent As stated several times above, the Effective Lagrangian Model we
use is the broken HLS (BHLS) model developped in \cite{ExtMod3}. In this  Reference,
the BHLS model is also applied to all data samples collected in  scan mode, by the various
Collaborations which have run on the successive Novosibirsk $e^+e^-$ colliders. These
$e^+e^-$ annihilation samples  cover the $\pi^+\pi^-$, $\pi^0\gamma$, $\eta \gamma$, 
$\pi^+\pi^-\pi^0$, $K^+K^-$, $K^0\overline{K^0}$ final states and have been
discussed in detail in several previous studies \cite{ExtMod1,ExtMod2,ExtMod3}; for the sake of 
conciseness, we will not repeat this exercise here. As the  BHLS model also covers 
the $\tau$ decays from the early stages of its formulation \cite{taupaper}, the previous
studies include the dipion spectra collected
in the  $\tau^\pm \ra \pi^\pm \pi^0 \nu_\tau$ decay mode by ALEPH \cite{Aleph,AlephCorr},
Belle \cite{Belle} and CLEO \cite{Cleo}. Also included within the BHLS fit procedure are
some light meson decay partial widths not connected with the 
annihilation channels already listed,   like $K^{*0} \ra K^0 \gamma$,
$K^{*\pm} \ra K^\pm  \gamma$, $\eta^\prime \ra \omg  \gamma$ or
$\phi \ra \eta^\prime \gamma$. 

A second step has been to extend the study in  \cite{ExtMod3} to treat the
high statistics ISR data samples for $e^+e^- \ra \pi^+\pi^-$; this has been the purpose
of the study in  \cite{ExtMod4} where the KLOE08 \cite{KLOE08}
and KLOE10 \cite{KLOE10} data samples collected by the KLOE Collaboration 
and the data sample produced by BaBar \cite{BaBar} have been examined. Since then,
two new samples have been produced by the KLOE (KLOE12 \cite{KLOE12}) and
BESSIII \cite{BESS-III} Collaborations\footnote{The KLOE12 and KLOE08 data samples
are tightly correlated; actually, they mostly differ by their respective normalization
procedures. Comparing their respective behaviors within our global treatment is, therefore,
interesting.} 
Except otherwise stated, all the fit results presented in this paper have been obtained 
using the  Configuration B \cite{ExtMod3} ({\it
i.e.}  dropping out from the fit procedure the three pion data samples collected 
in the $\phi$ mass region).

The studies covered by  \cite{ExtMod3,ExtMod4,BM_paris_2013,BM_roma2013}
rely on minimizing a global $\chi^2$ function summing up partial
$\chi^2$'s, each associated with a  given data sample.  For each 
of the $\simeq 40\div 50$ data samples, the partial $\chi^2$ was 
(canonically) constructed
following the rules detailed in Section \ref{GF_Methods}.
However, as the fit was not iterated in the studies
\cite{ExtMod3,ExtMod4}, it is worth checking
to  which extent the  value of the muon HVP derived herefrom
is changed by the iteration procedure.

For the present study, a few coding bug fixes have been performed
and a piece missing in the expression for the 
$e^+e^- \ra \pi^0\gamma$ and $e^+e^- \ra \eta\gamma$ cross sections
has been included. So, when different, the results in the present paper
supersede those in \cite{ExtMod3,ExtMod4}.

As for the missing piece just mentioned~: In the 
amplitudes $\gamma^* \ra  \gamma P_0$ (Eq. (65) in \cite{ExtMod3})  and the 
cross section formulae $e^+e^- \ra \gamma P_0$ (Eq. (68) in \cite{ExtMod3}), 
the non--resonant piece should be modified as follows~:
\be
\displaystyle
(1-c_4) L_{P_0} \Rightarrow \left( 1-\frac{[c_3+c_4]}{2} \right) L_{P_0}~~.
\label{Eq10}
\ee
This implies that the single process which depends $separately$ 
on the FKTUY \cite{FKTUY} parameters $c_3$ and $c_4$  
 is the $e^+e^- \ra \pi^+\pi^-\pi^0$ annihilation. In this case
 both  $c_3+c_4$ and $c_3-c_4$ combinations come in, while
 all others quantities only involve the $c_3+c_4$ 
combination\footnote{The studies
\cite{ExtMod3,ExtMod4} have been performed fixing $c_3=c_4$. The 
BHLS fit recovers a good fit quality by modifying 
 the value for $c_1-c_2$ as will be seen below.}. We apologize 
 for the inconvenience.

\section{BHLS Global Fit Method~: Iterating with NSK Data Only} 
\label{nsk_iter}
\indent \indent 
In this Section, we report on global fits using the data reminded in the preceding Section
and discussed in \cite{ExtMod3}; as for the pion form factor data, we focus for the present exercise
on using  only the most recent scan data collected by CMD2 and SND  \cite{CMD2-1995corr,CMD2-1998-1,CMD2-1998-2,SND-1998}, 
excluding the older data samples from OLYA and CMD \cite{Barkov}.
\begin{table}[ph!] 
\begin{tabular}{|| c  || c  || c  | c ||c ||c ||}
\hline
\hline
\hhhc
\hhhd  $\chi^2/N$ & \hhhv $A=m$   &\multicolumn{2}{|c||}{  \hhhv  Iteration Method} 
&\multicolumn{2}{|c||}{  \hhhv $A=M$ varying}\\
\hhhd ~~~~ & \hhhv  \cite{ExtMod3}   &  \hhhv$ A=M_0$ &   \hhhv$ A=M_1$ &  \hhhv  $A_{start}=M_1$&  \hhhv  $A_{start}=M_x$\\
\hline
\hline
 \hhhv Decays    			& 8.16/10    	& 8.01/10	& 8.03/10 	& 8.01/10 	& 8.02/10 	\\
\hline
 \hhhv New Timelike $\pi^+ \pi^- $ 	& 121.54/127 	& 121.75/127 	& 121.75/127	& 121.74/127 	& 121.75/127	\\
\hline
 \hhhv $\pi^0 \gamma$ 			& 63.84/86   	&63.98/86 	& 63.96/86 	& 63.98/86 	& 63.96/86	\\
\hline
 \hhhv $\eta \gamma$ 			& 120.87/182   	&120.84/182	  & 120.84/182	& 120.84/182	&120.83/182	\\
\hline
 \hhhv $\pi^+ \pi^- \pi^0$ 		& 101.82/99  	& 102.49/99  	 &102.43/99	& 102.49/99     &102.43/99       \\
\hline
 \hhhv $K^+ K^- $ 			& 29.87/36 	& 29.77/36  	 & 29.78/36	& 29.78/36	&29.78/36	\\
\hline
 \hhhv $K^0 \overline{K}^0 $ 		& 119.21/119	& 119.21/119   	& 119.18/119	& 119.20/119	&119.19/119	\\
\hline
 \hhhv ALEPH 				& 19.67/37      & 19.73/37 	 & 19.71/37	& 19.72/37	&19.70/37	\\
\hline
 \hhhv Belle 				& 28.24/19      & 28.27/19 	 & 28.29/19 	& 28.27/19	&28.29/19	\\
\hline
 \hhhv CLEO				& 34.96/26      & 34.82/29 	 & 34.82/29 	& 34.84/29	&34.84/29	\\
\hline
\hline
  \hhht $\chi^2/dof$  & 648.16/719       & 648.85/719       &648.78/719    &648.85/719    &648.78/719\\
  \hhhv Global Fit Probability & 97.2\%  &  97.1\%          &  97.1\%      &  97.1\%   & 97.1 \%\\
\hline
\end{tabular}
\caption {
\label{Table:T1} Global fit $\chi^2$ results derived by using only the data from \cite{CMD2-1998-1,CMD2-1998-2,SND-1998}
for the  $e^+e^-\ra \pi^+ \pi^-$ annihilation. See   the
discussion and comments in Section \ref{nsk_iter}.
}
\end{table}

The CMD2 data samples are reported to carry constant bin--to--bin correlated uncertainties of
0.6\% (\cite{CMD2-1995corr}), 0.8\% (\cite{CMD2-1998-1}) and 0.7\% (\cite{CMD2-1998-2}), while
SND reports a  1.3\% constant scale uncertainty \cite{SND-1998} -- except for their first 2 data points  
where it is 3.2\%. For these data samples, the partial
$\chi^2$'s are essentially given by expressions like Eq. (\ref{Eq4}). For the other data samples,
we performed as in \cite{ExtMod3}.

The first data column in Table \ref{Table:T1} displays the results of the fit 
 performed by setting  $A=m$  in the  $\chi^2$ 
 associated with each experimental data spectrum generically named $m$.  The form factor returned by this ($A=m$) global fit 
  is named $M_0$ and is used to perform the first iterated ($A=M_0$) global fit;  the results of this fit
  are shown in the data column \#2; this iteration \#1 global fit returns the solution named  $M_1$. 
  The iterated  \#2 fit is then performed by setting $A=M_1$ in the $\chi^2$ expressions of the
 pion form factor data samples,   leading to another ($M_2$) solution; the fit results are displayed in the third data
 column in Table \ref{Table:T1}. 
 
 One clearly observes a quite tiny change in the first iteration~: 0.2 unit in the $\chi^2$ value of 
 the $\pi^+ \pi^-$ data samples; also the global $\chi^2$ changes by only 0.7 unit. When going from
 the first to the second iteration, the changes are almost invisible.  This corresponds for experimental data to 
 the effect reported in Subsection \ref{iter_2} for our Monte Carlo data.  As derived quantity, let us report on 
 the leading order (LO) contribution $a_\mu(\pi \pi)$ derived by integrating Eq. (\ref{Eq1})
 between 0.63 GeV/c and 0.958  GeV/c; using obvious  notations, the previously reported fits  yield~:
\be
\left  \{
\begin{array}{lll}
A=m~~~~~~ &:& a_\mu(\pi \pi,[0.63,0.958])= 358.95  \pm  1.63\\
A=M_0~~~~ &:&  a_\mu(\pi \pi,[0.63,0.958])= 360.00 \pm  1.78\\
A=M_1~~~~ & :& a_\mu(\pi \pi,[0.63,0.958])= 359.99 \pm  1.79
\end{array}
\right .
\label{Eq11}
\ee
in  units of $10^{-10}$. So, one observes a tiny effect while iterating once (0.3\% for the central value) and
no effect when iterating twice. In the present case, where 
the former data from \cite{Barkov} have been dropped out from
the fit,  the "experimental" estimate is $a_\mu(\pi \pi,[0.63,0.958])= 361.26 \pm  2.66$ (see Table 7 in \cite{ExtMod3}).

Another way to account for the scale uncertainty is to set $A=M(\vec{a})$  (which depends
on the parameters under fit) and perform the fit. A starting value   for $A$  must be chosen
(denoted $A_{start}$) but its value changes at each  step
of the minimization procedure. In this case,
the fit convergence time is much larger than previously but the results are almost identical 
to those already obtained by iterating. The last 2 columns in Table  \ref{Table:T1} display the fit 
results $starting$  with $A_{start}=M_1$ and also those starting from the fit solution derived herefrom
(denoted $M_x$). As for $a_\mu(\pi \pi,[0.63,0.958])$, the values derived 
in these last fits numerically coincide with the iterated cases displayed above. 
 
 Therefore, one may indeed conclude,  as can be inferred from the Monte Carlo studies reported
in Appendix \ref{MC_test}, that the   HVP value reached without iterating is very close to the
HVP derived from the once iterated solution. One also observes, as expected, that  iterating only
once  already leads to the final result; indeed, from iteration \#1 to iteration \#2, the changes 
for $a_\mu(\pi \pi)$ are at the level of a few $10^{-12}$.

As for the fit quality reflected by the $\chi^2$ values at minimum and the corresponding fit 
probabilities, the last line in Table \ref{Table:T1} indicates that,  whatever is the way one treats 
the vector $A$, they are all alike. This, once more, corresponds to expectations, as can
be checked with the discussion in Subsection \ref{iter_2}  and expecially the properties
of Figure \ref{Fig:scale_2b}.  Nevertheless, it is useful to check that the twice iterated solution
does not modify the result derived from the once iterated solution in a significant way.

\section{BHLS Global Fit Method~: Iterating Scan and ISR Data} 
\label{global_iter}
\indent \indent It remains to introduce the other  $\pi^+\pi^-$ data samples collected at
$e^+e^-$ colliders using the ISR mechanism. Reference \cite{ExtMod4} has already done this work 
with the data samples then available using the method described in Subsection 
\ref{scale_err_2} without, however, iterating the procedure. The conclusion reached was that the KLOE08
 \cite{KLOE08} and
BaBar \cite{BaBar} data samples have difficulties to accomodate -- within the  BHLS framework --
the whole set of data samples covering the  channels already reminded in Section \ref{Status}.
In contrast, the KLOE10 \cite{KLOE10} data sample was found to fit well the BHLS expectations.  
Complementing preliminary works  \cite{BM_paris_2013,BM_roma2013}, we revisit here the 
issue with the  two new  data samples provided by KLOE (KLOE12) and BESSIII.

\subsection{The  $\tau$+PDG Analysis} 
\label{tau_pdg}
\indent \indent In Ref. \cite{ExtMod4}, it has been shown that the BHLS fitter can be run without
explicitly using definite $e^+ e^- \ra \pi^+\pi^-$ data samples besides the non $\pi^+\pi^-$ channels.
Indeed, on general grounds, one expects  that some limited isospin breaking (IB)  information 
specific of this annihilation channel can make the job together with the $\tau$ dipion spectra.
It has been shown that the partial widths $\Gamma(\omg/\phi \ra  \pi^+\pi^-)$ and 
$\Gamma(\rho^0 \ra  e^+ e^-)$, together with the products   ($V=\omg,~\phi$)
$\Gamma(V \ra  \pi^+\pi^-) \times \Gamma(V \ra  e^+ e^-)$ represent
an amount of information sufficient to reconstruct -- within BHLS --
 the pion form factor in the $e^+ e^- $ channel.

\vspace{-2cm}
\begin{figure}[!ht]
\hspace{-1.5cm}
\begin{minipage}{1.1\textwidth}
\begin{center}
\resizebox{1.1\textwidth}{!}
{\includegraphics*{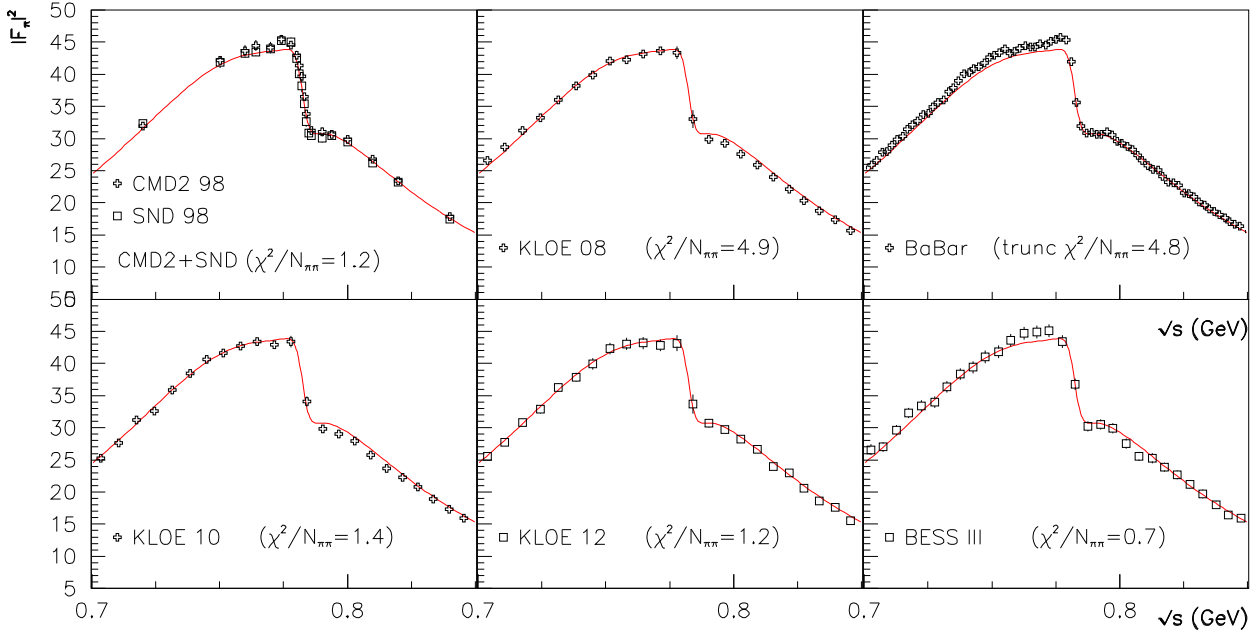}}
\end{center}
\end{minipage}
\begin{center}
\vspace{-1.5cm}
\caption{\label{Fig:tau_pdg} The $\tau$+PDG prediction (red curve) of the pion form factor 
in $e^+ e^- $ annihilations in the $\rho-\omg$ interference region. The various superimposed
 data samples are  not fitted; also displayed  are the average $\chi^2$
distances of each of the $e^+ e^- \ra \pi^+\pi^-$ data samples to the common $\tau$+PDG prediction.
}
\end{center}
\end{figure}

\vspace{-1.cm}

Before going on, it deserves noting that the decay information used to run the $\tau$+PDG 
method has been extracted from the Review of Particle  Properties (RPP) \cite{RPP2012} and
that the above mentioned pieces of information are in no way influenced
by the data collected  by KLOE, BaBar or BESSIII; actually, they are almost
100 \% determined by the data samples from the CMD--2 and SND experiments. 
On the other hand, the $\tau$+PDG analysis is not influenced by the global scale
issue which mostly motivates the present work. 

We have performed the $\tau$+PDG run using all annihilation data mentioned 
in the above Sections (configuration A \cite{ExtMod3}). The fit returns $\chi^2_{\tau}/N_{\tau}=82.1/85=0.97$. 
 The best fit solution allows to reconstruct the predicted invariant mass distribution of
 the pion form factor in the  $e^+ e^- \ra \pi^+\pi^-$  annihilation;  this prediction
 is expected valid over the whole BHLS range as shown by Figure 2 in \cite{ExtMod4}. It is worth
 showing here  the  mass range from 0.70 to 0.85 GeV;  Figure \ref{Fig:tau_pdg}
 displays the $\tau$+PDG prediction on this range together 
  with the available $ \pi^+\pi^-$ data superimposed ({\it  and not fitted});
we have calculated the $\chi^2$ distance of each sample over its full range\footnote{
 For BaBar,  the  computed $\chi^2$ referred to here is computed on its spectrum up to 1 GeV, 
 but  truncated  from the drop--off region (0.76$\div$ 0.80 GeV).}. The average $\chi^2$ per data point
 is  indicated inside the corresponding pannel.
  
Figure \ref{Fig:tau_pdg} indicates that the average $\chi^2$ distances
for the NSK (CMD--2 \& SND), KLOE10, KLOE12
and BESSIII samples   are small enough to claim a success of the $\tau$+PDG
method. One can conclude that they fulfill  the consistency
issue discussed in Section \ref{EF_method} with the full set of data 
and channels covered by BHLS. One should note that the description of
the BESSIII sample (which is not a fit) is as good as the fit published
by the BESSIII Collaboration \cite{BESS-III}. For KLOE08 and BaBar, we reach the 
same conclusion as in \cite{ExtMod4}; nevertheless, one can now compare
the behavior the twin\footnote{They mostly differ by the normalization method
used to reconstruct the spectrum from the same  collected data.} samples KLOE08 and KLOE12~:
We have $\overline{\chi^2}_{KLOE08} = 4.8$ while  $\overline{\chi^2}_{KLOE12} = 1.2$ 
clearly reflecting a better understanding of the error covariance matrix, while the
central values are almost unchanged, as clear from Figure \ref{Fig:tau_pdg}.

Stated otherwise, the issue met with  KLOE08 and BaBar   is confirmed
but the two new data samples published since \cite{ExtMod4} are both found
is good correspondence with expectations.

\subsection{The  Iterative Method~: Global Fit Properties} 
\label{global_iter_1}
\indent \indent 
The issue is now to report on the  behavior of the global fits  performed using the iterated method
when the $\pi^+\pi^-$ ISR   $and$  scan data are considered simultaneously; this 
complements the work  already presented in Section \ref{nsk_iter} 
when using the scan data only. Except otherwise stated, the $\tau$ data samples are always   
included into the fit procedure.
On the other hand, as the behavior of the global fit for data/channels other than $\pi^+\pi^-$ 
does not differ sensitively  from the information already displayed in Table \ref{Table:T1}, 
this will not be repeated.

\begin{table}[!ht] 
\hspace{-1.5cm}
\begin{tabular}{|| c  || c  || c  | c  | c |c ||c ||}
\hline
\hline
\hhhc
\hhhd  Fit Configuration & \multicolumn{6}{|c||}{  \hhhv  Iteration Method} \\
\hline
\hhhv ~ $[\chi^2_{\pi^+\pi^-}/N_{\pi^+\pi^-}]$ & \hhhv  KLOE08   &  \hhhv KLOE10 &   \hhhv KLOE12  &\hhhv NSK  &  \hhhv BESSIII--III & \hhhv BaBar~~~~ \\
\hline
 \hhhv  $N_{\pi^+\pi^-}$ &  \hhhv  (60) &  \hhhv  (75)&  \hhhv  (60) &  \hhhv  (127/[209]) &  \hhhv  (60)& \hhhv  (270) \\
\hline
\hline
 \hhht Fits in Isolation    		& 1.64    	& 0.96	  	& 1.02  & 0.96[0.83]  & 0.56	  	& 1.25	\\
 \hhht Global fit prob.    		& 59\%   	& 97\%		& 97\%	& 97\%[99\%]  & 99\%	& 40\% 	\\
\hline
 \hline
  \hhht Fit Combination 1    		& ~~~   	& 1.02 	  	& 1.48  & 1.18[0.96] 	& 0.56	&1.36(*) \\
\hhht  $\chi^2_{\pi^+\pi^-}/N_{\pi^+\pi^-}$ \& Gl. fit prob: &~~   		& \multicolumn{5}{|c||}{  \hhhv 1.21 \& 22\%  } 	\\
\hline
 \hline
 \hhht Fit Combination 2    		& ~~~   	& 1.00	  	& 1.05  & 1.11[0.89] & 0.61		&~~~  \\
\hhht  $\chi^2_{\pi^+\pi^-}/N_{\pi^+\pi^-}$ \& Gl. fit prob: &~~   		& \multicolumn{4}{|c||}{  \hhhv 0.98 \& 99\% }	& ~~~ 	 \\
\hline
\hline
 \hhht Fit Combination 3    		& ~~~   	& 1.02 	  	& 1.05  & 1.10[0.89] &~~~	&~~~  \\
\hhht  $\chi^2_{\pi^+\pi^-}/N_{\pi^+\pi^-}$ \& Gl. fit prob: &~~   		& \multicolumn{3}{|c||}{  \hhhv 1.06 \& 97\% }	& ~~~ 	& ~~~ \\
\hline
 \hline
\end{tabular}
\caption {
\label{Table:T2} Global fit results as function of the $e^+e^-\ra \pi^+ \pi^-$ data sample content.
Each entry displays the $[\chi^2_{\pi^+\pi^-}/N_{\pi^+\pi^-}]$ value returned by the global fit. 
The data samples involved can be tracked from
the column titles, the following line giving the corresponding data point numbers 
$[N_{\pi^+\pi^-}]$ in the range up to 1 GeV.
The value flagged by *  has been obtained using a BaBar sample truncated from the energy region 
$[0.76,~0.80]$ GeV  (250 data points). 
}
\end{table}

Table \ref{Table:T2} displays our main results using the scan and ISR  $e^+e^-\ra \pi^+ \pi^-$ 
annihilation data. They correspond to the iteration \# 1 fit (denoted above  $A=M_0$), however  the 
 previously called $A=m$ or $A=M_1$ solutions gives almost identical fit quality 
 results\footnote{As regard to the fit parameter values and uncertainties~:  The $A=M_0$ and  $A=M_1$
 solutions differ unsignificantly; the $A=m$ exhibits some small departure commented below.}. 

 The first data line displays the global fit properties with the indicated $e^+e^-\ra \pi^+ \pi^-$ 
data samples  used each in  isolation within the global BHLS context, together with all 
other data samples covering the rest of the encompassed physics (see Section \ref{Status}). 

One observes that the average (partial) $\chi^2$ per data point 
$\chi^2_{\pi^+\pi^-}/N_{\pi^+\pi^-}$
is of the order 1 or (much) better and the probability  high when running with any of
 the  KLOE10, KLOE12, NSK\footnote{NSK here denotes the collection of data samples from CMD2
\cite{CMD2-1995corr,CMD2-1998-1,CMD2-1998-2}, SND \cite{SND-1998} (127 data points in total) 
as well as the former (82 data points) samples
collected by OLYA and CMD  \cite{Barkov}. The numbers in Table \ref{Table:T2}  given within square 
brackets include the contributions from these former samples.} and BESSIII
data samples; as in \cite{ExtMod4}
the picture is not as good for
KLOE08 and BaBar. 

Performing a global BHLS fit  using  the data samples from KLOE10, KLOE12, BESSIII, NSK and BaBar 
(amputated\footnote{We remind that this removal is motivated by a possible mismatch in the energy
calibration in the $\rho^0-\omg$ 
interference region between BaBar and the other $\pi^+ \pi^-$ data samples submitted to the same 
global framework. In contrast, when running with the $\pi^+ \pi^-$  BaBar sample in isolation, 
its full spectrum is considered.} 
from the energy region $[0.76,~0.80]$ GeV) leads to results given at the entry 
lines flagged by "Fit Combination 1"; as the correlations between the KLOE08 and KLOE12 data samples
are strong and  their content not explicitely stated\footnote{Some work in this field seems ongoing 
\cite{KLOEComb}.}, it is more cautious to avoid dealing with
the KLOE08 and KLOE12 samples simultaneously. Despite the removal of the drop--off region in the
BaBar $\pi^+ \pi^-$ spectrum, the global fit quality looks poorer.

\begin{figure}[!ph]
\begin{center}
\resizebox{12cm}{9cm}
{\includegraphics*{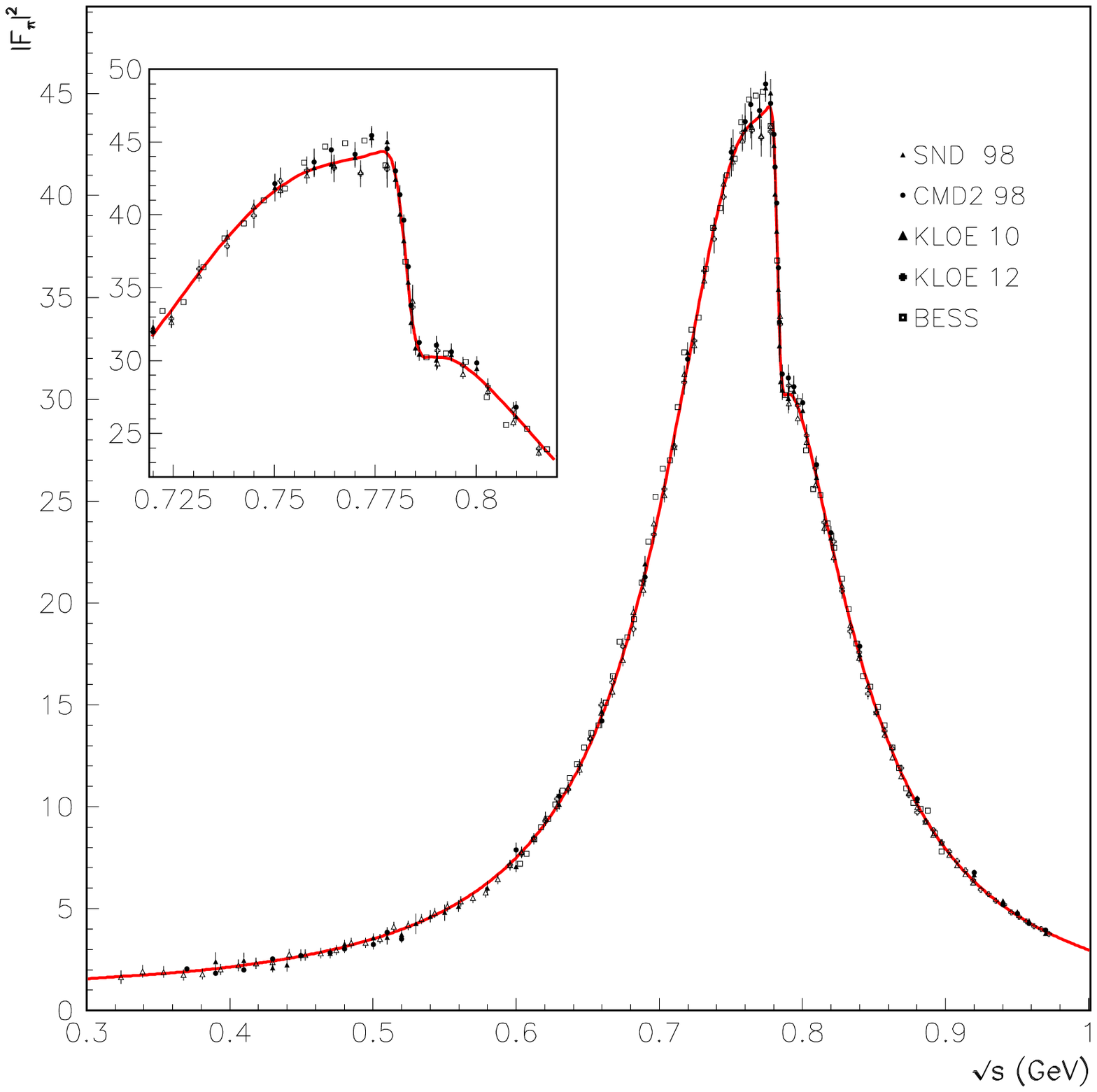}}
\resizebox{12cm}{8.5cm}
{\includegraphics*{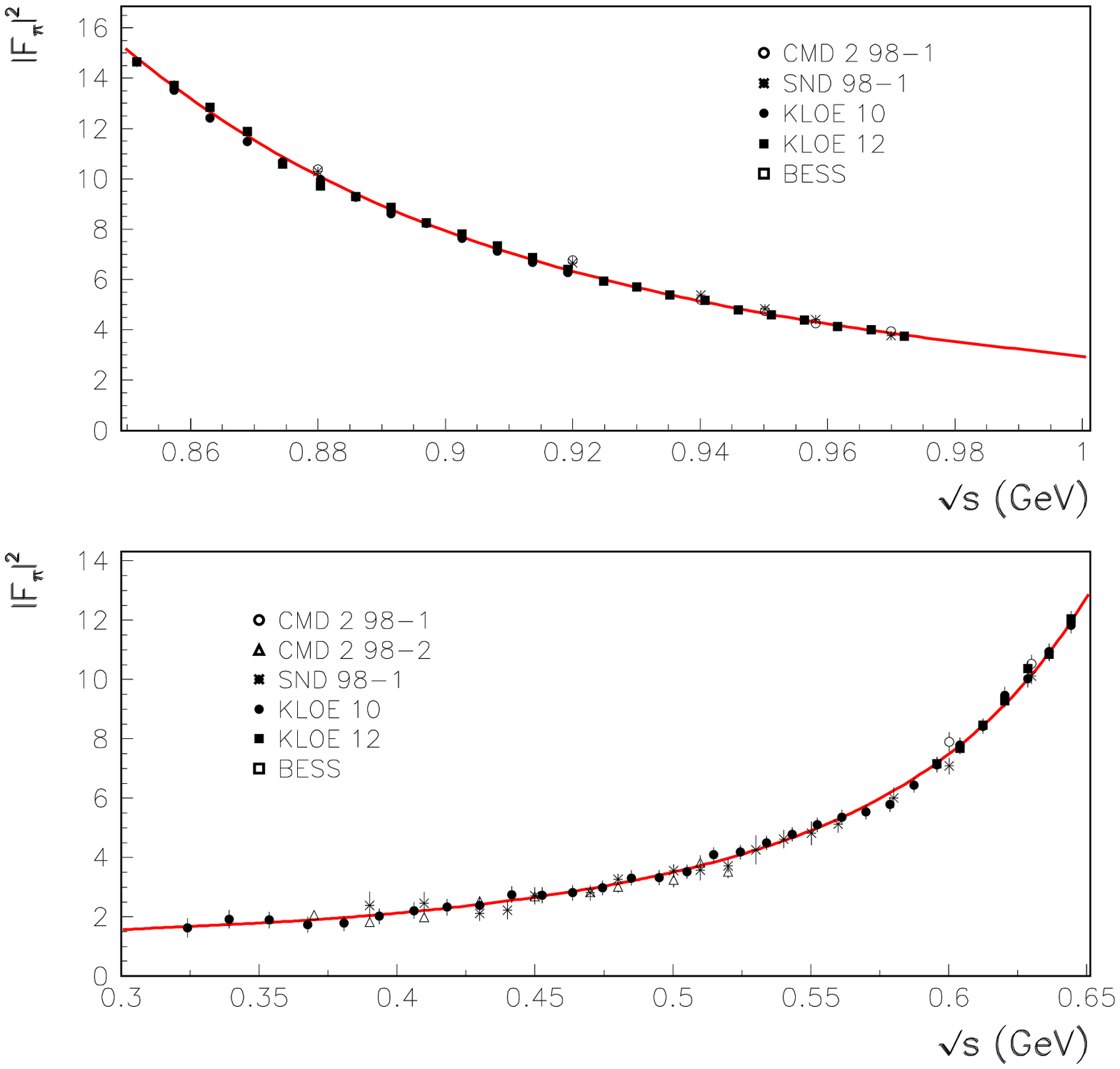}}
\end{center}
\begin{center}
\vspace{-1.cm}
\caption{\label{Fig:global_pipi} The pion form factor data and fit corresponding to the
iteration \# 1 BHLS global fit.   The $e^+e^-\ra \pi^+ \pi^-$ data samples 
are those shown in the  entry  "Fit Combination 2" in Table \ref{Table:T2}. The inset in the top 
panel magnifies the $\rho^0-\omg $ peak region. The dowmost panels magnify the behavior in
both distribution wings. See Section
\ref{global_iter_1} for further comments.}
\end{center}
\end{figure}

The results obtained when using  the KLOE10, KLOE12, NSK samples within the fit procedure are displayed
at the Entry "Fit Combination 2"  when BESSIII data are also included and "Fit Combination 3" when they are not;
 the data and fit corresponding to  the "Fit Combination 2"  are shown in Figure \ref{Fig:global_pipi}. 
Both Fit Combination 2 and Fit Combination 3 are  clearly satisfactory. 
 
Therefore, this proves that  the scan data from CMD2 and SND are consistent with
the KLOE10,  KLOE12 and BESSIII data samples and that all these are fully consistent with the other
data spectra introduced in the global fit procedure as indicated by the global fit probability.
One should also remark that the systematic uncertainties provided for KLOE12  
lead to a satisfactory global fit, in contrast with KLOE08, as already noted in
the previous Subsection.

Except otherwise stated, the fit parameter values presented from now on are derived
using the $e^+e^- \ra \pi^+ \pi^-$ data samples corresponding to the "Fit Combination 2"
(see Table \ref{Table:T2}); the fit results are those derived after the first iteration
and they do not differ significantly from the corresponding results at iteration \# 2.
The fit quality for the non--$\pi^+ \pi^-$ data samples are almost
undistinguishable from the numbers already given in the second data column from 
Table \ref{Table:T1}; they are not repeated for the sake of brevity.

\subsection{The  Iterative Method~: Updating The Model Parameter Values} 
\label{global_iter_2}
\indent \indent Beside improving the fits by mean of the iterative method,
the present work accounts for an error and a couple of bugs affecting  
 our \cite{ExtMod3,ExtMod4}. Moreover, the present work includes
 the new  KLOE12 data sample within the fit procedure; this is not harmless as  KLOE12
constrains the fit conditions more severely than the KLOE10 sample. Therefore, the present results 
 update  and supersede the corresponding ones previously given in  \cite{ExtMod3,ExtMod4}.
\subsubsection{The HLS--FKTUY Parameters}
\label{global_iter_2_s1}
\indent \indent 
The non--anomalous HLS Lagrangian (broken or not) can be written~:
\be
\displaystyle
{\cal L}_{HLS} = {\cal L}_{A} + a_{HLS}{\cal L}_{V} 
\label{Eq12}
\ee
The unbroken expression for ${\cal L}_{HLS}$  can be found in  \cite{HLSRef} and its
broken expression (BHLS) is given in  \cite{ExtMod3}.  The covariant derivative
which allows to construct both pieces of ${\cal L}_{HLS}$ introduces the fundamental parameter
 $g$,  known as universal vector coupling. The coefficient $a_{HLS}$ is a specific feature
of the HLS model, expected close to 2 in standard VMD approaches; however,  phenomenology
 rather favors $a_{HLS} \simeq 2.5$ since the early applications 
of the HLS model to pion form factor studies \cite{Heath1998,ffOld,WZWChPT}.

On the other hand, the anomalous (FKTUY) sector \cite{FKTUY} of the HLS model \cite{HLSRef}
consists of 5 pieces (see also Appendix D in  \cite{ExtMod3}), each weighted by a 
specific numerical parameter not fixed by the theory. Using common notations
\cite{HLSRef,ExtMod3} and  factoring out, for convenience, the weighting factors,
the FKTUY Lagrangian collecting all the anomalous couplings can be 
written\footnote{Actually, the erratum involved in  Eq. (\ref{Eq10})
comes from having missed the contribution of the $(c_4-c_3)$ term displayed
in Eq. (\ref{Eq13}) which actually turned out to impose $c_4=c_3$. 
 As already stated, after  correction, all the anomalous decay
couplings and the amplitudes for $e^+e^- \ra (\pi^0/\eta) \gamma$ anihilations
only depend on the combination $(c_4+c_3)/2$ and the single place where 
the difference $(c_4-c_3)$ occurs is  the $e^+e^- \ra \pi^0 \pi^+ \pi^-$
annihilation amplitude. In \cite{ExtMod3,ExtMod4} where $(c_4-c_3)$
was absent, its physical effect was absorbed by  $(c_1-c_2)$ 
to recover good fit qualities; so $(c_4-c_3)$ and $(c_1-c_2)$
should carry an important correlation.
\label{fktuy} }~:
\be
\displaystyle
{\cal L}_{FKTUY}=c_3{\cal L}_{VVP}+(c_4-c_3){\cal L}_{AVP}+(1-c_4){\cal L}_{AAP}
+(c_1-c_2-c_3){\cal L}_{VPPP}+(c_1-c_2+c_4){\cal L}_{APPP}
\label{Eq13}
\ee
where $P$ and $V$ indicate the basic pseudoscalar and vector meson nonets and 
$A$ the electromagnetic field. As ${\cal L}_{HLS}$, ${\cal L}_{FKTUY}$
depends on the universal vector coupling $g$. 
 
At iteration \# 1, the global BHLS fit returns~:
\be
\left \{
\begin{array}{lll}
\displaystyle
c_+ \equiv  \frac{(c_4+c_3)}{2} &= ~~~0.956 \pm 0.004\\[0.2cm]
\displaystyle
c_- \equiv  \frac{(c_4-c_3)}{2} &= -0.166 \pm 0.021\\[0.2cm]
\displaystyle
c_1-c_2  &=~~~0.915 \pm 0.052\\[0.2cm]
\displaystyle
g&=~~~5.507 \pm 0.001\\[0.2cm]
\displaystyle
a_{HLS}&=~~~2.479 \pm 0.001
\end{array}
\right .
\label{Eq14}
\ee
with correlation coefficients never larger than the percent level, except for
$<\delta g ~\delta a_{HLS}>=-0.30$ and $<\delta[c_1-c_2 ] ~\delta[(c_4-c_3)/2]>=+0.86$.
The sign of the $(g,a_{HLS})$ correlation term is easy to understand 
as the vector meson coupling to a pion or kaon pair 
rather depends on the product $g^\prime=a_{HLS} g$. The large value of the
$([c_1-c_2 ],~[c_4-c_3 ])$ correlation is also not surprising (see footnote \ref{fktuy}). 
The numerical values for $g$ and $a_{HLS}$ are in the usual ball park and do not call for more
comments than in \cite{ExtMod3,ExtMod4}.

Our value for $c_+$  agrees with the estimates
derived in \cite{HLSRef} from the $\pi ^0 \gamma \gamma^*$) 
form factor ($c_+=1.06 \pm 0.13$) and from
the $\omg \ra \pi^0  \gamma$  partial width ($c_+=0.99 \pm 0.16$)
with a much smaller uncertainty due to the large amount of data influencing
the (global) fit. After the bug fixing, $c_-$ is found small but non--zero
with  a  large significance and $(c_1-c_2)$ becomes closer to 1.
Using the full  $25\times25$ parameter error covariance matrix returned by
the global fit, we have computed  separately  $c_4$ and  $c_3$ by a Monte--Carlo sampling. 
This gives $c_3=1.124 \pm 0.022$ and $c_4=0.789 \pm 0.021$.

Among the  numbers displayed in Eq. (\ref{Eq14}), some are appealing~: The nearness to 1
of  the fitted $c_1-c_2$ and $c_+$  parameters, their customary
 guessed value   \cite{HLSRef}, should be noted
and deserves confirmation with more precise data on the anomalous annihilations and 
light meson radiative decays than those presently available. 

\subsubsection{The  Iterative Method~: Pseudoscalar Meson Mixing and Decay Parameters} 
\label{global_iter_2_s2}
\indent \indent The BHLS symmetry breaking of the  Lagrangian piece ${\cal L}_{A}$
leads to pseudoscalar physical fields constructed as linear combinations of their
bare partners. The  mechanism involved is the BKY 
mechanism extended so as to account for both Isospin and SU(3) symmetry breakings \cite{ExtMod3};
it can be complemented by the pseudoscalar nonet symmetry breaking scheme generated by
 the t'Hooft determinant terms \cite{tHooft}. The main effect of these
determinant terms is to provide the bare Lagrangian with a correction to the PS singlet
kinetic energy term  governed by a parameter $\lambda$  expected small
(see Eq. (7) in \cite{ExtMod3}). 
\begin{table}[!ht!]
\begin{center}
\begin{tabular}{|| c  || c  | c ||}
\hline
\hline		
\hhhd ~~~~  		& \hhhv General Fit   &  \hhhv Constrained Fit  		  \\
\hline
\hline
 \hhhv $\theta_0$  &    $2.77^\circ \pm 0.41^\circ$    &  {\bf 0}			  \\
\hline
 \hhhv  $\theta_8$ &    $-25.95^\circ \pm 0.35^\circ$   &  $-25.52^\circ \pm 0.20^\circ$  \\
\hline
 \hhhv	$\theta_P$ &    $-15.29^\circ \pm 0.32^\circ$   &  $-13.96^\circ \pm 0.16^\circ$  \\
\hline
 \hhhv $\lambda$   &    $(2.91 \pm 3.35) ~10^{-2}$   	&  $(1.86 \pm 1.17) ~10^{-2}$	  \\
\hline
 \hhhv $\varepsilon_0$ &    $(4.12 \pm 0.33) ~10^{-2}$  &  $(4.00 \pm 0.33) ~10^{-2}$	  \\
\hline
\hline
 \hhhv $\varepsilon(\eta)$   &    $(5.85 \pm 0.48) ~10^{-2}$   	&  $(5.57 \pm 0.47) ~10^{-2}$	  \\
\hline
 \hhhv $\varepsilon^\prime(\eta^\prime)$ &    $(1.46 \pm 0.13) ~10^{-2}$  &  $(1.36 \pm 0.12) ~10^{-2}$	  \\
\hline
\hline
 \hhhv $\chi^2/N_{dof}$ &    $887.5/994$  &  $892.5/995$	  \\
 \hhhw Probability &    99.3\%  &  99.1 \% \\	  
\hline
\hline
\end{tabular}
\end{center}
\caption {
\label{Table:T3}
Some parameter values derived when leaving free $\theta_P$ and $\lambda$ (first data column)
or when relating them by imposing $\theta_0 =0$ to the fit (second data column).
}
\end{table}

The BHLS model connects to (Extended) ChPT \cite{leutwb,leutw}, especially its two angle 
$\theta_0$ and $\theta_8$ mixing scheme; in particular, it relates these angles to the
singlet--octet mixing angle traditionally denoted $\theta_P$, together with the BKY breaking 
parameters $z_A$, $\Delta_A$  and to $\lambda$  \cite{ExtMod3}. 

The upper part of Table \ref{Table:T3} displays in its first data column our fit results
in the general case. The fit value for $\theta_8$ is in good agreement with other expectations 
\cite{leutw}  as well as that for  $\theta_0$. The  smallness of this has led us to impose 
$\theta_0=0$  within fits which leads to the results shown in the second data column. The value
for $\lambda$ undergoes a severe correction compared with \cite{ExtMod3,ExtMod4} and,
presently, because of its large uncertainty, could be neglected without any real degradation 
in fit qualities.

 BHLS also allows for some additional contribution to the
 $\pi^0-\eta-\eta^\prime$ mixing based on some possible aspects of Isospin breaking 
 not already accounted for by the extended BKY scheme developped in \cite{ExtMod3}.
 This turns out to redefine the physical (observable) fields (right--hand side)
 in terms of the (BHLS) renormalized (left--hand side) fields by \cite{leutw96}~:
\be
\left \{
\begin{array}{lll}
\displaystyle 
\pi_R^3= \pi^0-\varepsilon ~\eta-\varepsilon^\prime ~\eta^\prime\\[0.5cm]
\displaystyle 
\eta_R^8= \cos{\theta_P} (\eta+\varepsilon ~\pi^0)+
\sin{\theta_P} (\eta^\prime+\varepsilon^\prime ~\pi^0) \\[0.5cm]
\displaystyle 
\eta_R^0 = -\sin{\theta_P} (\eta+\varepsilon ~\pi^0)+
\cos{\theta_P} (\eta^\prime+\varepsilon^\prime ~\pi^0) 
\end{array}
\right.
\label{eq15}
\ee
Inspired by \cite{leutw96}, one can lessen the number of free parameters by stating~:
\be
\left \{
\begin{array}{ll}
\displaystyle
\varepsilon =& \displaystyle \epsilon_0 \cos{\theta_P}
\frac{\sqrt{2}\cos{\theta_P}-\sin{\theta_P}}{\sqrt{2}\cos{\theta_P}+\sin{\theta_P}} \\[0.5cm]
\displaystyle
\varepsilon^\prime =& \displaystyle -2 \epsilon_0 \sin{\theta_P}
\frac{\sqrt{2}\cos{\theta_P}+\sin{\theta_P}}{\sqrt{2}\cos{\theta_P}-\sin{\theta_P}} 
\end{array}
\right.
\label{eq16}
\ee
and fit $\epsilon_0$. Then, using the fit results (parameter central values and  
error covariance matrix), one can reconstruct the value for $\varepsilon$ and 
$\varepsilon^\prime$. The updated values are given
in Table \ref{Table:T3} still  indicate a $\pi^0-\eta$ mixing
much larger than the $\pi^0-\eta^\prime$ mixing (a factor of 4).

Before closing this Subsection, we mention that the Monte Carlo sampling method
allows to reconstruct the decay constant ratio $f_K/f_\pi=1.265 \pm 0.009$ which becomes
$f_K/f_\pi=1.295 \pm 0.002$ when constraining the fit with  $\theta_0=0$. 

\section{The Muon LO--HVP~: Evaluations From Iterated Fits} 
\label{global_iter_hvp}
\indent \indent 
\vspace{-1.5cm}
\begin{figure}[!ht]
\begin{minipage}{\textwidth}
\begin{center}
\resizebox{0.8\textwidth}{!}
{\includegraphics*{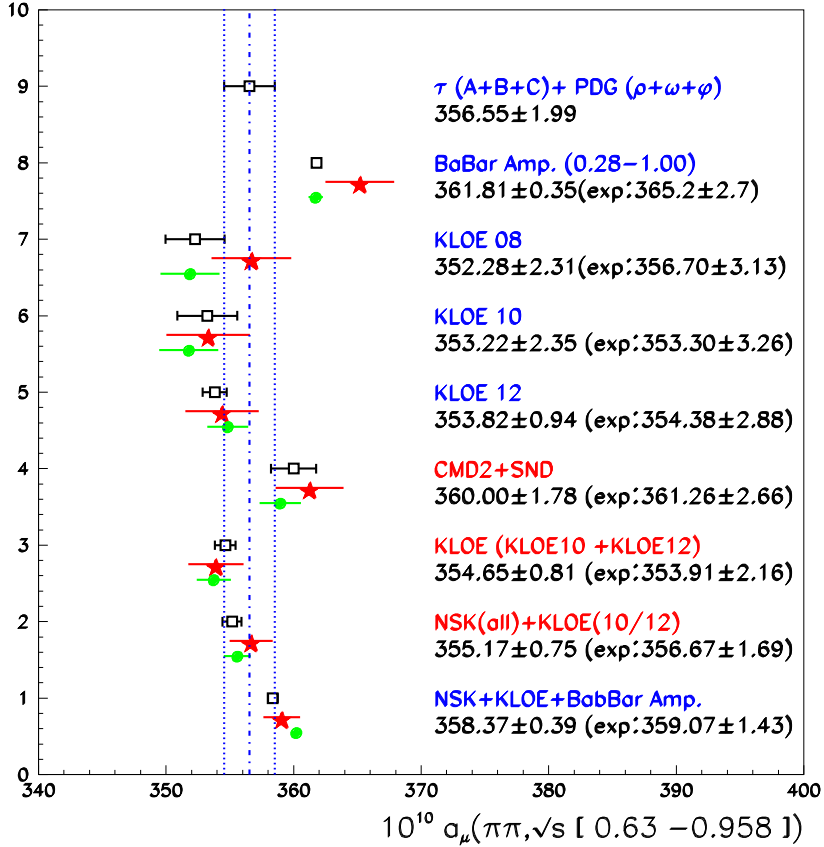}}
\end{center}
\end{minipage}
\begin{center}
\vspace{-1.cm}
\caption{\label{Fig:a_mu_ref} Values for $a_\mu(\pi \pi,[0.63,0.958])$ in units
of 10$^{-10}$ derived from global fits using the indicated $e^+e^-\ra \pi^+ \pi^-$ 
data samples or combinations; the $\tau$ dipion spectra are always used.
The full green circles are the results obtained from the $A=m$ fit (no iteration) and 
the  black empty squares are the results obtained from the $A=M_0$
fit (first iteration). The values derived by integrating the experimental spectra are 
 indicated by red stars. See Subsection \ref{hvp_1} for comments.}
\end{center}
\end{figure}
The main aim of the present study is to produce improved  estimates of the muon LO--HVP
\cite{ExtMod3,ExtMod4} by means of the  $iterated$ global fit method expected to 
cancel out possible biasing effects which could affect the $A=m$ ({\it i.e. non--iterated)} solution.
 The validity of the iterated method is supported  by the Monte Carlo study outlined 
in Appendix \ref{MC_test}, which clearly indicates that the iterated method cancels out possible biases
and returns, correctly estimated, the fit parameter uncertainties. Therefore, 
building on the conclusions collected in Subsection 
\ref{scale_err_3} one can produce  bias free evaluations of the muon LO--HVP. 
The effects of iterating\footnote{$M_0$ is the solution 
to the fit performed under the approximation  already named in short $A=m$ ({\it i.e.} each of the various 
$\pi^+ \pi^-$  experimental spectra is used for its individual contribution
to the global $\chi^2$).}  from $M_0$  to
$M_1$ --  the solution derived using  $A=M_0$  within the fit procedure
-- will be especially emphasized. To be complete, 
this update  also takes into account the new KLOE12  \cite{KLOE12} 
and BESSIII \cite{BESS-III}  $\pi^+ \pi^-$  data samples --
which happen to be very constraining --  and also corrects for some bugs.
Therefore the present numerical results supersede the corresponding ones in \cite{ExtMod3,ExtMod4}.

\subsection{Various Evaluations Of $a_\mu(\pi \pi,[0.63,0.958]~$GeV)}
\label{hvp_1}
\indent \indent
The point at top of Figure \ref{Fig:a_mu_ref} is the so--called $\tau$+PDG \cite{ExtMod4}
value for $a_\mu(\pi \pi,[0.63,0.958]~$GeV) derived by switching off the contributions of the
various $e^+e^- \ra \pi^+ \pi^-$ data samples from the minimized $\chi^2$, replacing them
by decay information extracted from the Review of Particle Properties (RPP) \cite{RPP2012} 
as emphasized in Subsection \ref{tau_pdg}.

In order to get the other points displayed in Figure \ref{Fig:a_mu_ref}, one 
always uses all the channels covered by BHLS, including the $\tau$ spectra from ALEPH, 
CLEO and Belle. As for the $e^+e^- \ra \pi^+ \pi^-$ data samples, one uses 
each of the BaBar, KLOE08, KLOE10 and KLOE12 samples in isolation  as indicated 
within the Figure (see also Table \ref{Table:T2} and  Subsection \ref{global_iter_1}). 
The point flagged by
 CMD2+SND is obtained from a fit to the so--called \cite{ExtMod3} 
 new timelike data from CMD2 and SND
 \cite{CMD2-1995corr,CMD2-1998-1,CMD2-1998-2,SND-1998},  leaving aside the older data from OLYA and CMD collected
 in \cite{Barkov} (see  Table \ref{Table:T2} and Section \ref{nsk_iter} above).
  As for the BaBar spectrum,
 for reasons already stated, the fit is performed on the spectrum amputated from the drop off region
 ($\sqrt{s} \in [0.76,0.80]~$ GeV). Finally, as the published BESSIII spectrum ends up at
 0.9 GeV, one cannot produce an experimental value on the interval $[0.63,0.958]$ GeV.
 
As a general statement, Figure \ref{Fig:a_mu_ref}  clearly illustrates 
that the iterated ($M_1$) and the non--iterated ($M_0$) solutions provide
 quite similar  fit estimates for $a_\mu(\pi \pi,[0.63,0.958]~$GeV).
 One should nevertheless remark that the agreement between both fit solutions 
 and the numerical integral of the experimental data is less satisfactory
 for the data samples which exhibit poor fit qualities within the global 
 framework (KLOE08 and BaBar) than
 for  the others (KLOE10, KLOE12, CMD2+SND)  as can be inferred from the "fit in isolation"
 properties displayed in  Table \ref{Table:T2}. Finally, the weighted averages of the experimental results
 for KLOE10 and KLOE12 alone or together with all NSK data (the so--called new
 timelike data and the former samples \cite{Barkov}) are always well reproduced by the global fit
 and are supported by quite good probabilities (see Table \ref{Table:T2}).

 Using the NSK+KLOE(10/12) sample configuration, the iterated BHLS global fit
 gives a slightly smaller central value (by $\simeq 1.5~10^{-10}$) while the uncertainty
 is improved by a factor $\simeq 2$. It is also worth  pointing out the role of
 the $\tau$ spectra within the BHLS global fit framework. The following
 numbers illustrate how the constraints involved by the $\tau$ spectra allow BHLS to 
 yield a more precise fit estimate for $a_\mu(\pi \pi,[0.63,0.958]~$GeV). 
 Comparing the direct integration result to the values derived from fits, one
 indeed gets at iteration \# 1~:
 \be
\left  \{
\begin{array}{lll}
{\rm Direct~Integration}~~~~~~ &:& a_\mu(\pi \pi,[0.63,0.958])= 356.67  \pm  1.69\\
A=M_0 ({\rm fit~excl. }\tau)~~~~ &:& a_\mu(\pi \pi,[0.63,0.958])= 355.07 \pm  0.96\\
A=M_0 ({\rm fit~incl.} \tau)~~~~ &:&  a_\mu(\pi \pi,[0.63,0.958])= 355.17 \pm  0.75
\end{array}
\right .
\label{Eq17}
\ee
in units of $10^{-10}$. 

Finally, the downmost point in Figure \ref{Fig:a_mu_ref} displays the 
result derived using all data samples (except for KLOE08 as there is not enough published information
to account for its strong correlation with KLOE12); 
this estimate for $ a_\mu(\pi \pi,[0.63,0.958])$
which benefits from a very small uncertainty has, however,  a poor fit probability as clear
from  Table \ref{Table:T2}.

 \subsection{Contributions To The Muon LO--HVP Up To 1.05 GeV}
 \label{hvp_2}
\indent \indent  

\begin{table}[!ht!]
\begin{center}
\begin{tabular}{|| c  || c  | c || c ||}
\hline
\hline		
\hhhd  Channel	& \hhhv $A=m$   &  \hhhv $A=M_0$ &  Exp. Value\\
\hline
\hline
 \hhhv $\pi^+\pi^-$  &  $495.06 \pm 1.43$  &  $494.59 \pm 0.89$ &  $492.98 \pm 3.38$   \\
\hline
 \hhhv  $\pi^0\gamma$ & $4.53 \pm 0.04$    &  $4.54 \pm 0.04$  &   $3.67 \pm 0.11$ \\
\hline
 \hhhv	$\eta\gamma$ &    $0.64 \pm 0.01$    & $0.64 \pm0.01$   &   $0.56 \pm 0.02$ \\
\hline
 \hhhv $\pi^+\pi^-\pi^0$ & $40.83 \pm 0.57$   &  $40.84 \pm 0.57$&   $43.54 \pm 1.29$\\
\hline
 \hhhv $K_L K_S$   &    $11.56 \pm 0.08$   	&    $11.53 \pm 0.08$ 	&     $12.21 \pm 0.33$\\
\hline
 \hhhv $K^+ K^-$ &    $16.79 \pm 0.20$  &  $16.90 \pm 0.20$	&   $17.72 \pm 0.52$ \\
\hline
\hline
 \hhhv Total &    $569.41\pm 1.55 $  &    $569.04 \pm 1.08 $	&     $570.68 \pm 3.67 $\\
\hline
\hline
\end{tabular}
\end{center}
\caption {
\label{Table:T4} The contributions to the muon LO--HVP from the various channels covered by BHLS from their respective
thresholds to 1.05 GeV in units of $10^{-10}$ at start and after iteration. 
 The last column displays the direct numerical integration of the various spectra
 used within BHLS. The $\pi^+\pi^-$  data samples considered are those flagged by "Combination 2" in Table \ref{Table:T2}.
 }
\end{table}

The LO--HVP's integrated from their respective thresholds up to 1.05 GeV are displayed in Table \ref{Table:T4}; the central
value for $a_\mu(\pi \pi)$ includes final state radiation (FSR) effects. The first data
column shows the results from the fit solution $M_0$ derived from fitting with $A=m$;
the second data column displays the results corresponding to the solution $M_1$ derived by fitting with
$A=M_0$.   These two data columns report on the fits performed using all annhilation channels
 encompassed by BHLS $and$ the $\tau$ dipion spectra. Finally, the rightmost data
column provides the  direct numerical integration of the 
experimental spectra -- actually  only those feeding the BHLS fit procedure, including the 
KLOE10, KLOE12 and BESSIII data samples besides the scan data. 

As for the $\pi^+\pi^-$ channel, both fits -- which include the $\tau$ spectra -- provide
central values in agreement with each other and with the direct estimate within the 
quoted error\footnote{\label{bias_exp}
As for the central value 
of the experimental estimate which is the present concern, one can legitimately expect
that it should be affected by some bias ({\it a priori}, of unknown magnitude) of the same nature than the $A=m$ result.
Indeed, roughly speaking, the experimental cross section $\sigma_{exp}(s)$ is related with the underlying
theoretical cross section $\sigma_{th}(s)$ by a relation of the form 
$\sigma_{exp}(s)=\sigma_{th}(s) + \delta \sigma(s)$ and the $ \delta \sigma(s)$   correction
depends on the normalization uncertainties which just motivate the iterative method!
 Actually, this $\delta \sigma(s)$ is exactly the  scale dependent term in 
  Eqs. (\ref{Eq5}) and  (\ref{Eq8}). Obviously it cannot be estimated
without some fitting procedure. }. 
If the $A=m$ solution were (inherently) exhibiting
a bias, comparing the first two numbers in the first line of  Table \ref{Table:T4} indicates that this
does not exceed $\simeq 0.5\times 10^{-10}$ -- {\it e.g.} half a standard deviation. Therefore, real 
experimental data samples confirm the gain provided by a global fit procedure when samples with  
 normalization errors small compared to their statistical accuracies 
 are included; exploring this effect is the purpose of  Subsection \ref{iter_2} in Appendix \ref{MC_test}. 

 One should also remark that the unbiasing iterative procedure lessens significantly the uncertainty
 on $a_\mu(\pi^+\pi^-)$ compared with the $A=m$ solution and, over the whole range of validity
 of BHLS (up to 1.05 GeV),  one ends up with a factor of $\simeq$ 3  reduction of the uncertainty 
 compared to the direct numerical integration. The same kind of effect is 
 reported in \cite{Ball} concerning the spread of the parton density 
 functions\footnote{In particular, Figure 5 in this Reference, is quite  informative 
 about the variety of correction kinds revealed by unbiasing procedures.}.

Therefore, relying on the iterative procedure, one observes that the global fit 
does not produce significant shifts of the central values of the HVP contributions which
could be attributed to the normalization (scale) uncertainties strongly affecting some
data samples. Relying on the Monte Carlo studies outlined in Appendix \ref{MC_test}, this can be attributed  
 to the large number of data samples where the statistical uncertainties dominate over 
 the normalization uncertainty. Moreover, the uncertainty on the part of the LO--HVP
 derived from the BHLS fit (more than 80\% of the total LO--HVP) is very small and even marginal.

 \subsection{The Muon $g-2$ From BHLS Global Fit Procedure}
 \label{hvp_3}
\begin{table}[ph!]
\hspace{-1.cm}
\begin{tabular}{|| c  | c || c  || c  ||}
\hline
\hhhv Contribution from	&  Energy Range           &   LO--HVP (2014)              &   LO--HVP (2011)  \\
\hline
\hhhv missing channels 	&  threshold $\to$ 1.05   &  {$1.34 (0.03)(0.11)[0.11]$}  &  {$1.44 (0.40)(0.40)[0.57]$}\\    
\hline
\hhhv $J/\psi$ 		& ~~~  			  &  {$8.94(0.42)(0.41)[0.59]$}   &  {$8.51(0.40)(0.38)[0.55]$} \\
\hline 
\hhhv $\Upsilon$        & ~~~  			  &  {$0.11(0.00)(0.01)[0.01]$}   &   {$0.10(0.00)(0.01)[0.01]$} \\
\hline  
\hhhv hadronic		& (1.05, 2.00)		  &  {$60.45(0.21)(2.80)[2.80]$}  &  {$60.76(0.22)(3.93)[3.94]$} \\
\hline 
\hhhv hadronic		& (2.00, 3.10)		  &  {$21.63(0.12)(0.92)[0.93]$}  &  {$21.63(0.12)(0.92)[0.93]$} \\
\hline 
\hhhv hadronic		& (3.10, 3.60)		  &  {$3.77(0.03)(0.10)[0.10]$}   &  {$3.77(0.03)(0.10)[0.10]$} \\
\hline
\hhhv hadronic		& (3.60, 5.20)		  &  {$7.50(0.04)(0.05)[0.06]$}   &  {$7.64(0.04)(0.05)[0.06]$} \\
\hline
\hhhv  pQCD	        & (5.20, 9.46)		  &  {$6.27(0.00)(0.01)[0.01]$}   &  {$6.19(0.00)(0.00)[0.00]$}\\
\hline 
\hhhv hadronic     	& (9.46, 13.00)		  &  {$1.28(0.01)(0.07)[0.07]$}   &  {$1.28(0.01)(0.07)[0.07]$}\\
\hline
\hhhv  pQCD		& (13.00,$\infty$)	  &  {$1.53(0.00)(0.00)[0.00]$}   &  {$1.53(0.00)(0.00)[0.00]$}\\
\hline
\hhhv Total		& 1.05 $\to \infty$	  &  {$112.82 \pm 3.01_{tot}$}    &  {$112.96 \pm4.13_{tot}$} \\  
\hhhe ~~~		& + missing channels	  &  {~~~} &  {~~~}\\
\hline
\hline
\end{tabular}
\caption{
\label{Table:T5} LO--HVP contributions to $10^{10} a_\mu$ with FSR corrections included. The
statistical and systematic errors are given within brackets; the total uncertainty
is given within square brackets. Column  "LO--HVP (2011)"  displays the contributions
estimated using only the data samples available in 2011; Column  "LO--HVP (2014)" displays the corresponding
 values updated with the data samples published up to the end of 2014. 
 }
\end{table}
\indent \indent  In order to evaluate the muon LO--HVP from the fit results derived by means of the BHLS
global fit procedure, the numbers given in Table \ref{Table:T4} should be supplied with
several additional contributions which cannot be derived from within the BHLS framework but should 
be estimated by other means. This covers the channels opened below 1.05 GeV but remaining outside the 
present BHLS scope\footnote{For instance the 4, 5 of 6 pion annihilation
channels, or the $\omg \pi^0$ final state.} and, more importantly,  all hadronic contributions
covering the non--perturbative QCD region  above 1.05 GeV should be 
estimated via the direct integration method. 

Table \ref{Table:T5} summarizes these additional
contributions to be combined with the BHLS results to derive the muon LO--HVP; in this Table, one 
reminds the information available by end of 2011 and used in our previous
\cite{ExtMod3,ExtMod4}. The data column flagged by "LO--HVP (2014)" is the update
derived by taking into account
the data samples more recently collected (and published up to the end of 2014); these are the 
$e^+e^-\to 3(\pi^+\pi^-)$ data from CMD--3 \cite{Akhmetshin:2013xc}, the 
$e^+e^- \to \omega\pi^0 \to \pi^0\pi^0\gamma$ from SND \cite{Achasov:2013btb} and several
data samples collected by BaBar in the ISR mode\footnote{These cover the $p \bar{p}$, 
$K^+K^-$, $K_LK_S,\:K_LK_S\pi^+\pi^-$, $K_SK_S\pi^+\pi^-,K_S K_S K^+K^-$
annihilation final states.} \cite{Lees:2013ebn,Lees:2013gzt,Lees:2014xsh,Davier:2015bka}. 
These data samples highly increase the available statistics for the annihilation channels opened above
1.05 GeV and lead to significant improvements. One thus should note the important 
improvement these provide for the LO--HVP contribution from the $[1.05,~2.0]$ GeV region~:
its uncertainty is reduced by 25 \%, while its central value is almost unchanged. Despite this
improvement, the energy region $[1.05,~2.0]$ GeV still remains the dominant
uncertainty on the muon LO--HVP and this strongly limits the effect of gaining
further in precision  on the  part of the LO--HVP covered by BHLS.

Deriving the full HVP value also requires to account for
the higher order effets. This includes  the next--to--leading order contribution (NLO)  taken from \cite{Fred11} 
($[-9.97 \pm 0.09]\times 10^{-10}$) and the recently estimated next--to--next--to--leading order (NNLO)
effects which happen to be non--negligible ($[1.24 \pm 0.01]\times 10^{-10}$) \cite{NNLO}.

To compute the muon $g-2$, one should also include the light--by--light (LBL) contribution 
(here taken from \cite{LBL}), the QED contribution \cite{Passera06,FJ2013} and the electroweak
contribution (EW) \cite{Fred09}. The next--to--leading order contribution to the LBL
amplitude (NLO--LBL) has also been computed recently \cite{LbLNLO} but is clearly negligible
($[0.3 \pm 0.2] \times 10^{-10}$). Altogether,  the numerical values we use (see Table \ref{Table:T6}) 
are rather consensual \cite{Knecht:2014}.
\begin{table}[!ht]
\hspace{-0.75cm}
\begin{tabular}{|| c || c | c || c ||}
\hline
\hline
\hhhv $10^{10} \times a_\mu$  &    \multicolumn{2}{|c||}{Values (incl. $\tau$)}& Direct Integration   \\  
\hhhv ~~~   &   scan only &   scan $\oplus$ KLOE $\oplus$ BESSIII & scan $\oplus$ KLOE $\oplus$ BESSIII\\
\hline
\hhhv LO--HVP   &    $683.26  \pm 3.78$   &    $681.86 \pm 3.20 $ & $683.50 \pm 4.75 $   \\
\hline
\hhhv HO (NLO) HVP &   \multicolumn{3}{|c||}{ $-9.97  \pm 0.09 $~~~\cite{Fred11}} \\
\hline
\hhhv NNLO HVP &   \multicolumn{3}{|c||}{ $~1.24  \pm 0.01 $~~~\cite{NNLO}} \\
 \hline
\hhhv LBL &   \multicolumn{3}{|c||}{ $10.5 \pm 2.6$~~~\cite{LBL} }   \\
\hline
\hhhv NLO--LBL &   \multicolumn{3}{|c||}{ $0.3 \pm 0.2$~~~\cite{LbLNLO} }   \\
\hline
\hhhv QED &   \multicolumn{3}{|c||}{$11~658~471.8851 \pm 0.0036$~~~\cite{Passera06,FJ2013}}   \\
\hline 
\hhhv EW &   \multicolumn{3}{|c||}{$15.40\pm
0.10_{\rm had} \pm 0.03_{\rm Higgs,top,3-loop}$~~~\cite{Fred09} }   \\
\hline  
\hline 
\hhhv Total Theor.& $11~659~172.62 \pm  4.60 $ & $11~659~171.22 \pm 4.13 $ & $11~659~172.86 \pm 5.42 $ \\
\hline 
\hline 
\hhhv Exper. Aver. &  \multicolumn{3}{|c||}{$11~659~208.9 \pm 6.3 $ } \\
\hline 
\hline 
\hhhv $\Delta a_\mu$ & $36.28\pm 7.80 $  & $37.68\pm 7.53 $ & $36.04 \pm 8.31$  \\
\hline
\hhhv Significance ($n \sigma$) & $4.65 \sigma $ & $5.00 \sigma$ &  $4.38 \sigma$ \\
\hline
\hline
\end{tabular}
\caption{\label{Table:T6} 
\footnotesize The various contributions to $10^{10} a_\mu$. 
$\Delta a_\mu= a_\mu^{exp}-a_\mu^{th}$ is given in units of $10^{-10}$.
For the measured value $a_\mu^{exp}$,
 we have adopted  the value reported in the RPP which uses the updated value for
$\lambda=\mu_\mu/\mu_p$ recommended by the CODATA group~\cite{CODATA2012}. By KLOE, 
one means that the KLOE10 and KLOE12 $\pi^+ \pi^-$ data samples are introduced in the BHLS fit
procedure $and$ in the directly integrated spectra. }
\end{table}

The first data column in Table  \ref{Table:T6} reproduces (after our methodological update)
the muon anomalous moment estimate coming from the corresponding BHLS global fit where
only the scan data for the $\pi^+ \pi^-$ channel are considered while all ISR data are excluded.
This supersedes the corresponding information in \cite{ExtMod3}.
The sample combination preferred by the BHLS global fit gives the results displayed in the second data
column; it exhibits a  $4.9 \sigma$ significance for a non--zero
$\Delta a_\mu= a_\mu^{exp}-a_\mu^{th}$. The evaluation derived by direct
integration of the spectra used within the global fits are given in the third data column.
The new data, as a whole,   increase the discrepancy for $\Delta a_\mu$ which is always found above 
the $4 \sigma$ level; effects of additional and not still accounted for systematics will be examined
in the next Subsection.

\begin{figure}[!th]
\begin{center}
\resizebox{0.8\textwidth}{!}
{\includegraphics*{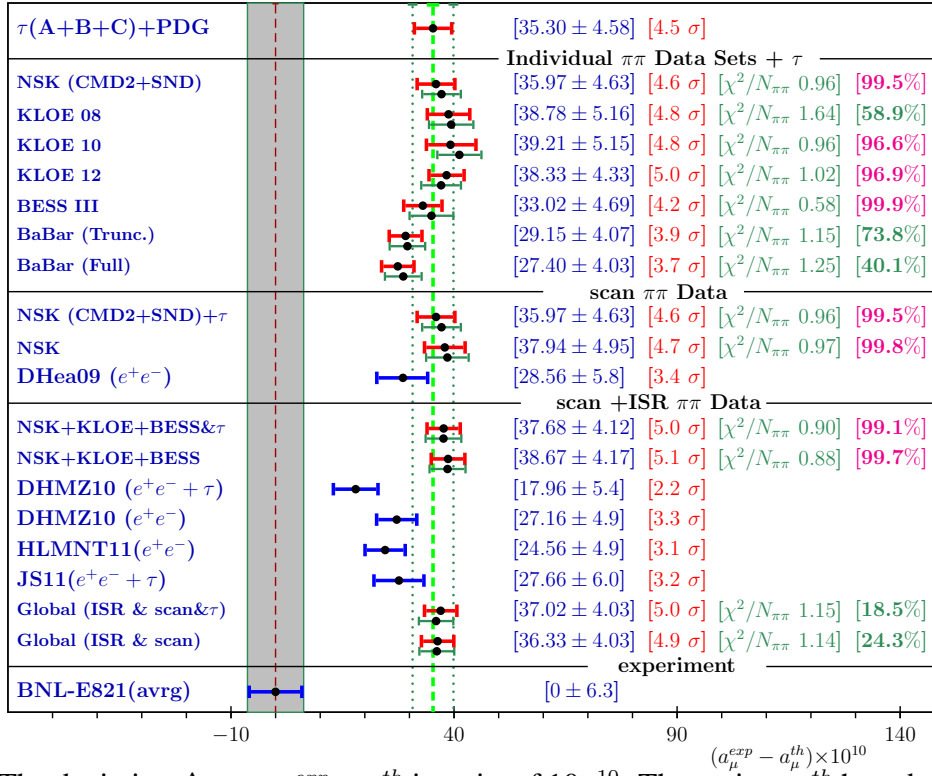}}
\end{center}
\begin{center}
\vspace{-1cm}
\caption{\label{Fig:gmoins2_ref} The deviation $\Delta a_\mu=a_\mu^{exp}- a_\mu^{th}$ in units
of 10$^{-10}$. The various $a_\mu^{th}$ have been derived from the global fit 
using the indicated $e^+e^-\ra \pi^+ \pi^-$ data samples and including/excluding the $\tau$ dipion spectra
as indicated. In red we display   $\Delta a_\mu$  corresponding to the iterated solution and in green
those corresponding to the $A=m$ (non--iterated) solution. In blue results from other studies are given  
corrected by
the recently evaluated next--to--next--to--leading order contribution \cite{NNLO}. 
See Section \ref{hvp_3} for comments.}
\end{center}
\end{figure}

Figure \ref{Fig:gmoins2_ref} displays the results for  $\Delta a_\mu$
derived using or not the $\tau$ data and various combinations of the available $\pi^+ \pi^-$
data samples introduced within the BHLS global fit procedure at first iteration. For comparison, one
also displays in this Figure the evaluations produced by other authors and flagged by
Dhea09 \cite{DavierHoecker}, DHMZ10 \cite{DavierHoecker3}, JS11 \cite{Fred11}
and HLMNT11 \cite{Teubner} -- corrected however for the recently calculated  NNLO--HVP and NLO--LBL
-- contributions as included in Table \ref{Table:T6}. 
{\it A priori}, the Dhea09 estimate compares exactly to 
our evaluations using scan data only;  the other results are derived using, beside the NSK samples,
 the BaBar, KLOE08 and KLOE10 samples. These may compare to the last couple of lines in 
Figure \ref{Fig:gmoins2_ref} where the scan data are supplemented with
the BaBar (not truncated), KLOE (10/12) and BESSIII samples.

The following comments are in order here~:
\begin{itemize} 
\item {\bf 1/} The difference between our estimates and those of other authors mainly
concerns  the estimated central value for $\Delta a_\mu$. Also, our uncertainties 
are now reduced because of the global fit method, but  also because of using
much more data samples than other authors; this is clear by comparing the  
errors shown in Figure \ref{Fig:gmoins2_ref}  with those  given in \cite{ExtMod4}.

When using only the scan data, the shift one observes should reflect the biasing effect certainly
present in the experimental data (see footnote \# \ref{bias_exp})  and corrected 
in our approach by the iterated fit method. When the ISR $\pi^+ \pi^-$ samples are also involved,
the issue just reminded is amplified because the weight of samples with large overall
scale uncertainties is much increased\footnote{$All$ ISR data samples are strongly dominated
by overall scale uncertainties, additionally $s$--dependent.}. The effect of the BaBar
data sample is no longer enough to balance the effect of the new data samples as clear
 by comparing the lines for
"NSK+KLOE+BESSIII" with the lines for "Global (ISR+scan)" which also include the (full) BaBar sample.
Nevertheless, one should note the large difference of the correponding probabilities.

\item {\bf 2/} When a comparison between a $\Delta a_\mu$ estimate derived
using the $\tau$ data and the corresponding one excluding these is possible,
ours exhibits the smallest difference ($1.12 \times 10^{-10}$ for NSK+KLOE+BESSIII,
$-0.7\times 10^{-10}$ for the Global fit including all the $\pi^+ \pi^-$ data samples). 
This is certainly due
to the vector meson mixing which defines the BHLS model. It is interesting to 
note  that the JS11  \cite{Fred11} value, which is based on the $\gamma-\rho^0$ mixing
by loop transitions\footnote{Within the BHLS model too, the $\gamma-\rho^0$ mixing is mediated by
loop effects.}, is the closest to ours.

\item {\bf 3/} Relying on the global fit properties, the BHLS model
favors the "NSK + KLOE10 + KLOE12 +BESSIII + $\tau$" as the largest consistent set of data samples.
This leads to   $\Delta a_\mu=(37.55 \pm 4.12)\times 10^{-10}$ which exhibits a $5. \sigma$
significance\footnote{If using the data from 2011 in Table \ref{Table:T5}, as in our previous
studies, this significance is "only" $4.8 \sigma$. This compares more directly to the results
from other authors displayed in Figure (\ref{Fig:gmoins2_ref}). The increased significance is
a pure consequence of the recent improvements of the hadronic contribution from the
$[1.05,~2.0]$ GeV region.}. 
Our estimate is expected to be free of biases  generated by the
overall scale uncertainties which dominate the ISR  $\pi^+ \pi^-$ data samples.
\end{itemize}

\subsection{Additional Systematics On The BHLS Estimate For The Muon $g-2$}
 \label{hvp_4}
\indent \indent A detailed study of additional systematics possibly affecting  the
BHLS evaluation of  $\Delta a_\mu$ has been already performed in \cite{ExtMod4}.
It concluded to an uncertainty of the LO--HVP central value for 
$\Delta a_\mu=a_\mu^{exp} -a_\mu^{th}$ in the range
 $[-1.3 \div 0.60]\times 10^{-10}$ coming from  $\pi^+ \pi^-$ contribution in the
$\phi$ mass region, where BHLS is weakly constrained. An uncertainty coming
from using the $\tau$ spectra has also been considered; it was argued that
the best motivated evaluation of this is the difference between fitting with the $\tau$
spectra and without them in the most constrained configuration. Presently, this
means that the BHLS preferred value ($\Delta a_\mu=(38.58 \pm 5.04)\times 10^{-10}$)
could be underestimated by  $\simeq 0.9\times 10^{-10}$.

\begin{figure}[ph!]
\begin{center}
\includegraphics[height=0.34\textheight]{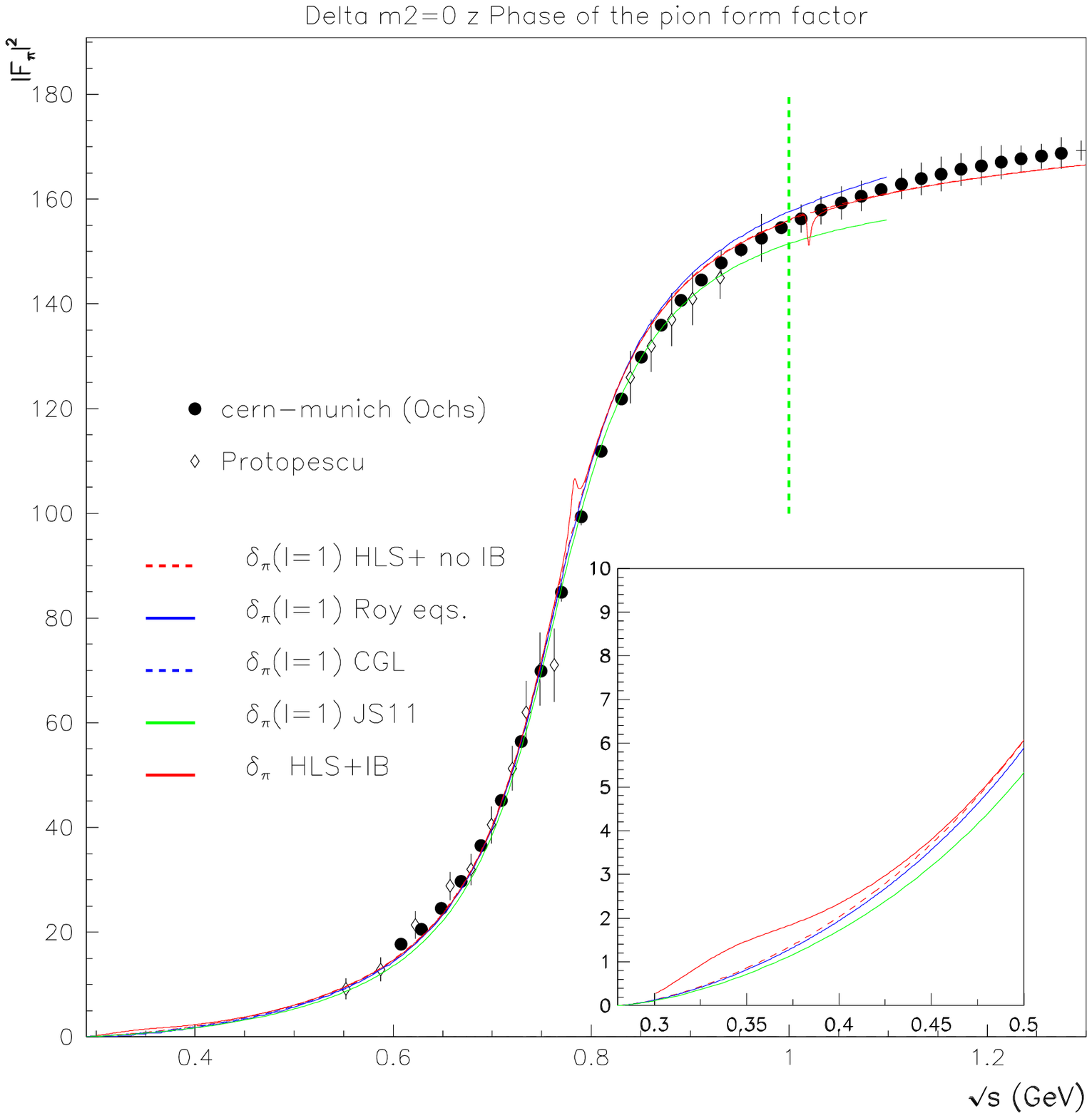}
\includegraphics[height=0.34\textheight]{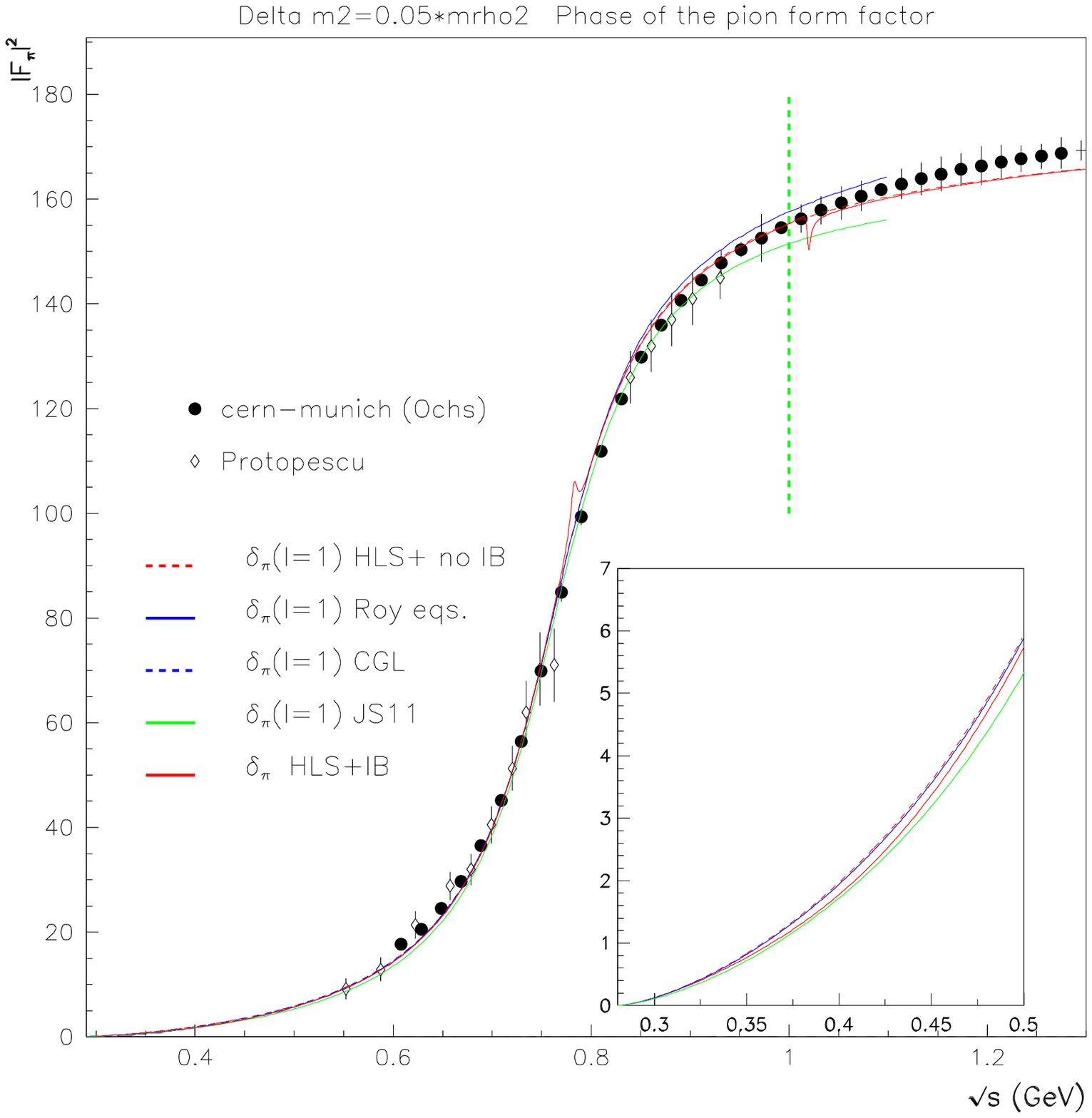}
\end{center}
\vspace{-0.3cm}
\caption{\label{Fig:PhaseShift} $P$--wave $\pi^+ \pi^-$ phase--shift data and predictions from
\cite{CGL} (CGL) and \cite{Fred11} (JS11) together with the BHLS phase--shift. The insets
magnify  the various behaviors close to threshold. See Subsection \ref{hvp_4} for further 
explanations.}
\end{figure}

Another mean to detect systematics is to compare with 
the accurate ChPT predictions on the $P$--wave $\pi^+ \pi^-$ phase--shift
\cite{CGL} and also with the available experimental data from the Cern--Munich
 \cite{Ochs} and Fermilab   \cite{Protopescu} groups. These are shown in Figure
\ref{Fig:PhaseShift}. Included also are the predictions derived from the Roy Equations
\cite{RoyEq}
  and from the phase of the pion form factor fit performed in \cite{Fred11} (JS11).

As for the BHLS predictions corresponding to using NSK+KLOE(10/12), we display in
this Figure the phase 
of the full  amplitude and those corresponding to dropping out the isospin breaking (IB) 
effects due to the vector meson mixing\footnote{This is obtained by cancelling out 
the "angles" $\alpha(s)$, $\beta(s)$ and $\gamma(s)$ from the full amplitude expression.}.
The $\tau$ spectra are included within the fit procedure.

The standard BHLS phase shift predictions are  displayed in the
left--hand side panel of Figure \ref{Fig:PhaseShift}. One clearly observes
a very good prediction of the phase--shift up to about 1.2 GeV, {\it i.e.}
much beyond our fitting range (from threshold to 1.0 GeV for the $\pi \pi $
data). Indeed the Cern--Munich data are very well accounted for and the BHLS predictions
are in accord with the other predictions. The inset, however, exhibits a (minor) issue 
for the full amplitude phase, a small bump of about $1^\circ$ close to threshold, 
absent from the IB amputated amplitude. 
 This can be tracked back to a peculiarity of the broken HLS model which does not 
 split up  the HK (Lagrangian) masses for the $\omg$ and $\rho^0$ mesons  and, consequently,
 the mixing angle $\alpha(s)$  does not exactly vanish at $s=0$ (see Figure
 6 in \cite{ExtMod1});  in contrast 
the other angles fulfill  $\beta(0)=\gamma(0)=0$. Indeed, one has~:  
\be
\displaystyle
\alpha(s) = \frac{\epsilon_1(s)}
{[m_\rho^{HK}]^2-[m_\omg^{HK}]^2+ \Pi_{\pi \pi}(s)}
\label{Eq18}
\ee
where \cite{ExtMod3}  $\epsilon_1(s)$ is the difference of the charged and neutral
kaon loops and $\Pi_{\pi \pi}(s)$ is the pion loop which both vanish at $s=0$. 
This assumption has been checked with  fits by imposing
$[m_\omg^{HK}]^2= (1+\eta)[m_\rho^{HK}]^2$ and choosing various fixed values for
$\eta$; the right--hand side panel in Figure \ref{Fig:PhaseShift} displays
the phase shift  for $\eta=5\%$ and, quite satisfactorily, its inset does not
 reveal a bump any longer.
A non--zero (HK) mass difference  $\eta~[m_\rho^{HK}]^2$
cannot be generated by the breaking mechanisms already implemented
within BHLS. However, a breaking of the nonet symmetry in the vector meson sector
(VNSB) enables such an effect; this turns out to modify the customary 
vector field matrix 
-- actually U(3) symmetric -- within the covariant derivatives of the HLS model 
\cite{HLSRef} by a perturbation term proportional to the singlet vector field
combination. The effect of VNSB  has been derived from 
 specific fit studies and indicates that $\Delta a_\mu$ might have to be 
lessened by about $1.4 \times 10^{-10}$.

Therefore, in total, the BHLS favored result  can be expressed, in units
of $10^{-10}$ as~: 
\be
\displaystyle
\Delta a_\mu = 37.68 + [^{+0.6}_{-1.3}]_\phi
+[^{+0.9}_{-0.0}]_\tau +[^{+0.0}_{-1.4} ]_{VNSB}
\pm 4.12_{th} \pm 6.3_{exp} 
\label{Eq19}
\ee
where the three additional contributions play as shifts on the central value. Adding them
up linearly, the maximum shift ($-2.7 \times 10^{-10}$) may reduce the central value 
to $34.85 \times 10^{-10}$ which has still a $4.6 \sigma$ significance. The effect of
these additional systematics is to reduce potentially by $\simeq 0.3 \sigma$ all the significances
displayed in Figure \ref{Fig:gmoins2_ref}. These 
are not due to overall scale uncertainties already accounted for by the
iterative method; they might be reduced by new annihilation data samples 
covering the region up to 1.05 GeV in all the physics channels in the realm of BHLS. 

\subsection{The HVP Slope At Origin In BHLS Fits }
\label{deRafael}
\indent \indent 
In the lattice QCD approach of calculating $\amuh$, extrapolation
methods have been developed (see e.g. contributions
to~\cite{BenayounMainz}) to overcome difficulties to reach the
physical point in the space of extrapolations. The low $Q^2$ behavior of
the euclidian electromagnetic current correlators on a lattice, which
exhibits a discrete momentum spectrum, poses a particular challenge
(see e.g.~\cite{Boyle:2011hu,Aubin:2013daa} and References just below). The
analysis of moments of the subtracted photon vacuum polarization
function $\Pi(Q^2)$ was particularly advocated in variants in
Refs.~\cite{deDivitiis:2012vs} and \cite{Aubin:2012me}. Recent lattice
calculations~\cite{Feng:2013xsa,Francis:2014dta,DellaMorte:2014rta,Malak:2015sla}
have been utilizing moment analysis techniques for a more precise
evaluation of $\amuh$. The leading moment is given by the slope of the
Adler function~\cite{deRafaelBell,deRafaelOrg}, the latter being given by~:
\be
D(Q^2)=Q^2\:\left[\int_{s_{min}}^{\infty}\frac{
R(s)}{(s+Q^2)^2}\D s\:\right]=\frac{3 \pi}{\alpha} Q^2 \frac{d}{dQ^2} \Delta\alpha_{\mathrm{had}}(-Q^2)
\label{Eq20}
\ee
where $R(s)$ is the {\it hadronic} spectral function\footnote{$R(s)=\sigma(e^+e^- \ra hadr.)/
\sigma(e^+e^- \ra \mu^+\mu^-)$ with $\sigma(e^+e^- \ra \mu^+\mu^-)=4 \pi \alpha^2/3s$
by neglecting the electron mass.} and
$s_{min}$  the smallest threshold energy squared ($s_{min}=m_{\pi^0}^2$ within  BHLS). 
Then, defining~:
\be
P_1=\int_{s_{min}}^{\infty}\,\frac{R(s)}{s^2}\,ds~~~~,
 \label{Eq21}
\ee
the HVP slope at the origin is given by~:
\be
\displaystyle
\left.\frac{d}{ds} \, \Delta  \alpha_{\mathrm{had}}(s)\right|_{-s\to +0}=
-\frac{\alpha}{3\pi}\,\int_{s_{min}}^{\infty}\,\frac{R(s)}{s^2}\,ds =
-\frac{\alpha}{3\pi}  P_1
\label{Eq22}
\ee

The constant 
$P_1$ can be directly estimated from data and partly from the BHLS fits. 
Therefore, one can proceed as done above with our evaluations of $\amuh$
 and derive the results gathered\footnote{The non--HLS part of $P_1$ amounts 
to $1.76 \pm 0.06$ GeV$^{-2}$.} in Table \ref{Table:slope_der}. Here,
one observes that the difference between the experimental and the HLS values 
for the HVP slope are at the percent
level (a $2~\sigma_{exp}$ effect)
and the uncertainty is scaled down by a factor of 10. However, to really feel the HLS
improvement on the slope, one needs once more an improved hadronic spectral function
at high energies.

\begin{table}[!ht]
\centering
{\footnotesize
\begin{tabular}{lc|c|c|c}
\hline\noalign{\smallskip}
moment &
{data direct} &
{HLS channels data} &
{HLS model} &
{HLS + non HLS}  \\
\noalign{\smallskip}\hline\noalign{\smallskip}
 $P_1$ (GeV$^{-2}$) & 11.83$\pm$    0.08&  10.07$\pm$ 0.05&  9.970$\pm$ 0.016 &  11.73 $\pm$ 0.06 \\[0.3cm]
 $ \displaystyle 10^2 \frac{d\Delta\alpha_{\mathrm{had}}}{\D s}(0)$&
  $-0.92\pm 0.01$&  $-0.78\pm 0.01$ &$-0.772 \pm 0.001$ & $-0.907 \pm$ 0.01\\
\noalign{\smallskip}
\hline
\end{tabular}
}
\caption{\label{Table:slope_der} 
\footnotesize The slope of the photon HVP at $s=0$.}
\end{table}

A lattice estimate of the Adler function slope $D'(0)$ has been
presented in~\cite{Francis:2013qna}. The result is 
${P}_1=5.8(5)~\gv^{-2}$, and has been compared with
${P}_1=9.81(30)~\gv^{-2}$, a result estimated
using a phenomenological toy-model
representation~\cite{Bernecker:2011gh} of the isovector
spectral function. The lattice results too include the isovector
part only and is missing higher energy contributions above 1 GeV.

In the study  \cite{Aubin:2012me}, the authors provide numerical values from
fits to Lattice data based on  Pad\'e approximants (PA). For this purpose, they
parametrize the HVP as~:
\be
\displaystyle
\Pi(Q^2)=\Pi(0) - Q^2 \left[a_0 + \sum_{n=1}^N 
\frac{a_n}{b_n+Q^2} \right]
\label{Eq23}
\ee
which thus leads to ~:
\be
\displaystyle
\frac{d\Delta\alpha_{\mathrm{had}}}{\D Q^2}(0)
= 4 \pi \alpha \frac{d\Pi}{\D Q^2}(0)=-4 \pi \alpha
\left[a_0 + \sum_{n=1}^N 
\frac{a_n}{b_n} \right]
\label{Eq24}
\ee
The parameters corresponding to the results they consider as optimal are given
in their Table 3. Using their notations, their fitted parameter values
lead\footnote{Assuming also the errors on the $a$'s and $b$'s parameters
are not correlated.}, for instance,
 to $(0.71 \pm 0.15)\times 10^{-2}$ (PA solution  [0,1]) or
 $(0.75 \pm 0.30)\times 10^{-2}$ (PA solution [1,1]). These compare reasonably well
 to the slope results reported in Table \ref{Table:slope_der} just above, taking into account
 the proviso expressed above about lattice data.

\section{Concluding Remarks} 
\label{conclusion}
\indent \indent The present study was motivated by the question which gives its title
to this paper. More precisely, the issue is whether the D'Agostini bias
\cite{D'Agostini,Blobel_2006} prevents to derive unbiased physical results from
global fits to experimental spectra affected by dominant overall scale 
uncertainties\footnote{We gratefully acknowledge G. Colangelo to have pointed out the issue 
for estimating the muon HVP using global fit methods. However, the bias issue is more general
as will be argued shortly.}. 

Actually, several issues are merged together. First, the effective {\it global} $\chi^2$ functions to 
be used in the minimization procedure should be appropriately defined.
 For the data samples where the statistical errors dominate the overall scale uncertainties, the
construction of the associated {\it partial} $\chi^2$'s is quite standard. The real issue starts
when the data samples are dominated by overall scale uncertainties. For each of them, 
substantially, the canonical partial $\chi^2$  has been reminded in Section \ref{EF_method}
and writes \cite{D'Agostini,Blobel_2003,Blobel_2006}~:
$$\chi^2=[m-M(\vec{a})-\lambda  A]^TV^{-1}[m-M(\vec{a})-\lambda  A],$$
  \noindent leaving aside the so--called "penalty term"  \cite{Blobel_2006}
proportional to $\lambda^2$. The (partial) $\chi^2$ being appropriately defined, 
another issue is the choice of the vector $A$.

In our former studies \cite{ExtMod3,ExtMod4}, beside the $\simeq 40$ data samples
dominated by statistical errors which follow the traditional treatment, the
data samples covering the  $e^+e^- \ra \pi^+ \pi^-$ annihilation channel are all, sometime
very strongly, dominated  by overall scale uncertainties; this especially refers
to the samples collected by the KLOE and BaBar Collaborations using the ISR production mode. Here,
for each sample, we chose for $A$ the experimental spectrum itself; this choice has been
referred to as  $A=m$ all along the paper. The guess behind was 
that all scale uncertainties affecting the different experimental spectra independently of each other
should smear out possible biases in the central values of the (common) theoretical form 
factor function parameters \cite{ExtMod4}.  

It happens that the results one can derive in this way from the BHLS global fit undergo
very small biases (compared to the errors  derived from the fit procedure); this is
shown in the present study\footnote{which also corrects for some coding bugs affecting our previous
studies.}. However, the guess just reminded was incorrect and the actual reason which explains
the almost bias free results is following~:
As shown in the Monte Carlo study presented in the Appendix,  there is no smearing out
of biases if $all$ the spectra submitted to fit undergo comparable strong scale uncertainties;
however, this study also shows that, if some  of the fitted spectra are dominated by 
(random) statistical errors rather than
global scale  uncertainties, the fit results can be strongly unbiased.

Nevertheless, a high level of unbiasing cannot be  taken as granted as the real weight of
the samples dominated by statistical errors within the full global fit procedure cannot be
ascertained beforehand. Basically, the choice $A=m$  potentially leads to biases of unknown magnitude;
this has been shown by G. D'Agostini \cite{D'Agostini} with a simple example
and more generally argued by V. Blobel \cite{Blobel_2006}. These authors also showed that
all biases vanish if, instead of  $A=m$, one makes the choice $A=M$, the "true" spectrum.
But this is just not possible within contexts like ours,  where fits are performed just in order  
to derive the "true" spectrum from data. Fortunately, iterative methods allow
to circumvent this difficulty by taking the path opened in \cite{Ball} in order to
derive the parton density function from data and correct for biases. The iterative method we propose has been
tested with the Monte Carlo study reported in the Appendix
 and shown  to produce unbiased results with a quite fast convergence speed; 
indeed,  only one iteration  is sufficient.

So, our main conclusion is indeed that global fit methods including a fast
iterative procedure  
are expected to produce reliable pieces of information as, methodologically,
 the central values are 
unbiased and the estimate for the uncertainties reliable; 
this especially applies to the part of the muon  leading order HVP derived
from $e^+ e^-$ annihilation cross sections.

\vspace{0.5cm}

Having shown that appropriate global fit methods should lead to results which 
can be trusted, a related remark is worth being expressed. Iterative global fits 
 allow to supply the BHLS Effective Lagrangian cross sections with reliable 
 and unbiased numerical central values for the fit parameters and a good estimate
of their error covariance matrix. Then, using these cross sections
 and the fit information,  Eq. (\ref{Eq1}) is expected to provide an unbiased estimate for $a_\mu(\pi \pi)$ as the
 ingredients are unbiased.
 
 On the other hand, when computing   $a_\mu(\pi \pi)$ by directly 
 integrating a dipion 
 spectrum in order to derive its  so--called experimental value,  one has to
 plug into Eq. (\ref{Eq1}) the experimentally measured cross section $\sigma_{exp.}(s)$. 
 However, as already noted in footnote \# \ref{bias_exp}, or as can be inferred from
 the canonical $\chi^2$ expression reminded just above, the experimental
 and model cross sections are related by~:
$$ \sigma_{exp.}(s) =\sigma_{theor.}(s) + \delta \sigma(s)$$
\noindent where the best estimate of the second term writes\footnote{In the case of a constant scale
uncertainty, as for the CMD2, SND and BESSIII data, there is only one scale factor $\lambda$.
For most ISR data samples, the expression is slightly more complicated but easy to derive (see also
the Appendix to \cite{ExtMod4}) and the conclusions are obviously likewise.} $\delta \sigma(s)=\lambda \sigma_{theor.}(s)$. As
obvious from  Eq. (\ref{Eq6}), the best estimate of the scale factor  
$\lambda$ equally depends  on the measured spectrum and on the "true" spectrum,
which can be identified with its (iterated) fit solution. 
 So,  using again self--explanatory notations, Eq. (\ref{Eq1})
leads to~:
$$a_\mu(\pi \pi, exp.)= a_\mu(\pi \pi, theor.) +\delta a_\mu(\pi \pi) $$
and thus $a_\mu(\pi \pi, exp.)$ looks intrinsically biased for any sample 
subject to  strong enough
overall scale uncertainties. This issue is also reflected by the residual plots which
are improved when plotting the corrected residuals $[m-(1+\lambda) M(\vec{a}) ]$ instead of the
raw ones $[m-M(\vec{a})]$, as can be seen in Figure 13 of \cite{ExtMod4}; this allows to infer
 that $\delta a_\mu(\pi \pi)$ is small but non--zero. It amounts to 
  $\delta a_\mu(\pi \pi) \simeq 2 \times 10^{-10}$
 in the case "NSK+KLOE10+KLOE12+BESSIII+$\tau$" favored by the BHLS model.

\vspace{0.5cm}

As for the physics conclusions, the present paper updates and corrects the results derived
by the global BHLS fit method which, following the considerings just summarized, has
been completed with an iteration procedure in order to cancel out possible biases. One thus
confirms that almost all of the existing data samples covering the annihilation
channels with the $\pi^0\gamma$, $\eta \gamma$, $\pi^+\pi^-\pi^0$, $K^+K^-$, $K^0\overline{K^0}$ 
final states and the  dipion spectra in the $\tau^\pm \ra \pi^\pm \pi^0 \nu$ decay accomodate perfectly
the BHLS framework. In the line of our previous works, one also finds that among the data samples
covering the $e^+e^- \ra \pi^+\pi^-$ annihilation, the data samples provided by CMD2 and SND, the KLOE10 
and now also the KLOE12 and BESSIII samples behave consistently with each other and with the other considered data
covering the various channels entering the BHLS scope. 

The present update, which also includes the recently published KLOE12 and BESSIII $\pi^+\pi^-$ 
samples, supersedes
our previous results; these are mostly given in Table \ref{Table:T3} and in Eqs. (\ref{Eq14}). From
a theoretical point of view, it is interesting to note the corrected values for the $c_i$'s
coefficients of the anomalous (FKTUY) terms of the HLS model \cite{FKTUY,HLSRef}~: The  combinations
$c_+=(c_4+c_3)/2$ and $c_1-c_2$ are found very close to the usually assumed value, {\it i.e.} 1; 
in contrast,  $c_-=(c_4-c_3)/2=-0.166 \pm 0.021$ is  non--zero with a $8\sigma$ significance.
 
 Figure \ref{Fig:a_mu_ref} displays the values for  $a_\mu(\pi \pi,[0.63,0.958])~$GeV derived
 from iterating the fits with the various available data samples. One observes a strong reduction
 of the uncertainty compared to the corresponding experimental value (about a factor of 2.5) and there
 is a close agreement between central values for all samples (or combinations of samples)
 which yield a good fit probability. The difference between the central values for the starting fit and
 the iterated one tends to indicate that biases are limited; this should be a consequence of
 also dealing with a large number of samples where the overall scale uncertainties are dominated
 by random statistical errors, as argued in  the Appendix.
  
Figure \ref{Fig:gmoins2_ref} exhibits the values for the muon $\Delta a_\mu=a_\mu^{exp}-a_\mu^{th}$ 
when various combinations of $e^+e^- \ra \pi^+\pi^-$ and $\tau^\pm \ra \pi^\pm \pi^0 \nu$ samples 
are used in the iterated global fit procedure.
The present study confirms that, within BHLS and because of its specific isospin breaking
mechanisms, one does not observe any serious mismatch between fits with only $e^+e^-$ annihilation
data and fits where these are supplemented with the $\tau$ dipion spectra. The central
values\footnote{the values for $a_\mu$ are given from now on in units of $10^{-10}$ for
convenience.}  for $a_\mu(e^+e^-)$ and $a_\mu(e^+e^-+ \tau)$ only differ by 2 units (NKS),
1 unit (NSK+KLOE+BESSIII+$\tau$) or 0.7 unit in the global fit of all data samples (including BaBar)
as can be seen  in Figure \ref{Fig:gmoins2_ref}.

 Figure \ref{Fig:gmoins2_ref} displays the value for $ \Delta a_\mu$ 
 derived using $all$  data samples except for KLOE08,
which can be written~:
$$ \Delta a_\mu = 37.02 + [^{+0.6}_{-1.3}]_\phi
+[^{+0.9}_{-0.0}]_\tau +[^{+0.0}_{-1.4} ]_{VNSB}
\pm 4.03_{th} \pm 6.3_{exp} ,$$
\noindent where an estimate of the magnitude of possible uncertainties 
coming from outside the BHLS framework is proposed. This  exhibits a $5 \sigma$ 
significance (which may reduce to $4.6 \sigma$ 
-- in the least favorable case --  if the additional systematics are added linearly
and assumed to play as a shift). One should note however that the fit probability is poor.

The most probable value for the muon $\Delta a_\mu$ is obtained by using
the  CMD2, SND, KLOE10, KLOE12 and BESSIII samples -- and the $\tau$ spectra; this leads to~:
$$ \Delta a_\mu = 37.68 + [^{+0.6}_{-1.3}]_\phi
+[^{+0.9}_{-0.0}]_\tau +[^{+0.0}_{-1.4} ]_{VNSB}
\pm 4.12_{th} \pm 6.3_{exp} .$$
\noindent This BHLS preferred estimate exhibits a $5. \sigma$ significance for a non--zero
$\Delta a_\mu$, which may reduce to $4.7 \sigma$ if one takes into account, as just above, 
the possible additional
systematics. This solution is associated with a 99\% fit probability.

As a summary, even complemented with an iterative procedure shown in the Appendix
to remove biases, the BHLS approach  favors a significance for $\Delta a_\mu$ above
the $\simeq 4.5 \sigma$ level; this value is a lower bound obtained by including possible additional
systematics added linearly. New data expected soon may further clarify the picture.
The uncertainties now become sharply dominated by the region above 1.05 GeV, {\it i.e.}
outside the BHLS scope.

\section*{Acknowledgements}
\indent \indent
We would like to acknowledge the Mainz Institute for Theoretical Physics (MITP) for its
hospitality which has allowed exchanges leading to a better understanding of the effects
of normalization uncertainties within global fit frameworks; this has allowed improving
the numerical methods for $g-2$ evaluation.

\clearpage
\appendix
\section{Appendix~: Monte Carlo Tests of the Iterative Procedure}
\label{MC_test}
\subsection{The Test Method}
\label{test_method}
\indent \indent In order to  test the iterative method, one has developped a minimization
code which deals with spectra generated from a given underlying function $M_{true}(s)$
where the parameters $\{a_i\}$ (which, of course,  are known at the generation level) are 
fitted within the code. The "experimental" spectra feeding this code are generated
using the true distribution smeared by introducing $gaussian$ uncertainty distributions.
Indeed, for the purpose of testing our analysis method,  it is  certainly the most appropriate 
to rely on "perfect" data samples, with perfectly known properties.
 
For sake of simplicity, at the generation level, 
any "experimental"  spectrum $E$ is chosen to carry 100  "measurements" $m_i^E$, performed 
at 100 equally spaced energy squared $s_i$ points ($s_i \in [0,~1]$ GeV$^2$), 
the same sequence for all spectra. 
The "measurements" are derived
by smearing the theoretical values $M_{true}(s_i)$ in the following way~:
For each spectrum $E$, one assumes the "measurements"  are
sampled out from gaussian distributions in the following way~:
\be
m_i^E= M_{true}(s_i) [1 + \sigma \varepsilon_{scale}^{E}(0,1)+ \eta \varepsilon_{stat}^{i,E}(0,1)]
 ~~~~,~~~i=1,\cdots,100
\label{A1}
\ee
where $\varepsilon_{stat}^{i,E}(0,1)$ indicates the $i^{th}$ sampling on a gaussian
distribution of $0$ mean and unit standard deviation generating the statistical error;
 it varies independently 
from "measurement" to "measurement" and from spectrum to spectrum. $\eta M_{true}(s_i)$ 
denotes the statistical error common  to all $m_i$, $\eta$ being some fixed fraction
of the order of a few percents, chosen the same for all the "measurements" in the spectrum $E$. 

On the other
hand,  $\lambda_E=\sigma \varepsilon_{scale}^{E}(0,1)$
 is the scale uncertainty affecting specifically the spectrum $E$; as indicated by its
 definition, it is sampled out from a gaussian distribution
of zero mean and $\sigma$  standard deviation. The overall scale uncertainty
affecting $E$ is obtained via $one$ sampling of $\varepsilon_{scale}^{E}(0,1)$ which, thus,
carries the same value for all the "measurements" $m_i^E$ in  the spectrum  $E$. Of course,
when going from a spectrum $E$ to another $E^\prime$,  another sampling of 
$\varepsilon_{scale}^{E}(0,1)$ should be performed. For specific tests, the overall scale uncertainty
can be switched off ($\sigma=0$).

One defines  $N_{rep}$ replicas (generally 1000) of $N_{exp}$ (generally 5) 
experimental spectra constructed as shown in Eq. (\ref{A1}) and submitted to
a global fit where the parameters entering  $M_{true}(s)$ are just the
parameters to be derived from the fit.
The "true" statistical error covariance matrix 
$V_{ij}=[\eta M_{true}(s_i)]^2 \delta_{ij}$ is practically approximated by
$V_{ij}=[\eta m_i^E]^2 \delta_{ij}$; we have avoided the unessential complication of non--diagonal
covariance matrix.
The fit results derived for each replica are stored and then used to construct
the statistical plots -- true residuals and pulls --with the help of  the known
parameter  "true" values. 

Therefore, we are just in the conditions described in Subsection \ref{scale_err}.
One should note that the {\sc minuit} code we have built performs 
the minimization of the $N_{exp}$ samples $and$ runs sequentially to treat
the $N_{rep}$ replicas within the same job. 

So, for each replica, the global $\chi^2$  minimized by our Monte Carlo {\sc minuit} procedure
is simply a sum of $N_{exp}$ terms like Eq. (\ref{Eq4}):
\be
\displaystyle \chi^2=\sum_{E=1}^{E=N_{exp}} \chi^2_E
\label{A2}
\ee 

When initializing the iteration procedure, one uses $A_E=m_E$, {\it i.e.} the spectrum
$E$ serves to construct its $\chi^2_E$; so $A_E$ differs from some other $A_{E^\prime}$
 by statistical fluctuations. When iterating, at first or higher order,  they
become identical as $A_E=A_{E^\prime}=M_{fit} \simeq M (\vec{a}_{fit})$.

Obviously, each such run provides  simultaneously all the information allowing to examine 
the statistical properties of
the iterative method corresponding to a given theoretical choice $M_{true}(s)$. 
The computer code also allows an easy  change of the functional form of $M_{true}(s)$ 
in order to examine the behavior of various kinds of non--linear parameter dependences.

The behavior of the fit parameters compared to truth is, of course, the
subject of the analysis; however, those of "physics quantities" derived
from them are as important. For this purpose, we chose to examine the 
ratios\footnote{Remind that $0$ and $1$ GeV$^2$ are the energy squared limits
 of the generated spectra.}~:
\be
\displaystyle  {\cal I}= \frac{\int_{0}^{1} M_{fit}(s)ds
}
{\int_{0}^{1} M_{true}(s)ds
}
\label{A3}
\ee 
which has properties similar to those of the $a_\mu({\cal H}_i)$'s, as the
weighting factor $K(s)$ in Eq. (\ref{Eq1}) is an unessential 
complication while looking for possible methodological biases
of the iterative method.
\subsection{The Test Results}
\indent \indent The aim of the present Appendix is to report on
 numerical analyses performed in various configurations in order to examine 
 how overall (global) normalization uncertainties and biases are related and
 whether non--linearities in the model parameters to be fitted lead
 to significant incorrect estimates of errors. As Reference \cite{Ball}
 which is faced with the same kinds of issues as the present work, we do not plan to
 establish rigorously general theorems on these topics -- assuming the
 scope of the issues would permit it. Nevertheless, one can think that 
 studying methods by relying on  Monte Carlo technics is an 
 acceptable way to check its (practical) validity under common conditions.
 After all, the fact that Eq. (\ref{Eq3}) with $A=M$ (the theoretical function) 
 is considered free from biases is not weakened by the fact that  the general (formal) 
 proof of this property -- if established -- is not commonly referred to.

 \subsubsection{Analytical Shape of the True Distributions}
 \label{Shape}
\indent \indent In order to use confidently  fit results derived using
the iterative method, one should examine the effects 
 of non--linear dependences upon the fit parameters within contexts similar
 to our physics distributions. The lineshape of the pion form factor as
 a function of $s$ on a given interval
 can be qualitatively reproduced using polynomials, ratios of polynomials, exponential
 of polynomials, sums of a Breit--Wigner function with polynomials etc \ldots
 with appropriate numerical parameter values.    

We have applied the method outlined in  Subsection \ref{test_method} to perform
fits relying on an intensive use of the tools provided by {\sc minuit} taking 
various kinds of functions $M_{true}(s)$, resembling 
-- sometimes weakly -- the pion form factor. Running in sequence
{\sc migrad/Hesse} and {\sc minos}, we did not observe significant departures
(beyond statistical fluctuations) from equality between parabolic and {\sc minos}
errors; as the issue was to examine effects of non--linear parameter dependences
this exercise was performed assuming statistical uncertainties only. Therefore,
this led us to conclude that, for the kind of experimental
distributions one deals with, non--linear effects are not generally significant.
For instance, using~:
\be
\displaystyle  
M_{true}(s)= \frac{g}
{(s-a)^2+b^2} + c +d~s +e~s^2~~~,
\label{A4}
\ee 
$\eta=3\%$ and no scale uncertainty (to discard any need for iterating),
 the probability distribution was
observed flat and the parameter pulls consistent with normal gaussians $G(m=0,\sigma=1)$; the
 distribution of the ratio ${\cal I}$ for the 1000 replicas was also found well 
 centered at 1 (actually its mean is 1.0001 and its standard deviation
 $1.62 \times 10^{-3}$ from a gaussian fit with $\chi^2/N_{points}=8.9/11$).
 So, except for pathological cases which may always occur, non--linear 
 dependences do not look practically an issue. 
  
 From now on, we limit ourselves to reporting on using $M_{true}(s)$
 as given by Eq. (\ref{A4}). Moreover, for sake of succinctness, we may only
 mention the fit parameter residual and pull distribution properties qualitatively
 and concentrate on discussing the distribution of the ratios ${\cal I}$
 which, {\it in fine} carries -- summarized --  the relevant information.
 Each value of ${\cal I}$ entering this distribution is computed from a 
 {\sc minuit} fit of $N_{exp}=5$ data samples and this is done  for
 $N_{rep}=1000$ replicas to construct numerically its distribution.

\subsubsection{Normalization Uncertainty and Iterative Method}
\label{iter_1}
\begin{figure}[!ph]
\begin{minipage}{\textwidth}
\begin{center}
 \vspace{-1.cm}
\resizebox{\textwidth}{!}
{\includegraphics*{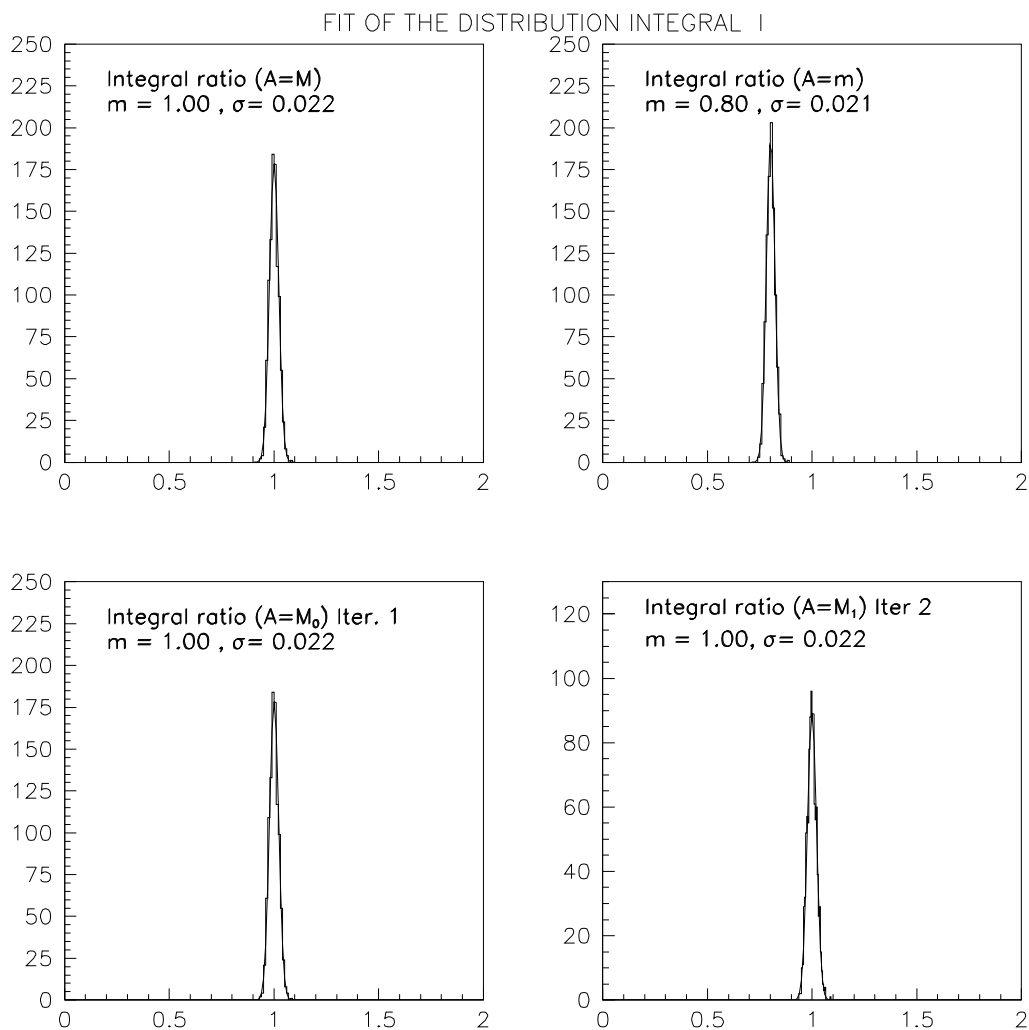}}
\end{center}
\end{minipage}
\begin{center}
\vspace{-1.5cm}
\caption{\label{Fig:scale_1} Distributions of the ratios ${\cal I}$ derived by varying
the function $A$ in the $\chi^2$ expression as indicated in each panel. The choice
$A=m^E$ ({\it i.e.} the "measured" data sample) exhibits a 20\% bias while the other choices
are unbiased. For more
comments, see Subsection \ref{iter_1}.
 }
\end{center}
\end{figure}

\indent \indent We first examined the results derived  by fit of spectra
with data points generated as in Eq. (\ref{A1}) with a statistical uncertainty
$\eta=3\%$ and generating the scale uncertainty $\lambda$ with $\sigma=5\%$;
so $\eta$ is smaller than $\sigma$. In this case, the interesting plots are 
gathered in Figure \ref{Fig:scale_1}.

As one knows $M_{true}(s)$, one can construct the $N_{exp}$ partial $\chi^2$'s
with $A=M_{true}(s)$ (see Eq.(\ref{Eq3})) and minimize their sum using {\sc minuit}.
In this case, no bias is expected \cite{D'Agostini,Blobel_2003,Blobel_2006} and 
this is indeed confirmed by the top left panel in Figure \ref{Fig:scale_1} where 
the distribution of the  $N_{rep}$ values for ${\cal I}$ is displayed. 

When, instead, one uses $A=m$ (the data spectrum), the results are shown in
the  top right panel of Figure \ref{Fig:scale_1}, where one observes a shift of the central
value by as large as 20\%~! Denoting the result of the  corresponding fit  by $M_0$,
one restarts fitting the same data by setting $A=M_0$, this -- first -- iteration 
leads to the distribution shown in the bottom left panel of Figure \ref{Fig:scale_1} 
which looks identical to having used $A=M_{true}$. Denoting the fit solution of
this first iteration by $M_1$, one restarts fitting the same data by setting $A=M_1$,
and get the step 2 solution $M_2$ which correponds to  the bottom right panel 
of Figure \ref{Fig:scale_1}, which clearly indicates no change for the  ${\cal I}$ 
distribution. 

So, one may conclude that the iterative procedure has already 
converged at the first iteration and so, we have $M_1=M_{true}$. This fortunate 
high convergence speed has also been observed by \cite{Ball} and it is
quite remarkable that this has allowed to recover from\footnote{The numerical
importance of this bias is intimately related with the ratio $\sigma/\eta=5/3$;
if instead one works with $\sigma/\eta=1$, the bias coming out from fitting with $A=m$
would only be 4\%.} 
a 20\% bias! 

Fit residuals are observed unbiased and pulls consistent with normal centered gaussians
for $A=M_{true}$, $A=M_0$ and $A=M_1$. As for the $\chi^2$ probability distributions, for $A=m$,
it exhibits a huge spike at 1, while it is consistent with flatness  (mean $\simeq 0.5$ 
and r.m.s. $\simeq 1/\sqrt{12}$) for all the other cases.

This already indicates that starting with $A=m$ (the measured data spectrum) and
iterating only once allows to give up knowing the theoretical function $M$ beforehand
to drop out biases in physics quantity estimates. Moreover, as the parameter pulls
are  centered gaussians of unit standard deviations, the uncertainties
derived from from the fit parameter error covariance matrix are reliable.
\subsubsection{Effects of Subsamples Free from Normalization Uncertainties}
\label{iter_2}
\indent \indent In the specific problem of globally fitting a large number of
experimental data samples, one is faced with as many as 40 to 50 spectra
to be treated \cite{ExtMod3,ExtMod4,BM_roma2013,BM_paris_2013}.  Within this ensemble
of data samples, one observes several configurations concerning uncertainties~:
some samples have  statistical errors dominated by scale uncertainties (the ISR collected
data samples), while, in contrast, some others are reported with scale uncertainties 
marginal compared to statistical errors (the $e^+e^-\ra \gamma P$ data, for instance); sometimes,
no specific  information is reported concerning scale uncertainties, as for the 
the $\tau$ dipion spectra \cite{Aleph,Cleo,Belle}.

\begin{figure}[!ph]
\begin{minipage}{\textwidth}
\begin{center}
\resizebox{\textwidth}{!}
{\includegraphics*{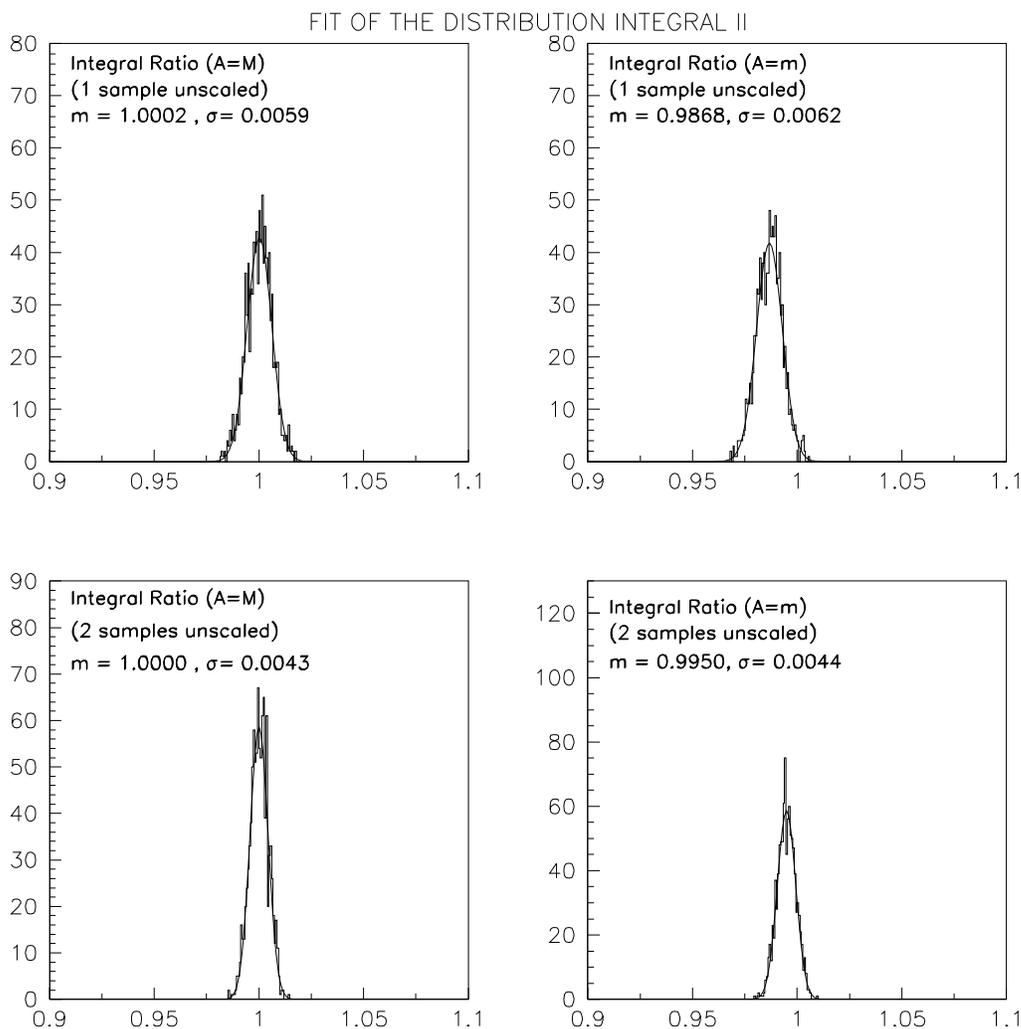}}
\end{center}
\end{minipage}
\begin{center}
\vspace{-1.5cm}
\caption{\label{Fig:scale_2} Effect of having 1 (top panels) or 2 (downmost panels)
data sample(s) among the fitted $N_{exp}=5$ samples simultaneously fitted. Left plots
report on fitting with $A=M$ (the truth), right plots on fitting with $A=m^E$  (the
measured spectra); in the former case no bias is observed, in the latter case, the 
bias happens to be much limited. See text for more details.
}

\end{center}
\end{figure}

This makes interesting to examine configurations mixing samples of both kinds.
In this paragraph, one summarizes the results obtained by running $N_{rep}$ replicas
of ensembles of 4 data sets where, as before, the scale error is 
$\sigma =5\%$ and  the statistical error $\eta =3\%$, together with 1 data set with
$\sigma =0\%$ (no scale uncertainty) and $\eta =6\%$  (twice worse statistical precision).
This (4,1) combination will be supplemented with a (3,2) combination with the same 
characteristics. The main results are shown in Figure \ref{Fig:scale_2}. Here we do not report 
on iterating the fit procedure, as obviously the results will  follow the pattern shown in
 Figure \ref{Fig:scale_1}.

\begin{figure}[!ph]
\begin{minipage}{\textwidth}
\begin{center}
\resizebox{\textwidth}{!}
{\includegraphics*{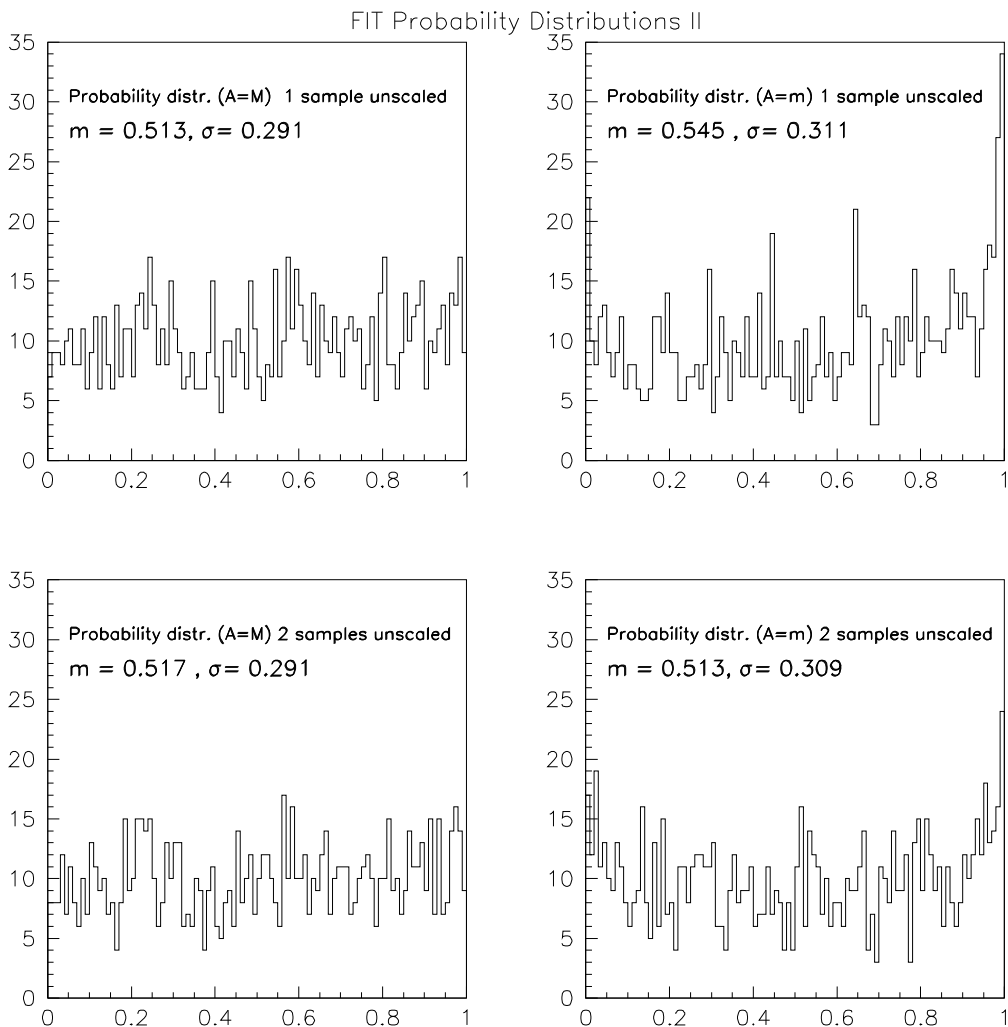}}
\end{center}
\end{minipage}
\begin{center}
\vspace{-1.5cm}
\caption{\label{Fig:scale_2b}
Probability distributions when fitting with $A=M_{truth}$ (left panels)
or $A=m_E$ (right panels). The top panel plots correspond to the case when among 
the $N_{exp}=5$ fitted spectra, one is systematically free from normalization uncertainty;
in the downmost panels  2 of the 5 fitted spectra are 
free from normalization uncertainty.}

\end{center}
\end{figure}

The top panels in Figure \ref{Fig:scale_2} display distributions of the ratios
${\cal I}$  in the (4,1) configuration. The left plot shows the case when 
the  $N_{rep}$ replicas are fitted using $A=M_{truth}$ in the $\chi^2$ expressions.
In this case, the absence of any bias is confirmed by the gaussian fit 
result shown within this plot.
While using $A=m^E$, the top right panel exhibits a 1.3\% bias. Therefore, the effect
of a single spectrum free from scale uncertainty out of 5 is enough to lessen
dramatically  the observed bias~: It reduces from 20\% to 1.3\%. 
  
The downmost panels in Figure \ref{Fig:scale_2}  display the corresponding
results when fitting   $N_{rep}$ replicas of  (3,2) combinations. In this case,
using $A=m^E$ in the minimized $\chi^2$ expression, leads to an even smaller bias 
(0.5\%). 

So, even if they carry  a poor statistical precision, having some 
spectra free from a (significant) scale uncertainty is quite helpfull to
strongly limit the real magnitude of a possible bias for a derived quantity.
It is a quite interesting property to observe that some spectra with degraded statistical
quality supplementing  other spectra dominated by scale uncertainties might be  enough
to avoid the need of an iteration procedure to unbias physics pieces of  information.

As for the probability distributions, comparing of the
corresponding left and right panels in Figure \ref{Fig:scale_2b} clearly shows
that the departures from uniformity ({\it i.e.} average=0.5 and r.m.s.=0.289) due to using
$A=m^E$ are quite limited. 

Nevertheless, when dealing with true experimental data (and thus
unknown truth), one cannot take as granted that the number of samples with negligible
scale uncertainties compared to statistical errors is sufficient to ascertain that
biases are negligible. 
Therefore, in the practical case of the global fit of real experimental 
data performed within BHLS, secure results can only be ascertained by iterating until 
the change of $a_\mu(H_i)$  is small enough.

\newpage

                   \bibliographystyle{h-physrev}
                     \bibliography{vmd4}

\begin{thebibliography}{100}

\bibitem{GL1}
J.~Gasser and H.~Leutwyler,
\newblock Annals Phys. {\bf 158}, 142 (1984),
\newblock {Chiral Perturbation Theory to One Loop}.

\bibitem{GL2}
J.~Gasser and H.~Leutwyler,
\newblock Nucl. Phys. {\bf B250}, 465 (1985),
\newblock {Chiral Perturbation Theory: Expansions in the Mass of the Strange
  Quark}.

\bibitem{Colangelo_final}
S.~Aoki {\em et~al.},
\newblock Eur. Phys. J. {\bf C74}, 2890 (2014), 1310.8555,
\newblock {Review of lattice results concerning low-energy particle physics}.

\bibitem{Lattgm2-1}
ETM, F.~Burger {\em et~al.},
\newblock JHEP {\bf 1402}, 099 (2014), 1308.4327,
\newblock {Four-Flavour Leading-Order Hadronic Contribution To The Muon
  Anomalous Magnetic Moment}.

\bibitem{Lattgm2-2}
F.~Burger, G.~Hotzel, K.~Jansen, and M.~Petschlies,
\newblock (2015), 1501.05110,
\newblock {Leading-order hadronic contributions to the electron and tau
  anomalous magnetic moments}.

\bibitem{BNL}
Muon G-2, G.~W. Bennett {\em et~al.},
\newblock Phys. Rev. {\bf D73}, 072003 (2006), hep-ex/0602035,
\newblock {Final report of the muon E821 anomalous magnetic moment measurement
  at BNL}.

\bibitem{BNL2}
B.~L. Roberts,
\newblock Chin. Phys. {\bf C34}, 741 (2010), 1001.2898,
\newblock {Status of the Fermilab Muon $(g-2)$ Experiment}.

\bibitem{LeeRoberts}
Fermilab P989 Collaboration, B.~Lee~Roberts,
\newblock Nucl.Phys.Proc.Suppl. {\bf 218}, 237 (2011),
\newblock {The Fermilab muon (g-2) project}.

\bibitem{Fermilab_gm2}
Muon g-2 Collaboration, J.~Grange {\em et~al.},
\newblock (2015), 1501.06858,
\newblock {Muon (g-2) Technical Design Report}.

\bibitem{Iinuma}
J-PARC New g-2/EDM experiment Collaboration, H.~Iinuma,
\newblock J.Phys.Conf.Ser. {\bf 295}, 012032 (2011),
\newblock {New approach to the muon g-2 and EDM experiment at J-PARC}.

\bibitem{Ecker2}
G.~Ecker, J.~Gasser, H.~Leutwyler, A.~Pich, and E.~de~Rafael,
\newblock Phys.Lett. {\bf B223}, 425 (1989),
\newblock {Chiral Lagrangians for Massive Spin 1 Fields}.

\bibitem{Ecker1}
G.~Ecker, J.~Gasser, A.~Pich, and E.~de~Rafael,
\newblock Nucl.Phys. {\bf B321}, 311 (1989),
\newblock {The Role of Resonances in Chiral Perturbation Theory}.

\bibitem{HLSRef}
M.~Harada and K.~Yamawaki,
\newblock Phys. Rept. {\bf 381}, 1 (2003), hep-ph/0302103,
\newblock {Hidden local symmetry at loop: A new perspective of composite gauge
  boson and chiral phase transition}.

\bibitem{HLSOrigin}
M.~Bando, T.~Kugo, and K.~Yamawaki,
\newblock Phys. Rept. {\bf 164}, 217 (1988),
\newblock {Nonlinear Realization and Hidden Local Symmetries}.

\bibitem{FKTUY}
T.~Fujiwara, T.~Kugo, H.~Terao, S.~Uehara, and K.~Yamawaki,
\newblock Prog. Theor. Phys. {\bf 73}, 926 (1985),
\newblock {Nonabelian Anomaly and Vector Mesons as Dynamical Gauge Bosons of
  Hidden Local Symmetries}.

\bibitem{BKY}
M.~Bando, T.~Kugo, and K.~Yamawaki,
\newblock Nucl. Phys. {\bf B259}, 493 (1985),
\newblock {On the Vector Mesons as Dynamical Gauge Bosons of Hidden Local
  Symmetries}.

\bibitem{BGP}
A.~Bramon, A.~Grau, and G.~Pancheri,
\newblock Phys. Lett. {\bf B345}, 263 (1995), hep-ph/9411269,
\newblock {Effective chiral lagrangians with an SU(3) broken vector meson
  sector}.

\bibitem{BGPbis}
A.~Bramon, A.~Grau, and G.~Pancheri,
\newblock Phys. Lett. {\bf B344}, 240 (1995),
\newblock {Radiative vector meson decays in SU(3) broken effective chiral
  Lagrangians}.

\bibitem{Heath}
M.~Benayoun and H.~B. O'Connell,
\newblock Phys. Rev. {\bf D58}, 074006 (1998), hep-ph/9804391,
\newblock {SU(3) breaking and hidden local symmetry}.

\bibitem{Hashimoto}
M.~Hashimoto,
\newblock Phys. Rev. {\bf D54}, 5611 (1996), hep-ph/9605422,
\newblock {Hidden local symmetry for anomalous processes with isospin/SU(3)
  breaking effects}.

\bibitem{ODonnell}
P.~J. O'Donnell,
\newblock Rev. Mod. Phys. {\bf 53}, 673 (1981),
\newblock {Radiative decays of mesons}.

\bibitem{rad}
M.~Benayoun, L.~DelBuono, S.~Eidelman, V.~N. Ivanchenko, and H.~B. O'Connell,
\newblock Phys. Rev. {\bf D59}, 114027 (1999), hep-ph/9902326,
\newblock {Radiative decays, nonet symmetry and SU(3) breaking}.

\bibitem{WZWChPT}
M.~Benayoun, L.~DelBuono, and H.~B. O'Connell,
\newblock Eur. Phys. J. {\bf C17}, 593 (2000), hep-ph/9905350,
\newblock {VMD, the WZW Lagrangian and ChPT: The third mixing angle}.

\bibitem{leutw}
H.~Leutwyler,
\newblock Nucl. Phys. Proc. Suppl. {\bf 64}, 223 (1998), hep-ph/9709408,
\newblock {On the 1/N-expansion in chiral perturbation theory}.

\bibitem{leutwb}
R.~Kaiser and H.~Leutwyler,
\newblock in Adelaide 1998, Nonperturbative methods in quantum field theory ,
  15 (2001), hep-ph/9806336,
\newblock {Pseudoscalar decay constants at large $N_c$}.

\bibitem{Fred11}
F.~Jegerlehner and R.~Szafron,
\newblock Eur. Phys. J. {\bf C71}, 1632 (2011), 1101.2872,
\newblock {$\rho^0-\gamma$ mixing in the neutral channel pion form factor
  $|F_\pi|^2$ and its role in comparing $e^+ e^-$ with $\tau$ spectral
  functions}.

\bibitem{taupaper}
M.~Benayoun, P.~David, L.~DelBuono, O.~Leitner, and H.~B. O'Connell,
\newblock Eur. Phys. J. {\bf C55}, 199 (2008), hep-ph/0711.4482,
\newblock {The Dipion Mass Spectrum In $e^+e^-$ Annihilation and tau Decay: A
  Dynamical ($\rho^0$, $\omega$, $\phi$) Mixing Approach}.

\bibitem{DavierPrevious1}
M.~Davier, S.~Eidelman, A.~Hocker, and Z.~Zhang,
\newblock Eur. Phys. J. {\bf C27}, 497 (2003), hep-ph/0208177,
\newblock {Confronting spectral functions from $e^+ e^-$ annihilation and tau
  decays: Consequences for the muon magnetic moment}.

\bibitem{DavierHoecker}
M.~Davier {\em et~al.},
\newblock Eur. Phys. J. {\bf C66}, 127 (2010), 0906.5443,
\newblock {The Discrepancy Between $\tau$ and $e^+e^-$ Spectral Functions
  Revisited and the Consequences for the Muon Magnetic Anomaly}.

\bibitem{Eidelman}
S.~I. Eidelman,
\newblock (2009), 0904.3275,
\newblock {Standard Model Predictions for the Muon $(g-2)/2$}.

\bibitem{Fred09}
F.~Jegerlehner and A.~Nyffeler,
\newblock Phys. Rept. {\bf 477}, 1 (2009), 0902.3360,
\newblock {The Muon g-2}.

\bibitem{ExtMod1}
M.~Benayoun, P.~David, L.~DelBuono, and O.~Leitner,
\newblock Eur. Phys. J. {\bf C65}, 211 (2010), 0907.4047,
\newblock {A Global Treatment Of VMD Physics Up To The $\phi$: I. $e^+e^-$
  Annihilations, Anomalies And Vector Meson Partial Widths}.

\bibitem{ExtMod2}
M.~Benayoun, P.~David, L.~DelBuono, and O.~Leitner,
\newblock Eur. Phys. J. {\bf C68}, 355 (2010), 0907.5603,
\newblock {A Global Treatment Of VMD Physics Up To The phi: II. $\tau$ Decay
  and Hadronic Contributions To g-2}.

\bibitem{ExtMod3}
M.~Benayoun, P.~David, L.~DelBuono, and F.~Jegerlehner,
\newblock Eur.Phys.J. {\bf C72}, 1848 (2012), 1106.1315,
\newblock {Upgraded Breaking Of The HLS Model: A Full Solution to the $\tau^- -
  e^+e^-$ and $\phi$ Decay Issues And Its Consequences On g-2 VMD Estimates}.

\bibitem{ExtMod4}
M.~Benayoun, P.~David, L.~DelBuono, and F.~Jegerlehner,
\newblock Eur.Phys.J. {\bf C73}, 2453 (2013), 1210.7184,
\newblock {An Update of the HLS Estimate of the Muon g-2}.

\bibitem{KLOE08}
KLOE, G.~Venanzoni {\em et~al.},
\newblock AIP Conf. Proc. {\bf 1182}, 665 (2009), 0906.4331,
\newblock {A precise new KLOE measurement of $|F_\pi|^2$ with ISR events and
  determination of $\pi\pi$ contribution to $a_\mu$ for $0.592 < M_{\pi\pi} <
  0.975$ GeV}.

\bibitem{KLOE10}
KLOE, F.~Ambrosino {\em et~al.},
\newblock Phys.Lett. {\bf B700}, 102 (2011), 1006.5313,
\newblock {Measurement of $\sigma(e^+ e^- \to \pi^+ \pi^-$) from threshold to
  0.85 $GeV^2$ using Initial State Radiation with the KLOE detector}.

\bibitem{KLOE12}
KLOE Collaboration, D.~Babusci {\em et~al.},
\newblock Phys.Lett. {\bf B720}, 336 (2013), 1212.4524,
\newblock {Precision measurement of $\sigma(e^+e^-\rightarrow
  \pi^+\pi^-\gamma)/ \sigma(e^+e^-\rightarrow \mu^+\mu^-\gamma)$ and
  determination of the $\pi^+\pi^-$ contribution to the muon anomaly with the
  KLOE detector}.

\bibitem{BaBar}
BABAR, B.~Aubert {\em et~al.},
\newblock Phys. Rev. Lett. {\bf 103}, 231801 (2009), 0908.3589,
\newblock {Precise measurement of the $e^+e^- \rightarrow \pi^+ \pi^- (\gamma)$
  cross section with the Initial State Radiation method at BABAR}.

\bibitem{BaBar2}
BABAR Collaboration, J.~Lees {\em et~al.},
\newblock Phys.Rev. {\bf D86}, 032013 (2012), 1205.2228,
\newblock {Precise Measurement of the $e^+ e^- \to \pi^+\pi^- (\gamma)$ Cross
  Section with the Initial-State Radiation Method at BABAR}.

\bibitem{BESS-III}
BESSIII, M.~Ablikim {\em et~al.},
\newblock (2015), 1507.08188,
\newblock {Measurement of the $\mathrm e^+\mathrm
  e^-\rightarrow\mathrm\pi^+\mathrm\pi^-$ Cross Section between 600 and 900 MeV
  Using Initial State Radiation}.

\bibitem{D'Agostini}
G.~D'Agostini,
\newblock Nucl.Instrum.Meth. {\bf A346}, 306 (1994),
\newblock {On the use of the covariance matrix to fit correlated data}.

\bibitem{Peelle}
R.~W. Peelle,
\newblock Informal Memorandum, Oak Ridge National Laboratory, TN, USA  (1987),
\newblock {Peelle's Pertinent Puzzle}.

\bibitem{Chiba}
S.~Ciba and D.~Smith,
\newblock Nuclear Data and Measurements Series, Argonne National Laboratory,
  Argonne, IL, USA ANL/NDM--121  (1991),
\newblock {A suggested procedure for resolving an anomaly in Least--squares
  data analysis known as 'Peelle's Pertinent Puzzle' and the general
  implications for nuclear data evaluation}.

\bibitem{Blobel_2003}
V.~Blobel,
\newblock eConf {\bf C030908}, MOET002 (2003),
\newblock {Some Comments on $\chi^2$ Minimization Applications}.

\bibitem{Blobel_2006}
V.~Blobel,
\newblock Banff International Research Station {\bf Statistical Inference
  Problems in High Energy Physics}, http://www.desy.de/~blobel/banff.pdf
  (2006),
\newblock {Dealing with systematics for chi--square and for log likelihood
  goodness of fit}.

\bibitem{Ball}
NNPDF, R.~D. Ball {\em et~al.},
\newblock JHEP {\bf 5}, 075 (2010), 0912.2276,
\newblock Fitting Parton Distribution Data with Multiplicative Normalization
  Uncertainties.

\bibitem{Ball_2}
R.~D. Ball {\em et~al.},
\newblock Nucl.Phys. {\bf B838}, 136 (2010), 1002.4407,
\newblock {A first unbiased global NLO determination of parton distributions
  and their uncertainties}.

\bibitem{Aleph}
ALEPH, S.~Schael {\em et~al.},
\newblock Phys. Rept. {\bf 421}, 191 (2005), hep-ex/0506072,
\newblock {Branching ratios and spectral functions of tau decays: Final ALEPH
  measurements and physics implications}.

\bibitem{Cleo}
CLEO, S.~Anderson {\em et~al.},
\newblock Phys. Rev. {\bf D61}, 112002 (2000), hep-ex/9910046,
\newblock {Hadronic structure in the decay $\tau^- \to \pi^- \pi^0
  \nu_{\tau}$}.

\bibitem{Belle}
Belle, M.~Fujikawa {\em et~al.},
\newblock Phys. Rev. {\bf D78}, 072006 (2008), 0805.3773,
\newblock {High-Statistics Study of the $\tau^- \to \pi^- \pi^0 \nu_{\tau}$
  Decay}.

\bibitem{CMD2-1998-1}
CMD-2, R.~R. Akhmetshin {\em et~al.},
\newblock Phys. Lett. {\bf B648}, 28 (2007), hep-ex/0610021,
\newblock {High-statistics measurement of the pion form factor in the rho-meson
  energy range with the CMD-2 detector}.

\bibitem{CMD2-1998-2}
R.~R. Akhmetshin {\em et~al.},
\newblock JETP Lett. {\bf 84}, 413 (2006), hep-ex/0610016,
\newblock {Measurement of the $e^+ e^- \to \pi^+ \pi^-$ cross section with the
  CMD-2 detector in the 370-MeV - 520-MeV cm energy range}.

\bibitem{SND-1998}
M.~N. Achasov {\em et~al.},
\newblock J. Exp. Theor. Phys. {\bf 103}, 380 (2006), hep-ex/0605013,
\newblock {Update of the $e^+ e^- \to \pi^+ \pi^-$ cross section measured by
  SND detector in the energy region 400-MeV $< \sqrt{s} <$ 1000-MeV}.

\bibitem{DavierHoecker2}
M.~Davier, A.~Hoecker, B.~Malaescu, C.~Z. Yuan, and Z.~Zhang,
\newblock Eur. Phys. J. {\bf C66}, 1 (2009), 0908.4300,
\newblock {Reevaluation of the hadronic contribution to the muon magnetic
  anomaly using new $e^+ e^- \to \pi^+ \pi^-$ cross section data from BABAR}.

\bibitem{BM_paris_2013}
M.~Benayoun,
\newblock PoS {\bf Photon2013}, 048 (2013),
\newblock {Effective Lagrangians : A New Approach to g - 2 Evaluations}.

\bibitem{BM_roma2013}
M.~BENAYOUN,
\newblock Int.J.Mod.Phys.Conf.Ser. {\bf 35}, 1460416 (2014),
\newblock {Impact of the Recent KLOE data samples on the estimate for the muon
  $g - 2$}.

\bibitem{DavierHoecker3}
M.~Davier, A.~Hoecker, B.~Malaescu, and Z.~Zhang,
\newblock Eur. Phys. J. {\bf C71}, 1515 (2011), 1010.4180,
\newblock {Reevaluation of the Hadronic Contributions to the Muon g-2 and to
  alpha(MZ)}.

\bibitem{Teubner2}
K.~Hagiwara, R.~Liao, A.~D. Martin, D.~Nomura, and T.~Teubner,
\newblock J.Phys. {\bf G38}, 085003 (2011), 1105.3149,
\newblock {$(g-2)_\mu$ and $\alpha(M_Z^2)$ re-evaluated using new precise
  data}.

\bibitem{Teubner}
T.~Teubner, K.~Hagiwara, R.~Liao, A.~D. Martin, and D.~Nomura,
\newblock Chin. Phys. {\bf C34}, 728 (2010), 1001.5401,
\newblock {Update of g-2 of the muon and Delta alpha}.

\bibitem{RPP2012}
Particle Data Group, J.~Beringer {\em et~al.},
\newblock Phys.Rev. {\bf D86}, 010001 (2012),
\newblock {Review of Particle Physics (RPP)}.

\bibitem{Marciano}
W.~J. Marciano and A.~Sirlin,
\newblock Phys. Rev. Lett. {\bf 71}, 3629 (1993),
\newblock {Radiative corrections to pi(lepton 2) decays}.

\bibitem{Cirigliano1}
V.~Cirigliano, G.~Ecker, and H.~Neufeld,
\newblock (2001), hep-ph/0109286,
\newblock {Isospin violation and the magnetic moment of the muon}.

\bibitem{Cirigliano2}
V.~Cirigliano, G.~Ecker, and H.~Neufeld,
\newblock Phys. Lett. {\bf B513}, 361 (2001), hep-ph/0104267,
\newblock {Isospin violation and the magnetic moment of the muon}.

\bibitem{Cirigliano3}
V.~Cirigliano, G.~Ecker, and H.~Neufeld,
\newblock JHEP {\bf 08}, 002 (2002), hep-ph/0207310,
\newblock {Radiative tau decay and the magnetic moment of the muon}.

\bibitem{Mexico2}
F.~Flores-Baez, A.~Flores-Tlalpa, G.~Lopez~Castro, and G.~Toledo~Sanchez,
\newblock Phys. Rev. {\bf D74}, 071301 (2006), hep-ph/0608084,
\newblock {Long-distance radiative corrections to the di-pion tau lepton
  decay}.

\bibitem{Mexico1}
A.~Flores-Tlalpa, F.~Flores-Baez, G.~Lopez~Castro, and G.~Toledo~Sanchez,
\newblock Nucl. Phys. Proc. Suppl. {\bf 169}, 250 (2007), hep-ph/0611226,
\newblock {Model-dependent radiative corrections to $\tau^- \to \pi^- \pi^0
  \nu$ revisited}.

\bibitem{Mexico4}
F.~Flores-Baez, G.~L. Castro, and G.~Toledo~Sanchez,
\newblock Phys.Rev. {\bf D76}, 096010 (2007), 0708.3256,
\newblock {The Width difference of rho vector mesons}.

\bibitem{Ghozzi}
S.~Ghozzi and F.~Jegerlehner,
\newblock Phys.Lett. {\bf B583}, 222 (2004), hep-ph/0310181,
\newblock {Isospin violating effects in e+ e- versus tau measurements of the
  pion form-factor $|F_\pi(s)|^2$}.

\bibitem{Fruehwirth}
R.~Fruehwirth, D.~Neudecker, and H.~Leeb,
\newblock EPJ Web of Conferences {\bf 27}, 00008 (2012),
\newblock {Peelle's Pertinent Puzzle and its Solution}.

\bibitem{minuit}
F.~James and M.~Roos,
\newblock Comput. Phys. Commun. {\bf 10}, 343 (1975),
\newblock {Minuit: A System For Function Minimization And Analysis Of The
  Parameter Errors And Correlations}.

\bibitem{Benayoun_isr}
M.~Benayoun, S.~Eidelman, V.~Ivanchenko, and Z.~Silagadze,
\newblock Mod.Phys.Lett. {\bf A14}, 2605 (1999), hep-ph/9910523,
\newblock {Spectroscopy at B factories using hard photon emission}.

\bibitem{AlephCorr}
M.~Davier, A.~Höcker, B.~Malaescu, C.-Z. Yuan, and Z.~Zhang,
\newblock Eur.Phys.J. {\bf C74}, 2803 (2014), 1312.1501,
\newblock {Update of the ALEPH non-strange spectral functions from hadronic
  $\tau$ decays}.

\bibitem{CMD2-1995corr}
CMD-2, R.~R. Akhmetshin {\em et~al.},
\newblock Phys. Lett. {\bf B578}, 285 (2004), hep-ex/0308008,
\newblock {Reanalysis of hadronic cross section measurements at CMD- 2}.

\bibitem{Barkov}
L.~M. Barkov {\em et~al.},
\newblock Nucl. Phys. {\bf B256}, 365 (1985),
\newblock {Electromagnetic Pion Form-Factor in the Timelike Region}.

\bibitem{KLOEComb}
KLOE KLOE-2, V.~De~Leo,
\newblock Acta Phys.Polon. {\bf B46}, 45 (2015), 1501.04446,
\newblock {Measurement of hadronic cross section at KLOE/KLOE-2}.

\bibitem{Heath1998}
M.~Benayoun, H.~B. O'Connell, and A.~G. Williams,
\newblock Phys. Rev. {\bf D59}, 074020 (1999), hep-ph/9807537,
\newblock {Vector meson dominance and the $\rho$ meson}.

\bibitem{ffOld}
M.~Benayoun, P.~David, L.~DelBuono, P.~Leruste, and H.~B. O'Connell,
\newblock Eur. Phys. J. {\bf C29}, 397 (2003), nucl-th/0301037,
\newblock {The pion form factor within the Hidden Local Symmetry model}.

\bibitem{tHooft}
G.~'t~Hooft,
\newblock Phys. Rept. {\bf 142}, 357 (1986),
\newblock {How Instantons Solve the U(1) Problem}.

\bibitem{leutw96}
H.~Leutwyler,
\newblock Phys.Lett. {\bf B374}, 181 (1996), hep-ph/9601236,
\newblock {Implications of $\eta -\eta^\prime$ mixing for the decay $\eta
  \rightarrow 3 \pi$}.

\bibitem{Akhmetshin:2013xc}
CMD-3, R.~Akhmetshin {\em et~al.},
\newblock Phys.Lett. {\bf B723}, 82 (2013), 1302.0053,
\newblock {Study of the process $e^+e^-\to 3(\pi^+\pi^-)$ in the c.m.energy
  range 1.5--2.0 gev with the cmd-3 detector}.

\bibitem{Achasov:2013btb}
M.~Achasov {\em et~al.},
\newblock Phys.Rev. {\bf D88}, 054013 (2013), 1303.5198,
\newblock {Study of $e^+e^- \to \omega\pi^0 \to \pi^0\pi^0\gamma$ in the energy
  range $1.05-2.00$ GeV with SND}.

\bibitem{Lees:2013ebn}
BaBar, J.~Lees {\em et~al.},
\newblock Phys.Rev. {\bf D87}, 092005 (2013), 1302.0055,
\newblock {Study of $e^+e^- \to p \bar{p}$ via initial-state radiation at
  BABAR}.

\bibitem{Lees:2013gzt}
BaBar, J.~Lees {\em et~al.},
\newblock Phys.Rev. {\bf D88}, 032013 (2013), 1306.3600,
\newblock {Precision measurement of the $e^+e^- \to K^+K^-$ cross section with
  the initial-state radiation method at BABAR}.

\bibitem{Lees:2014xsh}
BaBar, J.~Lees {\em et~al.},
\newblock Phys.Rev. {\bf D89}, 092002 (2014), 1403.7593,
\newblock {Cross sections for the reactions $e^+ e^-\to K_S^0 K_L^0$, $K_S^0
  K_L^0 \pi^+\pi^-$, $K_S^0 K_S^0 \pi^+\pi^-$, and $K_S^0 K_S^0 K^+K^-$ from
  events with initial-state radiation}.

\bibitem{Davier:2015bka}
BaBar, M.~Davier,
\newblock Nucl.Part.Phys.Proc. {\bf 260}, 102 (2015),
\newblock {$e^+e^-$ results from BABAR and their impact on the muon $g-2$
  prediction}.

\bibitem{NNLO}
A.~Kurz, T.~Liu, P.~Marquard, and M.~Steinhauser,
\newblock Phys.Lett. {\bf B734}, 144 (2014), 1403.6400,
\newblock {Hadronic contribution to the muon anomalous magnetic moment to
  next-to-next-to-leading order}.

\bibitem{LBL}
J.~Prades, E.~de~Rafael, and A.~Vainshtein,
\newblock (2009), 0901.0306,
\newblock {Hadronic Light-by-Light Scattering Contribution to the Muon
  Anomalous Magnetic Moment}.

\bibitem{Passera06}
M.~Passera,
\newblock Phys.Rev. {\bf D75}, 013002 (2007), hep-ph/0606174,
\newblock {Precise mass-dependent QED contributions to leptonic g-2 at order
  $\alpha^2$ and $\alpha^3$}.

\bibitem{FJ2013}
F.~Jegerlehner,
\newblock Acta Phys.Polon. {\bf B44}, 2257 (2013), 1312.3978,
\newblock {Application of Chiral Resonance Lagrangian Theories to the Muon
  $g-2$}.

\bibitem{LbLNLO}
G.~Colangelo, M.~Hoferichter, A.~Nyffeler, M.~Passera, and P.~Stoffer,
\newblock Phys.Lett. {\bf B735}, 90 (2014), 1403.7512,
\newblock {Remarks on higher-order hadronic corrections to the muon g\u22122}.

\bibitem{Knecht:2014}
M.~Knecht,
\newblock (2014), 1412.1228,
\newblock {The Muon Anomalous Magnetic Moment}.

\bibitem{CODATA2012}
P.~J. Mohr, B.~N. Taylor, and D.~B. Newell,
\newblock Rev. Mod. Phys. {\bf 84}, 1527 (2012), 1203.5425,
\newblock {CODATA Recommended Values of the Fundamental Physical Constants:
  2010}.

\bibitem{CGL}
G.~Colangelo, J.~Gasser, and H.~Leutwyler,
\newblock Nucl.Phys. {\bf B603}, 125 (2001), hep-ph/0103088,
\newblock {$\pi \pi$ scattering}.

\bibitem{Ochs}
G.~Grayer {\em et~al.},
\newblock Nucl. Phys. {\bf B75}, 189 (1974),
\newblock {High Statistics Study of the Reaction $\pi^- p \to \pi^- \pi^+ n$:
  Apparatus, Method of Analysis, and General Features of Results at 17-GeV/c}.

\bibitem{Protopescu}
S.~D. Protopopescu {\em et~al.},
\newblock Phys. Rev. {\bf D7}, 1279 (1973),
\newblock {$\pi \pi$ Partial Wave Analysis from Reactions $\pi^+ p \to \pi^+
  \pi^- \Delta^{++}$ and $\pi^+ p \to K^+ K^- \Delta^{++}$ at 7.1-GeV/c}.

\bibitem{RoyEq}
B.~Ananthanarayan, G.~Colangelo, J.~Gasser, and H.~Leutwyler,
\newblock Phys. Rept. {\bf 353}, 207 (2001), hep-ph/0005297,
\newblock {Roy equation analysis of pi pi scattering}.

\bibitem{BenayounMainz}
M.~Benayoun {\em et~al.},
\newblock (2014), 1407.4021,
\newblock {Hadronic contributions to the muon anomalous magnetic moment
  Workshop. $(g-2)_{\mu}$: Quo vadis? Workshop. Mini proceedings}.

\bibitem{Boyle:2011hu}
P.~Boyle, L.~Del~Debbio, E.~Kerrane, and J.~Zanotti,
\newblock Phys. Rev. {\bf D85}, 074504 (2012), 1107.1497,
\newblock {Lattice Determination of the Hadronic Contribution to the Muon $g-2$
  using Dynamical Domain Wall Fermions}.

\bibitem{Aubin:2013daa}
C.~Aubin, T.~Blum, M.~Golterman, and S.~Peris,
\newblock Phys. Rev. {\bf D88}, 074505 (2013), 1307.4701,
\newblock {Hadronic vacuum polarization with twisted boundary conditions}.

\bibitem{deDivitiis:2012vs}
G.~M. de~Divitiis, R.~Petronzio, and N.~Tantalo,
\newblock Phys. Lett. {\bf B718}, 589 (2012), 1208.5914,
\newblock {On the extraction of zero momentum form factors on the lattice}.

\bibitem{Aubin:2012me}
C.~Aubin, T.~Blum, M.~Golterman, and S.~Peris,
\newblock Phys. Rev. {\bf D86}, 054509 (2012), 1205.3695,
\newblock {Model-independent parametrization of the hadronic vacuum
  polarization and g-2 for the muon on the lattice}.

\bibitem{Feng:2013xsa}
X.~Feng {\em et~al.},
\newblock Phys. Rev. {\bf D88}, 034505 (2013), 1305.5878,
\newblock {Computing the hadronic vacuum polarization function by analytic
  continuation}.

\bibitem{Francis:2014dta}
A.~Francis {\em et~al.},
\newblock (2014), 1411.3031,
\newblock {Lattice QCD Studies of the Leading Order Hadronic Contribution to
  the Muon $g-2$}.

\bibitem{DellaMorte:2014rta}
M.~Della~Morte {\em et~al.},
\newblock PoS {\bf LATTICE2014}, 162 (2014), 1411.1206,
\newblock {Study of the anomalous magnetic moment of the muon computed from the
  Adler function}.

\bibitem{Malak:2015sla}
Budapest-Marseille-Wuppertal, R.~Malak {\em et~al.},
\newblock PoS {\bf LATTICE2014}, 161 (2015), 1502.02172,
\newblock {Finite-volume corrections to the leading-order hadronic contribution
  to $g_\mu-2$}.

\bibitem{deRafaelBell}
J.~S. Bell and E.~de~Rafael,
\newblock Nucl. Phys. {\bf B11}, 611 (1969),
\newblock {Hadronic vacuum polarization and g(mu)-2}.

\bibitem{deRafaelOrg}
E.~de~Rafael,
\newblock Phys. Lett. {\bf B322}, 239 (1994),
\newblock {Hadronic contributions to the muon g-2 and low-energy QCD}.

\bibitem{Francis:2013qna}
A.~Francis, B.~Jaeger, H.~B. Meyer, and H.~Wittig,
\newblock Phys. Rev. {\bf D88}, 054502 (2013), 1306.2532,
\newblock {A new representation of the Adler function for lattice QCD}.

\bibitem{Bernecker:2011gh}
D.~Bernecker and H.~B. Meyer,
\newblock Eur. Phys. J. {\bf A47}, 148 (2011), 1107.4388,
\newblock {Vector Correlators in Lattice QCD: Methods and applications}.

\end{thebibliography}

\end{document}